 \documentclass[12pt]{article}

 \pdfoutput=1

\usepackage[round]{natbib}   
\usepackage{color}

\usepackage{graphicx} 
\usepackage{amsmath}

\usepackage{enumitem}

\usepackage{float}
\usepackage{ifthen}
\usepackage{hyperref}

\usepackage{titlesec}

\titlespacing*{\section}{0pt}{0.2\baselineskip}{\baselineskip}

\usepackage[english]{babel}

\usepackage{mathtools}
\usepackage{empheq}

\begin{document}



\begin{center}
{\bf \Large 
A summary of several meteorological properties
}
\\
\vspace*{2mm}
{\bf \Large 
of the moist-air entropy variables $\theta_s$ and $PV(\theta_s)$.
}
\\
\vspace*{3mm}
{\large by Pascal Marquet \\
 \vspace*{2mm}
 (retired from Météo-France, 
 CNRM/GMAP, Toulouse, France)} \\ 
\vspace*{3mm}
{\large Version 3 of \today.}\\ 
\vspace*{3mm}
{\large \it E-mail: pascalmarquet@yahoo.com} \\
\vspace*{1mm}
\end{center}
\vspace{0mm}


 \section{\underline{\large Introduction - Motivations${}_{ }$}}
 \label{==INTRO==}
\vspace{-3mm}



The aim of this note is to describe several meteorological properties shown in \cite{Marquet_2022} for the moist-air specific entropy, the associated potential temperature ($\theta_s$) defined in \cite{Marquet_QJ_2011} and the associated potential vorticity $PV(\theta_s)$ defined in \cite{Marquet_QJ_2014}.

The three Tables~\ref{table_const1} to \ref{table_const3} provide the numerical values of all thermodynamic constants used in this note. 
\vspace*{-5mm}
\begin{table}[!h]
\caption{Dry- and moist-air thermodynamic constants} 
\centering 
\vspace*{1mm}
\begin{tabular}{|c |c |c |c |c |c |c |c |c |c|}
\hline
$R_d$ & 
$R_v$ & 
${c}_{pd}$ & 
${c}_{pv}$ & 
${c}_l$ & 
${c}_i$ &
$L_v(T_0)$ &
$L_s(T_0)$
\\
\hline
$287.06$ & 
$461.52$ & 
$1004.7$ & 
$1846.1$ &
$4218$ &
$2106$ &
$2501$ &
$2835$
\\
\hline 
J/K/kg & 
J/K/kg & 
J/K/kg & 
J/K/kg & 
J/K/kg & 
J/K/kg & 
kJ/kg & 
kJ/kg
\\
\hline
\end{tabular}
\label{table_const1} %
\end{table}
\vspace*{-7mm}
\begin{table}[!h]
\caption{Standard and reference values}
\centering 
\vspace*{1mm}
\hspace*{-2mm}
\begin{tabular}{|c | c | c |c ||c |c |c |c |c |c |c|}
\hline
\!$T_0=T_r$\! &
\!$p_0=p_r$\! &
\!$s_{d0}(T_0,p_0)$\! & 
\!$s_{v0}(T_0,p_0)$\! &
\!$e_r=e_s(T_r)$\! & 
\!$p_{dr}=p_r\!-\!e_r$\! &
\!$s_{dr}(T_r,p_{dr})$\! & 
\!$s_{vr}(T_r,e_r)$\! &
$r_r$
\\
\hline
$273.15$ &
 $1000$  &
 $6775$ & 
$10320$ &
\!$6.1064$\! & 
\!$993.89$\! & 
 $6777$ & 
$12673$ &
$3.82$
\\
\hline 
K &
hPa &
J/K/kg & 
J/K/kg &
 hPa & 
 hPa & 
J/K/kg & 
J/K/kg &
\!\!g/kg\!\! 
\\
\hline
\end{tabular}
\label{table_const2} %
\end{table}
\vspace*{-7mm}
\begin{table}[!h]
\caption{Non-dimensional derived constants} 
\vspace*{1mm}
\centering 
\begin{tabular}{|c | c | c |c |c |c |c |c|}
\hline
$\eta$ & 
$\varepsilon$ & 
$\gamma$  & 
$\kappa$ & 
$\lambda$ & 
$\delta$& 
$\Lambda_r$ \\
\hline
$1.608$ & 
$0.622$ & 
$0.4594$ & 
$0.2857$ & 
$0.8375$ & 
$0.6078$ &  
$5.869 \pm 0.03$
\\
\hline
\end{tabular}
\label{table_const3} %
\end{table}

The way the old dry-air potential temperature ($\theta$) was defined in the nineteenth century is recalled in section~\ref{==Dry-POT-TEMPE==}.
The definitions of other moist-air potential temperatures 
($\theta$, $\theta'_w$, $\theta_e$, 
 $\theta_v$, $\theta_l$, $\theta_{es}$,
 $\theta_{il}$, $\theta'_e$) are provided in section~\ref{==Moist-POT-TEMPE==}.
The definition of the new moist-air entropy potential temperature ($\theta_s$) published in \cite{Marquet_QJ_2011} is recalled in section~\ref{==Moist-ENROPY-POT-TEMPE==}, before to be compared to the other potential temperatures in section~\ref{==Compare_ThetaX==}.

The moist-air entropy potential vorticity $PV(\theta_s)$ published in \cite{Marquet_QJ_2014} is recalled in section~\ref{==Moist-PV==}, together with the other previous definitions $PV(\psi)$ with 
$\psi = \theta$, $\theta'_w$, $\theta_e$, 
 $\theta_v$ or $\theta_{es}$. 
The section~\ref{==Compare_PV==} shows comparisons between the potential vorticity functions $PV(\theta)\approx PV(\theta_v)$, $PV(\theta_s)$ and $PV(\theta_e)$, with in particular regions of negative values of $PV(\theta_s)$ close to the cold fronts that seems to indicate regions of symmetric instability.

General conclusions are provided in section~\ref{==Conclusions==}.


\newpage
 \section{\underline{\large The old dry-air potential temperature ($\theta$)}}
 \label{==Dry-POT-TEMPE==}
\vspace{-3mm}

\cite{Bezold_1888b} with \cite{Helmholtz_1888} coined the term ``potential temperature'' for the dry-air quantity:
\vspace{-2mm}
\begin{equation}
\boxed{ \;
\theta \,(T,p)
\; = \; T \:
\left( \frac{p_0}{p} \right)^{R_d/c_{pd}}
\; = \; T \:
\left( \frac{p_0}{p} \right)^{\kappa}
\; }
\: ,
\end{equation}
where $T$ is the absolute temperature, 
$p$ the pressure, 
$p_0=1000$~hPa a standard value,
$R_d$ the dry-air gas constant,
$c_{pd}$ the dry-air specific heat at constant pressure
and $\kappa = R_d/c_{pd}$.

The link between the dry-air potential temperature ($\theta \,$) and the dry-air specific entropy ($s_d$) was derived by \citet{Bauer_1908,Bauer_1910}, through a relationship which can be written as
\begin{align}
& \boxed{ \;
s_d \; = \; c_{pd} \: \ln\left( \theta \right) 
\;} \;
\: + \: 
\left[ \: 
 s_{d0}(T_0,p_0) 
 \: - \: 
 c_{pd} \: \ln\left( T_0 \right) 
 \: \right]
\; = \;
c_{pd} \: \ln\left( \frac{\theta}{\theta_0} \right)
\: + \: 
 s_{d0}(T_0,p_0) 
\: 
\label{eq_def_s_theta} 
\\
\mbox{and thus} \;\; &
\boxed{\;\;
s_d \; = \;
c_{pd} \: 
\ln\left({\theta}\right)
\: + \: s_{0}(p_0) 
 \;\;}
\;,\;\;\;\mbox{where}\;\;\;
s_0(p_0) =
 s_{d0}(T_0,p_0) 
 \: - \: 
 c_{pd} \: \ln\left( T_0 \right) 
\: .
\vspace{-2mm}
\label{eq_def_s_theta_bis} 
\end{align}
The dry-air specific heat $c_{pd}$, the reference (standard) temperature $T_0 = \theta_0 = \theta(T_0,p_0)$ and the standard dry-air entropy $s_{d0}(T_0,p_0)$ are all constant terms.
Moreover, the quantity $s_0(p_0)$ in (\ref{eq_def_s_theta_bis}) only depends on the reference standard pressure $p_0=1000$~hPa, because the changes of $s_{d0}(T_0,p_0)$ with $T_0$ are balanced (and cancelled) by the second term $- \: c_{pd} \: \ln\left( T_0 \right)$. 


 \section{\underline{\large The old moist-air potential temperatures ($\theta'_w$, $\theta_v$, $\theta_e$, $\theta_l$, $\theta_{es}$, $\theta_{il}$, $\theta'_e$)}}
 \label{==Moist-POT-TEMPE==}
\vspace{-3mm}

\cite{Bezold_1888a}  coined the definition of the ``pseudo-adiabatic'' processes, corresponding to the wet-bulb potential temperature ($\theta'_w$) and defined by integration of saturation moist-air differential equations, without analytic expression for 
\vspace{-2mm}
\begin{equation}
\boxed{ \;
\theta'_w 
\; }
\: .
\label{eq_theta_pw}
\end{equation}

The concept of ``virtual potential temperature'' was coined by \citet{Guldberg_Mohn_1876,Guldberg_Mohn_1910}, with the more general definition of it given by
\begin{equation}
\boxed{ \;
\theta_v 
\; = \; \theta 
\left[ \:
1 
+ \left( \frac{R_v}{R_d} -1 \right) q_v 
\: - \: (q_l+q_i) 
\: \right]   
\; }
\label{eq_theta_v}
\: ,
\end{equation}
where $R_v$ is the water-vapor gas constant and
$q_v$, $q_l$ and $q_i$ are the specific content for water vapor, liquid water and ice species, respectively.

The difference between the wet-bulb ($\theta'_w$) and equivalent ($\theta_e$) potential temperatures was studied by \citet{Normand_1921}, \citet{Stuve_1927}, \citet{Robitzsch_1928}, \citet{Rossby_1932}, \citet{Bleeker_QJ_1939}, among others, with the equivalent potential temperature defined by quantities close to the first-order approximation
\begin{equation}
\boxed{ \;
\theta_e  
\; \approx \; \theta \:
\exp\!\left( \frac{L_v \: q_v}{c_{pd} \: T} \right)
\; }
\: ,
\label{eq_theta_e}
\end{equation}
where $L_v$ is the latent heat of vaporization.

\citet{Hewson_1936,Hewson_1938} studied the application of the wet-bulb potential temperature ($\theta'_w$) to air mass analysis, with its meaning and significance studied in  \citet{Hewson_1946a,Hewson_1946b}.

The same differential equations derived in  \cite{Bezold_1888a} are used in \cite{Saunders_QJ_1957} to define the adiabatic and pseudo-adiabatic equations.

The three quantities $\theta'_w$, $\theta_v$ and $\theta_e$ were considered in \cite{Lilly_QJ_1968}

As a companion of $\theta_e$, the ``liquid-water potential temperature'' ($\theta_l$) was defined and studied in \cite{Betts_QJ_1973}, with the first-order approximation relationship
\begin{equation}
\boxed{ \;
\theta_l  
\; \approx \; \theta \:
\exp\!\left( - \: \frac{L_v \: q_l}{c_{pd} \: T} \right)
\; }
\: .
\label{eq_thetal_B73}
\end{equation}
Similar values of $\theta_l$ and/or $\theta_e$ were derived and studied by 
\cite{Emanuel_1994,
Pauluis_al_2008_Science,
Pauluis_Czaja_Korty_2010,
Laliberte_al_2013,
Pauluis_Mrowiec_2013,
Pauluis_2016,
Mrowiec_al_2016,
Pauluis_Zhang_2017,
Fang_Pauluis_Zhang_2017,
Dauhut_al_JAS_2017,
Fang_Pauluis_Zhang_2019}.

Empirical formula for the saturation pseudo-adiabats and the ``saturation equivalent potential temperature'' ($\theta_{es}$) were derived in
\cite{Betts_Dugan_1973}, leading to the first-order approximation relationship
\begin{equation}
\boxed{ \;
\theta_{es} 
\; \approx \; \theta \:
\exp\!\left( \frac{L_v \: r_{sw}}{c_{pd} \: T} \right)
\; }
\: ,
\label{eq_theta_es}
\end{equation}
where $r_{sw}$ is the saturation water-vapor mixing ratio.

The ``ice-liquid water potential temperature'' ($\theta_{il} $) defined in  \cite{Tripoli_Cotton_1981} can be considered as a generalization of the \cite{Betts_QJ_1973} ``liquid-water potential temperature'' ($\theta_l$) to the case of ice ($q_i \neq 0$), with the relationship
\begin{equation}
\boxed{ \;
\theta_{il}  
\; \approx \; 
\theta \:
\exp\!\left( - \: \frac{L_v \: q_l + L_s \: q_i}{c_{pd} \: T} \right)
\; }
\: ,
\label{eq_theta_il_a}
\end{equation}
where $L_s$ is the latent heat of sublimation.

The \cite{Betts_QJ_1973} conservative quantity $\theta_l$  is nowadays used in most of parameterization of atmospheric turbulence \citep[ex. based on:][etc]{Deardorff80,Mellor_Yamada_1982,CBR_2000},
and to study the subgrid-condensation schemes 
\citep[ex. those based on:][etc]{Sommeria_Deardorff_1977,Bougeault_1982}, 
with both $\theta_l$ and $\theta_e$ usually considered to study the cloud-top entrainment instability
\citep[ex. those based on:][etc]{Deardorff80,Mac_Vean_Mason_90,Lock_QJ_2004}.

The equivalent potential temperature computed in \cite{Bolton_1980}, \cite{Bryan_2008} and \cite{Davies_Jones_2009}, among others, are different from the ``adiabatic'' version  $\theta_e$ approximated by Eq.~(\ref{eq_theta_e}), and rather corresponds to the ``pseudo-adiabatic'' value $\theta'_w$, which is not analytically defined in Eq.~(\ref{eq_theta_pw}).
As an example, the (adjusted up to about $0.03$~K?) approximation derived in Eqs.~(21), (22), (24) and (39) of \cite{Bolton_1980} correspond to
\begin{align}
T_L & 
\: \approx \; 55 \: + \:
\frac{2840}{3.5 \: \ln(T) - \ln(e/100) - 4.805
\: } 
\; \approx \; 55 \: + \:
{\displaystyle
\frac{1}{\displaystyle
\frac{1}{T-55} \: - \: \frac{\ln(H_l)}{2840}
}
}
\: ,
\label{eq_theta_pe1}
\\
\theta_{DL} & \: \approx \; 
T
\left( \frac{p_0}{p-e} \right)^{0.2854}
\left( \frac{T}{T_L} \right)^{0.28\:r_v}
\: ,
\label{eq_theta_pe2}
\\
&
\boxed{ \;
\theta'_e 
\; \approx \; 
\theta_{DL} \:
\exp\!\left[ \: 
\left( \frac{3036}{T_L} - 1.78 \right)
\:
\left(1 + 0.448 \: r_v\right)
\:
r_v
\: \right]
\; }
\: ,
\label{eq_theta_pe3}
\end{align}
where $T$ is the absolute temperature (K),
$p$ the total pressure (Pa),
$e$ the water-vapor partial pressure (Pa),
$H_l=e/e_{sw}$ the relative humidity (Pa/Pa), and 
$r_v = q_v/(1-q_t)$ the water-vapor mixing ratio (kg/kg).

This formulation of \cite{Bolton_1980} is different from $\theta_e$ given by (\ref{eq_theta_e}), and especially in non-saturated regions where $H_l \neq 1$ and $T_L \neq T$, since:
\vspace*{0.mm}
\\ $\bullet$ 
$\theta_{DL} \neq  \theta = T \: (p_0/p)^\kappa
 \approx T \: (p_0/p)^{0.2857}$, because $T_L \neq T$, $p-e \neq p$ and $0.2854 \neq 0.2857$ in (\ref{eq_theta_pe2});
\vspace*{1mm}
\\ $\bullet$ 
${3036}/{T_L} - 1.78 
\neq
 \left[\:
 L_v(T_0)
 + (c_{pv}-c_l) (T-T_0)
 \:\right]/({c_{pd} \: T})
\approx
3134/T - 2.36
$ in (\ref{eq_theta_pe3}),
\vspace*{1mm}
\\ $\bullet$ 
even if $\left(1 + 0.448 \: r_v\right) \: r_v
\approx
[\:
 q_v
\:]
$
because
$r_v=q_v/(1-q_v)\approx q_v$ and 
$\left(1 + 0.448 \: r_v\right) \approx 1$.

 \section{\underline{\large The moist-air entropy potential temperature ($\theta_s$)}}
 \label{==Moist-ENROPY-POT-TEMPE==}
\vspace{-2mm}

The more recent ``moist-air entropy potential temperature'' ($\theta_s$) published in \cite{Marquet_QJ_2011} 
can be considered as an improvement of the quantity $\theta_s$ already derived in \cite{Hauf_Hoeller_1987} and \cite{Marquet_1993}.

The relationship between the moist-air specific entropy ($s$) and the corresponding potential temperature ($\theta_s$) was assumed in \cite{Marquet_QJ_2011} to be the same as the one derived by \citet{Bauer_1908,Bauer_1910} for the dry-air values ($s_d$ and $\theta$), leading to
\vspace{-2mm}
\begin{align}
&\boxed{\;\;
s \; = \;
c_{pd} \: 
\ln\left(\frac{\theta_{s}}{\theta_0}\right)
\: + \: s_{d0}(T_0,p_0) 
 \;\;}
\; = \; 
c_{pd} \: \ln\left( \theta_s \right) 
\: + \: 
\left[ \: 
 s_{d0}(T_0,p_0) 
 \: - \: 
 c_{pd} \: \ln\left( T_0 \right) 
 \: \right]
\: ,
\vspace{-2mm}
\label{eq_def_s_thetas} \\
\mbox{and} \;\; &
\boxed{\;\;
s \; = \;
c_{pd} \: 
\ln\left({\theta_{s}}\right)
\: + \: s_{0}(p_0) 
 \;\;}
\;,\;\;\;\mbox{where}\;\;\;
s_0(p_0) =
 s_{d0}(T_0,p_0) 
 \: - \: 
 c_{pd} \: \ln\left( T_0 \right) 
\: .
\vspace{-2mm}
\label{eq_def_s_thetas_bis} 
\end{align}
\noindent
The specific heat capacity (at constant pressure) $c_{pd}$, 
the reference (standard) temperature 
$T_0 = \theta_0 = \theta(T_0,p_0)$ and 
the reference dry-air entropy $s_{d0}(T_0,p_0)$ are all constant.
Moreover, the quantity $s_0(p_0)$ in (\ref{eq_def_s_thetas_bis}) only depends on the same reference standard pressure $p_0=1000$~hPa as the one introduced to define the dry-air potential temperature $\theta$ in (\ref{eq_def_s_theta_bis}), because the changes of $s_{d0}(T_0,p_0)$ with $T_0$ are balanced (and cancelled) by the second term $- \: c_{pd} \: \ln\left( T_0 \right)$. 

These properties (\ref{eq_def_s_thetas}) and (\ref{eq_def_s_thetas_bis}) show that $\theta_s$ becomes a true synonym for the specific moist-air entropy $s$.
This property did not hold in the previous formulations of 
$\theta_s$ and $\theta^{\ast}$ derived in \cite{Hauf_Hoeller_1987} and \cite{Marquet_1993},
nor in those of $\theta_l$ and $\theta_e$ derived in
\cite{Betts_QJ_1973}, \cite{Emanuel_1994} or \cite{Pauluis_Czaja_Korty_2010}.

A general version of $\theta_s$, which generalizes those of 
\cite{Marquet_QJ_2011,Marquet_JAS_2017}, can be written from (\ref{def_ths_e}) as:
\vspace{-2mm}
\begin{equation}
\boxed{
\begin{array}{rcl}
  {\theta}_s 
   & = & \: \theta \; \:
   {\displaystyle
    \exp\! \left[ \: 
    - \: \frac{L_v \: (q_l+q_{\rm rain}) 
             + L_s \: (q_i+q_{\rm snow})}{c_{pd} \: T} 
    \: \right]
       \:
    \exp\! \left(  \Lambda_r \: q_t  \right)
   }
\vspace{2mm}
\\
  & &  \; \; \; \times \;
   {\displaystyle
        \left( \frac{T}{T_r}\right)^{\!\!\lambda \,q_t}
  \left( \frac{p_r}{p}\right)^{\!\!\kappa \,\delta \,q_t}
  \left( \frac{r_r}{r_v} \right)^{\!\!\gamma\,q_t}
  \;  \frac{(1\!+\!\eta\,r_v)^{\,\kappa \, (1+\,\delta \,q_t)}}
           {(1\!+\!\eta\,r_r)^{\,\kappa \,\delta \,q_t}}
   }
\vspace{2mm}
\\
  & &  \; \; \; \times \;
   {\displaystyle
     \; {(H_l)}^{\, \gamma \, (q_l+q_{\rm rain})} \;
     \; {(H_i)}^{\, \gamma \, (q_i+q_{\rm snow})}
     \; {\left( \frac{T_{\rm rain}}{T} 
        \right)}^{\! c_l\, q_{\rm rain}/c_{pd}}
     \; {\left( \frac{T_{\rm snow}}{T} 
        \right)}^{\! c_i\, q_{\rm snow}/c_{pd}}
   }
\label{eq_thetas} \: 
\end{array}
}
\end{equation}
where $\theta = T \:(p_0/p)^{R_d/c_{pd}}$ is the dry-air potential temperature.
The relative humidities are noted $H_l = e/e_{sw}$ and $H_i = e/e_{si}$. 
The temperatures of precipitations (rain and snow) are noted $T_{\rm rain}$ and $T_{\rm snow}$.
See the Tables~\ref{table_const1} to \ref{table_const3} for the other thermodynamic constants.
\vspace*{-4mm}

 \begin{center} \rule[0mm]{7.cm}{0.1mm} \end{center}
\vspace*{-2mm}

Differently, the potential temperature $\theta_{s/\rm{HH87}}$  defined in
\citet[see Eqs.~3.23 to 3.25 page 2893]{Hauf_Hoeller_1987} 
was linked to the specific entropy through the relationship
\vspace*{-1mm}
\begin{align}
\!\!\!\!
 s 
 & \: = \;
q_d \; c^{\ast}_{pe}
 \;
\ln\!\left(\frac{\theta_{s/\rm{HH87}}}{T_0}\right)
\; + \;
q_d \; s^{\ast}
\: ,
\label{eq_HH87} \\
\!\!\!\!
\mbox{with} \; \; \;
  \theta_{s/\rm{HH87}}
   & = \:  \: T \;
 \; \left( \frac{p-e}{p_0}\right)^{\!\!-R_d/c^{\ast}_{pe}}
    \exp\! \left( 
        \frac{L_{v} \: r_v - L_{s} \: r_i}{c^{\ast}_{pe} \: T}
     \: \right)  \; \; 
     \; {\left( H_l \right)}^{\,-R_v\: r_v/c^{\ast}_{pe}}
   \;\; {\left( H_i \right)}^{\,R_v\: r_i/c^{\ast}_{pe}}
\label{eq_thetas_HH87} \: .
\end{align}
\citet[][p.2890]{Hauf_Hoeller_1987} explained that, in (\ref{eq_HH87}) and (\ref{eq_thetas_HH87}): 
``{\it $T_0$ and $p_0$ denote reference values of $\,T$ and $p$, and 
$s_d(T_0,p_0)$, $s_v(T_0,p_0)$, $s_l(T_0)$ and $s_i(T_0)$ 
are the values of dry-air and water entropies at this reference point.\/}''
This means that we must consider that $T_r=T_0$ and $p_r=p_0$ in \citet{Hauf_Hoeller_1987}, without justification.

In (\ref{eq_HH87}) both 
$q_d  \: c^{\ast}_{pe} = (1-q_t) \: c_{pd} \: + \: q_t \: c_l$ 
and 
$q_d \: s^{\ast} = \left[ \: 
(1-q_t) \: s_d(T_0,p_0) \: + \: q_t \: s_l(T_0) 
\: \right]
$
depend on $q_t$.
Therefore, the first drawback is that $\theta_{s/\rm{HH87}}$ cannot represent the moist-air entropy ($s$) unless the total-water specific content $q_t=r_t/(1+r_t)$ is a constant.
The words of \citet[][p.2893]{Hauf_Hoeller_1987} were:
 ``{\it The entropy temperature $\theta_{s/\rm{HH87}}$ is related to entropy ($s$) by the functional relationship (\ref{eq_HH87}), and thus can be considered as a measure of entropy. However, Eq.~(\ref{eq_HH87}) involved, beside entropy, the concentration of dry air $q_d = 1 - q_t$ and the total water mixing ratio $r_t=q_t/(1-q_t$).
 Therefore, a constant entropy does not generally imply that the entropy temperature  $\theta_{s/\rm{HH87}}$ is also a constant.\/}''

The other drawback is the term 
${( H_l )}^{\,-R_v\: r_v/c^{\ast}_p}$ in (\ref{eq_thetas_HH87})
that is different from $1$ in unsaturated regions, where the relative humidity $H_l = e/e_{sw} < 1$ but $r_v > 0$ is different from $0$.
The last term ${( H_i )}^{\,R_v\: r_i/c^{\ast}_{pe}}$ is most often equal to 1 for both unsaturated ($r_i=0$) and saturated regions ($H_i=1$).
This issue 
\citep[see][p.2893]{Hauf_Hoeller_1987} is due to
``{\it the definition of entropy teperature which presumes a system of cloudy air and especially the existence of liquid water, as can be seen from 
$q_d  \: c^{\ast}_{pe} = (1-q_t) \: c_{pd} \: + \: q_t \: c_l$ 
and 
$q_d \: s^{\ast} = \left[ \: 
(1-q_t) \: s_d(T_0,p_0) \: + \: q_t \: s_l(T_0) 
\: \right]
$,\/}'' 
which both depend on $c_l$ and $s_l(T_0)$.
``{\it 
Problems arise if no liquid water is present, either locally or in principle.\/}''
\vspace*{-2mm}

 \begin{center} \rule[0mm]{7.cm}{0.1mm} \end{center}
\vspace*{-2mm} 

The potential temperature $\theta^{\ast}_{\rm M93}$ defined in \citet[before Eq.42 and Eq.43]{Marquet_1993} partially solved the issues of \citet{Hauf_Hoeller_1987}, with $\theta^{\ast}_{\rm M93}$ linked to the specific entropy through the relationship
\vspace*{-1mm}
\begin{align}
 s 
 & \: = \;
q_d\; c^{\ast}_{pl}
 \:
\ln\!\left(\frac{\theta^{\ast}_{\rm M93}}{\theta^{\ast}_0}\right)
\; + \;
\left[ \: 
   (1-q_t)   \: s_d(T_0,p_0-e_0) 
 \: + \: q_t \: s_v(T_0,    e_0) 
\: \right]
\: ,
\label{eq_M93}
\end{align}
with $e_0=e_{sat}(T_0)$ and
\vspace*{-1mm}
\begin{align}
\!\!\!\!
  \theta^{\ast}_{\rm M93}
   & = \:   T 
 \; \left( \frac{p-e}{p_0}\right)^{\!\!-R_d/c^{\ast}_{pl}}
    \exp\! \left( 
        - \frac{L_{v} \: r_l + L_{s} \: r_i}{c^{\ast}_{pl} \: T}
     \: \right) 
 \; \left( \frac{e}{p_0}\right)^{\!\!-R_v\: r_t/c^{\ast}_{pl}}
     \; {\left( H_l \right)}^{\,R_v\: r_l/c^{\ast}_{pl}}
   \;\; {\left( H_i \right)}^{\,R_v\: r_i/c^{\ast}_{pl}}
\label{eq_theta_M93} \: .
\end{align}
Like in (\ref{eq_HH87}), both 
$q_d \; c^{\ast}_{pl} = (1-q_t) \; c_{pd} \: + \: q_t \; c_{pv}$
and the bracketed terms (with the reference entropies) depend on $q_t$ in (\ref{eq_M93}), and therefore $\theta_s^{\ast}$ cannot always represent the moist-air entropy ($s$) in (\ref{eq_M93}) unless the total-water specific content $q_t=r_t/(1+r_t)$ is a constant. 

However, the first advantage is the term  $c_l$ used in
\citet{Hauf_Hoeller_1987} that is replaced by $c_{pv}$ in $\theta^{\ast}_{\rm M93}$, with the more comfortable idea that there is always some water-vapor content in real parcels of moist air, whereas there was no guaranty of existing liquid water in the formulation of \citet{Hauf_Hoeller_1987}.

The other advantage is the symmetric use of the two mixing ratios ($r_l$, $r_i$) in the exponential, instead of the use of ($r_v$, $r_i$) with two different signs in \citet{Hauf_Hoeller_1987}, with also the possibility to neglect the last two terms 
${( H_l )}^{\,R_v\: r_l/c^{\ast}_p}$ and
${( H_i )}^{\,R_v\: r_i/c^{\ast}_p}$  
in (\ref{eq_theta_M93}) for both unsaturated (since $r_l=r_i=0$) and saturated regions (if $H_l=H_i=1$).
\vspace*{-2mm}

 \begin{center} \rule[0mm]{7.cm}{0.1mm} \end{center}
\vspace*{-2mm} 

In fact, the formulation (\ref{eq_thetas_HH87}) of 
\cite{Hauf_Hoeller_1987} corresponds to the equivalent potential temperature derived in \cite{Emanuel_1994} and \cite{Pauluis_Czaja_Korty_2010}, but with the ice content included.
Therefore : 
$\theta_{s/\rm{HH87}}
\approx 
\theta_{e/\rm{E94}}
\approx 
\theta_{e/\rm{P10}}
$,
but with no citation to the JAS paper of \cite{Hauf_Hoeller_1987} by K. Emanuel nor by O. Pauluis.

The words of \citet[][p.2895]{Hauf_Hoeller_1987} were: 
``{\it The equivalent potential temperature is a constant in a closed adiabatic system where water vapor is saturated.
If in a rising parcel of cloudy air the liquid water remains in the parcel and does not fall out, then $\theta_e$ is a constant.\/}''
And then \citep[][p.2896]{Hauf_Hoeller_1987}:
``{\it As a result, we see that for a saturated warm cloud the entropy temperature\/}'' ($\theta_{s/\rm{HH87}}$) ``{\it is identical with the equivalent potential temperature $\theta_e$\/}'' 
(i.e. this equivalence is valid only if $q_t$ is a constant).

Moreover 
$\theta_{s/\rm{HH87}} = \theta_{e/\rm{HH87}}$ can be approximated by $\theta_e$ of \cite{Betts_QJ_1973} given by (\ref{eq_theta_e}), which can be considered as a first-order approximation of the more general formulations of
\cite{Hauf_Hoeller_1987}, \cite{Emanuel_1994} and \cite{Pauluis_Czaja_Korty_2010}. 
This is the reason why the ``first-order approximation'' $\theta_e$ of \cite{Betts_QJ_1973} given by (\ref{eq_theta_e}) is used in almost all practical applications to study convection and is considered as a ``conservative variable'' in almost all moist-air turbulence schemes.
The words of \citet[][p.2895]{Hauf_Hoeller_1987} were:
 ``{\it other definitions of equivalent potential temperature exist \citep[e.g.][]{Betts_QJ_1973}, which arise from approximations of the definition of\/}'' $\theta_{s/\rm{HH87}}$.
 ``{\it The present one\/}'' (i.e. $\theta_{s/\rm{HH87}} \approx \theta_{e/\rm{HH87}}$ for constant $q_t$) ``{\it is the most general definition, as no further approximations or assumptions have been made.\/}'' 

The main result is that the equivalent potential temperatures $\theta_e$ derived in \cite{Betts_QJ_1973}, \cite{Hauf_Hoeller_1987}, \cite{Emanuel_1994} and \cite{Pauluis_Czaja_Korty_2010} cannot represent the moist-air entropy ($s$) unless the total-water specific content $q_t=r_t/(1+r_t)$ and the mixing ratio $r_t = q_t/(1-q_t)$ are constant. 
Only $\theta_s$ given by (\ref{eq_def_s_thetas}) is really synonymous with the specific moist-air entropy, due to the relationship (\ref{eq_thetas}) where both 
$c_{pd}$ and $s_{d0}(T_0,p_0)$ are constant.
\vspace*{-2mm}

 \begin{center} \rule[0mm]{7.cm}{0.1mm} \end{center}
\vspace*{-2mm}

The same is true for the liquid-water potential entropy $\theta_l$ that can be written, following the same method described in \cite{Hauf_Hoeller_1987}, as
\vspace*{-1mm}
\begin{align}
 s 
 & \: = \;
 q_d \: c^{\ast}_{pl}
 \;
\ln\!\left(\frac{\theta_{l}}{T_0}\right)
\; + \;
\left[ \: 
(1-q_t) \: s_d(T_0,p_0) \: + \: q_t \: s_v(T_0,e_0) 
\: \right]
\: ,
\label{eq_HH87_theta_l}
\end{align}
where
\begin{align}
  \theta_{l}
   & = \:  \: T \;
 \; \left( \frac{p-e}{p_0}\right)^{\!\!-R_d/c^{\ast}_{pl}}
    \exp\! \left( -\:
        \frac{L_{v} \: r_l + L_{s} \: r_i}{c^{\ast}_{pl} \: T}
     \: \right)  \; \; 
 \; \left( \frac{e}{e_0}\right)^{\!\!-r_t \:R_v/c^{\ast}_{pl}}
     \; {\left( H_l \right)}^{\,R_v\: r_l/c^{\ast}_{pl}}
   \;\; {\left( H_i \right)}^{\,R_v\: r_i/c^{\ast}_{pl}}
\label{eq_thetal_HH87} \: .
\end{align}
Both $(1-q_t) \; c^{\ast}_{pl} = 
(1-q_t) \; c_{pd} \: + \: q_t \; c_{pv}$ 
and the bracketed terms of (\ref{eq_HH87_theta_l}) depend on $q_t$, and both prevent $\theta_l$ to represent the moist-air entropy in all conditions if $q_t$ and $r_t$ are not constant. 

The last terms of (\ref{eq_HH87_theta_l})  ${( H_l )}^{\,R_v\: r_l/c^{\ast}_{pl}}$ 
and ${( H_i )}^{\,R_v\: r_i/c^{\ast}_{pl}}$ are most often equal to 1 for both unsaturated (because $r_l=r_i=0$) and saturated regions (if either $H_l=1$ or $H_i=1$).

However, the term
$\left( {e}/{e_0}\right)^{-r_t \:R_v/c^{\ast}_{pl}}$
may be significantly different from unity and was missing in the formulation of \cite{Betts_QJ_1973}.

The only difference between $\theta_l$ defined by (\ref{eq_thetal_HH87}) and $\theta^{\ast}_{\rm M93}$ defined by (\ref{eq_theta_M93}) is the term 
$\left( {e}/{e_0}\right)^{-r_t \:R_v/c^{\ast}_{pl}}$ 
in (\ref{eq_thetal_HH87}) replaced by 
$\left( {e}/{p_0}\right)^{-r_t \:R_v/c^{\ast}_{pl}}$ 
in (\ref{eq_theta_M93}), with a ratio of these terms 
$\left( {e_0}/{p_0}\right)^{-r_t \:R_v/c^{\ast}_{pl}}$
that is constant if $r_t$ is a constant. 
However, ${e}/{e_0}$ with $e_0=6$~hPa is likely more often closer to unity than ${e}/{p_0}$ with $p_0=1000$~hPa, and for this reason (\ref{eq_thetal_HH87}) is likely more relevant than (\ref{eq_theta_M93}).

\subsection{\underline{\large Establishment/demonstration of the formulation for $\theta_s$}}
 \label{==Moist-ENROPY-thetas_def==}
\vspace{-2mm}

The formulation (\ref{eq_thetas}) can be derived starting from the Appendix~B of \cite{Marquet_QJ_2011}, but:
\vspace{-2mm}
\begin{itemize}
\item 
\vspace{-2mm}
with a more direct method;
\item 
\vspace{-2mm}
with all computations shown;
\item 
\vspace{-2mm}
with the additional hypotheses of $T_l$ and $T_i$ different from $T$ for condensed species (precipitations at the wet-bulb temperature?); and
\item 
\vspace{-2mm}
with the possibility that $H_l \neq 1$ and $H_i \neq 1$ where $q_l \neq 0$ and $q_i \neq 0$ (unsaturated conditions with condensed species), or the possibility that $H_l > 1$ or $H_i > 1$ (surer-saturation or mixed-phase conditions).
\end{itemize}
\vspace{-3mm}

The specific moist-air entropy is first assumed to be a weighted sum of   the partial entropies for dry air $s_d(T,p-e)$, 
water vapor $s_v(T,e)$, 
liquid water $s_l(T)$, 
ice $s_i(T)$, 
rain $s_{\rm rain}(T_{\rm rain})$ and 
snow $s_{\rm snow}(T_{\rm snow})$ species, 
with $q_d+q_v+q_l+q_i + q_{\rm rain} + q_{\rm snow}
= q_d+q_t = 1$ 
the specific contents of these species (i.e. the mass concentrations $q_x=m_x/m$ with 
$m=m_d+m_v+m_l+m_i+m_{\rm rain}+m_{\rm snow}
= m_d+m_t$ 
the total mass of a small and homogeneous parcel of the moist-air fluid), leading to
\vspace*{-2mm}
\begin{align}
s & \: = \; q_d \; s_d(T,p-e) \: + \: q_v \; s_v(T,e) 
    \: + \: q_l \; s_l(T)     \: + \: q_i \; s_i(T)
  \nonumber  \\
  & \quad
    \: + \: q_{\rm rain} \; s_l(T_{\rm rain})   
    \: + \: q_{\rm snow} \; s_i(T_{\rm snow})
\label{eq_s_sum}
\: .
\end{align}
The moist-air entropy formulation (\ref{eq_s_sum}) 
depends on the partial pressure of dry air ($p-e$) and water vapor $(e)$.
This formulation (\ref{eq_s_sum}) can then be transformed by using 
$q_v=q_t-q_l-q_i-q_{\rm rain}-q_{\rm snow}$
in the first line,
and by inserting/removing
$s_l(T)$ and $s_i(T)$  
in the second line, leading to
\vspace*{-2mm}
\begin{align}
   s  & = \: q_d\;s_d(T,p-e) \: + \: q_t\;s_v(T,e)
 \: + \: q_l \: \left[ \: s_l(T)-s_v(T,e) \: \right]
 \: + \: q_i \: \left[ \: s_i(T)-s_v(T,e) \: \right]
  \nonumber \\
      & \quad
 \: + \: \: q_{\rm rain} \: 
      \left[ \: \: s_l(T_{\rm rain}) \: - s_l(T) \: \right]  
 \: + \: \: q_{\rm rain} \: 
      \left[ \: s_l(T) - s_v(T,e) \: \right]
  \nonumber \\
      & \quad
 \: + \: q_{\rm snow} \: 
      \left[ \: s_i(T_{\rm snow}) - s_i(T) \: \right]
 \: + \: q_{\rm snow} \: 
      \left[ \: s_i(T) - s_v(T,e) \: \right]
   \: ,
  \label{def_ths_a} 
  \\
   s  & = \: q_d\;s_d(T,p-e) \: + \: q_t\;s_v(T,e)
  \nonumber \\
      & \quad
 \: + \: \: q_{\rm rain} \: 
      \left[ \: \: s_l(T_{\rm rain}) \: - s_l(T) \: \right]  
 \: - \: \: (q_l+q_{\rm rain}) \: 
      \left[ \: s_v(T,e) - s_l(T)  \: \right]  
  \nonumber \\
      & \quad
 \: + \: q_{\rm snow} \: 
      \left[ \: s_i(T_{\rm snow}) - s_i(T) \: \right]
 \: - \: (q_i+q_{\rm snow}) \: 
      \left[ \: s_v(T,e) - s_i(T) \: \right]
   \: .
  \label{def_ths_b}
\end{align}
If the ``gas constants'' ($R_d$ and $R_v$) and the ``specific heat at constant pressure'' ($c_{pd}, c_{pv}, c_l, c_i$) are assumed to be constant within the range of atmospheric temperatures, the partial entropies can be computed from the relationships 
\vspace*{-1mm}
\begin{align}
s_d(T,p-e) & \: = \: 
c_{pd} \: \ln\!\left( \frac{T}{T_r} \right)
\; - \; 
R_d \: \ln\!\left( \frac{p-e}{p_r-e_r} \right)
\; + \; s_{dr}(T_r,p_r-e_r)
\label{eq_sd} \: , \\
s_v(T,e) & \: = \: 
c_{pv} \: \ln\!\left( \frac{T}{T_r} \right)
\; - \; 
R_v \: \ln\!\left( \frac{e}{e_r} \right)
\; + \; s_{vr}(T_r,e_r)
\label{eq_sv} \: , \\
s_l(T) & \: = \: 
c_{l} \: \ln\!\left( \frac{T}{T_r} \right)
\; + \; s_{lr}(T_r)
\label{eq_sl} \: , \\
s_i(T) & \: = \: 
c_{i} \: \ln\!\left( \frac{T}{T_r} \right)
\; + \; s_{ir}(T_r) 
\label{eq_si} \: ,
\end{align}
where, at this stage, the reference temperature $T_r \neq T_0$ and the reference total pressure $p_r \neq p_0$ are, a priori, different from the standard values $T_0=273.15$~K and $p_0=1000$~hPa.
The reference partial pressure for water vapor is assumed to be the saturation water-vapor pressure $e_r=e_s(T_r)$, which is the only possible value different from $0$ and depending on the reference temperature $T_r$.

The differences in entropy  $s_l(T_{\rm rain}) \: - s_l(T)$ and $s_i(T_{\rm snow}) - s_i(T)$ in (\ref{def_ths_a}) can be computed from (\ref{eq_srain}) and (\ref{eq_ssnow}), yielding
\vspace*{-2mm}
\begin{align}
s_l(T_{\rm rain}) \: - \:  s_l(T)  & \: = \:
c_{l} \: \ln\!\left( \frac{T_{\rm rain}}{T} \right)
\label{eq_srain} \: \\
\mbox{and} \;\;\;
s_i(T_{\rm snow}) \: - \:  s_i(T)  & \: = \: 
c_{i} \: \ln\!\left( \frac{T_{\rm snow}}{T} \right)\: ,
\label{eq_ssnow}
\end{align}
whereas the other differences in entropy can be computed in terms of the latent heats $L_v(T)$ and $L_s(T)$ and the relative humidities $H_l=e/e_{sw}$ and $H_i=e/e_{si}$ according to \citet[Eqs.~3.10 to 3.12, p.~2891]{Hauf_Hoeller_1987}, 
yielding
\vspace*{-2mm}
\begin{align}
   s_v(T,e) \: - \: s_l(T)  & 
     \: = \; \frac{L_v(T)}{T} 
       \: - \: R_v \: \ln \left(H_l\right)
   \: ,
  \label{defAPPslv} \\
   s_v(T,e) \: - \: s_i(T)  & 
     \: = \; \frac{L_s(T)}{T} 
       \: - \: R_v \: \ln \left(H_i\right)
   \: .
  \label{defAPPsiv}
\end{align}
These differences in water-species entropy  
(\ref{eq_srain})-(\ref{defAPPsiv})
can be inserted into (\ref{def_ths_b}), 
together with $s_d(T,p-e)$ and $s_v(T,e)$ 
given by (\ref{eq_sd})-(\ref{eq_sv}), 
to give
\vspace*{-2mm}
\begin{align}
\!\!\!
   s  & = \: q_d\;s_{dr}(T_r,p_r-e_r) 
     \: + \: q_t\;s_{vr}(T_r,e_r)
 \: - \: \frac{(q_l+q_{\rm rain}) \: L_v(T)
              +(q_i+q_{\rm snow}) \: L_s(T)}{T} 
  \nonumber \\
\!\!\!
      & \quad
   + \: (q_d\:c_{pd} \: + \: q_t\:c_{pv})  
   \, \ln\!\left( \frac{T}{T_r} \right)
  \: - \: q_d \: R_d \:\ln\!\left( \frac{p-e}{p_r-e_r} \right)
  \: - \: q_t \: R_v \:\ln\!\left( \frac{e}{e_r} \right)
  \nonumber \\
\!\!\!
      & \quad
  +  q_{\rm rain} \: 
      c_{l}  \ln\!\left(\frac{T_{\rm rain}}{T}\!\right) 
  +  q_{\rm snow} \: 
      c_{i}  \ln\!\left(\frac{T_{\rm snow}}{T}\!\right)
  +  (q_l+q_{\rm rain}) \, R_v \ln\!\left(H_l\right)
  +  (q_i+q_{\rm snow}) \, R_v \ln\!\left(H_i\right)
   .
  \label{def_ths_c}
\end{align}
The property $q_d=1-q_t$ can be used to rewrite
the first two terms of (\ref{def_ths_c}) as
\vspace*{-1mm}
\begin{align}
             q_d\;s_{dr}(T_r,p_r-e_r) 
     \: + \: q_t\;s_{vr}(T_r,e_r)
   & = 
   s_{dr}(T_r,p_r-e_r)  \: + \: c_{pd} \; \Lambda_r \; q_t
   \: ,
  \label{defAPPs5}
\end{align}
where the non-dimensional term
$\Lambda_r  = 
[\: s_{vr}(T_r, e_r) - s_{dr}(T_r , p_r - e_r) \:]
/ c_{pd}$
depends on the reference entropies for water vapor and dry air, to be computed for the reference temperature $T_r$ and partial pressures $p_r$ and $e_r$.
Similarly, the term in factor of the logarithm of $T$ can be rewritten as
\vspace*{-1mm}
\begin{align}
   q_d \: c_{pd} \: + \: q_t \:c_{pv}  
   & = 
   c_{pd} \: + \: c_{pd} \: \lambda \: q_t
   \: ,
   \;\; \mbox{where}\;\;\;\;
   \lambda = (c_{pv}-c_{pd})/c_{pd} \: .
  \label{defAPPs6}
\end{align}
The dry-air and water species reference entropies can be computed for any set of reference values ($T_r$, $p_r$, $e_r$) according to
\vspace*{-1mm}
\begin{align}
   s_{dr}(T_r,p_r-e_r)  & \: = \; s_{d0}(T_0,p_0)
    \:+ c_{pd} \: \ln\!\left( \frac{T_r}{T_0} \right)
    \:- R_{d} \: \ln\!\left( \frac{p_r-e_r}{p_0} \right)
   \: ,
  \label{defAPPsdr} \\
   s_{vr}(T_r, e_r)  & \: = \; s_{v0}(T_0, p_0)
    \:+ c_{pv} \: \ln\!\left( \frac{T_r}{T_0} \right)
    \:- R_{v} \: \ln\!\left( \frac{e_r}{p_0} \right)
   \:  ,
  \label{defAPPsvr} \\
   s_{lr}(T_r)  & 
     \: = \; \; s_{l0}(T_0)
    \:+ c_l \: \ln\!\left( \frac{T_r}{T_0} \right)
   \: ,
  \label{defAPPslr} \\
   s_{ir}(T_r)  & 
     \: = \; s_{i0}(T_0)
    \:+ c_i \: \ln\!\left( \frac{T_r}{T_0} \right)
   \: .
  \label{defAPPsir}
\end{align}
Eqs.~(\ref{defAPPs5}), (\ref{defAPPs6})
and (\ref{defAPPsdr}) 
can be inserted into (\ref{def_ths_c}), also with the term 
``$R_d \ln\!\left( {p_0}/{p} \right)$'' 
added and removed in the first 
line, to 
give$\,$\footnote{$\:$Many thanks to Almut Gassman who see a typo in the version~1 of the arXiv note for the term 
$- \,q_t \: R_v \,\ln\!\left({e}/{e_r}\right)$
at the end of the second line of (\ref{def_ths_d}), which was badly written $c_{pd} \: \gamma \: q_t  \,\ln\!\left({r_r}/{r_v}\right)$, in that wrongly anticipating the end of the third line of (\ref{def_ths_e}) and via a bad LaTeX copy/paste... sorry!} 
\vspace*{-2mm}
\begin{align}
\!\!\!
   s  & = \: s_{d0}(T_0,p_0) 
      +  c_{pd}\, \ln\!\left( \frac{T}{T_0} \right)
      +  R_d \, \ln\!\left( \frac{p_0}{p} \right)
  -  \frac{(q_l+q_{\rm rain}) \: L_v(T)
              +(q_i+q_{\rm snow}) \: L_s(T)}{T} 
  \nonumber \\
\!\!\!
      & \quad
     + c_{pd}\, \Lambda_r\: q_t
     + c_{pd} \: \lambda \: q_t 
       \: \ln\!\left( \frac{T}{T_r} \right)
   -  R_d \:\ln\!\left( \frac{p_r-e_r}{p} \right)
   - q_d \: R_d \,\ln\!\left( \frac{p-e}{p_r-e_r} \right)
   - q_t \: R_v \,\ln\!\left( \frac{e}{e_r} \right)
  \nonumber \\
\!\!\!
      & \quad
  +  q_{\rm rain} \: 
      c_{l}  \ln\!\left(\frac{T_{\rm rain}}{T}\!\right) 
  +  q_{\rm snow} \: 
      c_{i}  \ln\!\left(\frac{T_{\rm snow}}{T}\!\right)
  +  (q_l+q_{\rm rain}) \, R_v \ln\!\left(H_l\right)
  +  (q_i+q_{\rm snow}) \, R_v \ln\!\left(H_i\right)
   .
  \label{def_ths_d} 
\end{align}
An important property can be deduced from (\ref{def_ths_d}): because the first and third lines do not depend on the reference values $T_r$, $p_r$ and $e_r(T_r)$, the second line cannot depend on, and is thus independent on, these reference value.
This means that any changes in $T_r$, $p_r$ and $e_r(T_r)$ corresponds to changes in all terms that leave the whole second line unchanged.

The next step is to write the equations for the dry air, water vapor and moist air
\vspace{-0.15cm}
\begin{align}
p_d = p - e & 
 \: = \; {\rho}_d \: R_d \: T 
 \; = \; \rho \:\: q_d \: R_d \: T
\: , 
\\
 e  & 
 \: = \; {\rho}_v \: R_v \: T 
 \; = \; \rho \:\: q_v \: R_v \: T
 \: ,
\\
 p = p_d + e_r  & 
 \: = \; 
     \, \rho \:\: R \:\: T \:
 \; = \; 
 \rho \: (q_d\:R_d + q_v \: R_v)  \: T
 \; = \; 
 \rho \:\: q_d \: R_d \: (1 + \eta \: r_v)  \: T
 \: ,
\\
 \mbox{and thus with} \;\;\;
 p & 
 \: = \; 
 (1 + \eta \: r_v) \: (p - e)
 \: ,
\\
 \mbox{and} \;\;\;
 e  & 
 \: = \; 
 \rho \:\: q_d \: R_d \:\: \eta \:r_v \: T
 \; = \; 
 \eta \:r_v \: (p - e)
 \: ,
\end{align}
where $r_v=q_v/q_d={\rho}_v/{\rho}_d$ and 
$\eta=R_v/R_d$, leading to
\vspace{-0.15cm}
\begin{align}
  p - e   & = 
      \frac{1}{1+\eta\:r_v} \: p
  \: , \label{def0pd} \\
  e   & = 
      \frac{\eta\:r_v}{1+\eta\:r_v} \: p
  \: . \label{def0e}
\end{align}
The same relationship exist for the ``\,just saturated\,'' (i.e. with $q_{lr}=q_{ir}=0$ and thus  $q_{tr}=q_{vr}=q_{sat}(T_r,p_r)=q_r$) reference state
\vspace{-0.15cm}
\begin{align}
  p_r - e_r   & = 
      \frac{1}{1+\eta\:r_r} \: p_r
  \: , \label{def0pdr} \\
  e_r   & = 
      \frac{\eta\:r_r}{1+\eta\:r_r} \: p_r
     \; = \;
     \eta\:r_r\:(p_r-e_r)
  \: , \label{def0er} \\
 \mbox{and thus with} \;\;\;
  r_r   & = 
      \frac{e_r}{\eta\:(p_r-e_r) }
   \;\;\;\mbox{and} \;\;\;
  q_r    = 
      \frac{r_r}{1+r_r}
  \: . \label{def0qr}
\end{align}
The partial pressures relationships 
(\ref{def0pd}),
(\ref{def0e}),
(\ref{def0pdr}) and
(\ref{def0er}),
can then be inserted into (\ref{def_ths_d}) 
with 
$\eta = R_v/R_d$, 
$\gamma = R_v/c_{pd}$,
$\kappa=R_d/c_{pd}$ and
$\delta = R_v/R_d-1$,
leading to:
\vspace*{1mm}
\begin{align}
\!\!\!
   s  & \: = \; s_{d0}(T_0,p_0) 
      -  c_{pd}\, \ln\!\left( T_0 \right)
      +  c_{pd}\, \ln\!\left( T \right)
      +  c_{pd}\, \kappa \, \ln\!\left( \frac{p_0}{p} \right)
  \nonumber \\
\!\!\!
      & \quad
  -  c_{pd} 
  \left[\:
  \frac{(q_l+q_{\rm rain}) \: L_v(T)
       +(q_i+q_{\rm snow}) \: L_s(T)}
       {c_{pd}\:T}
  \:\right] 
  +  c_{pd}\, \Lambda_r \: q_t
  \nonumber \\
\!\!\!
      & \quad
  + c_{pd} \: \lambda \: q_t 
       \: \ln\!\left( \frac{T}{T_r} \right)
  + c_{pd} \: \kappa \: \delta \: q_t 
    \: \ln\!\left( \frac{p_r}{p} \right)
  + c_{pd} \: \gamma \: q_t 
    \: \ln\!\left( \frac{r_r}{r_v} \right)
  \nonumber \\
\!\!\!
      & \quad
  - c_{pd} \: \kappa \: \delta \: q_t 
       \: \ln\!\left( 1+ \eta \ r_r \right)
  + c_{pd} \: \kappa \: (1+\delta \: q_t) 
       \: \ln\!\left( 1+ \eta \ r_v \right)
  \nonumber \\
\!\!\!
      & \quad
  +  q_{\rm rain} \: 
      c_{l}  \ln\!\left(\frac{T_{\rm rain}}{T}\!\right) 
  +  q_{\rm snow} \: 
      c_{i}  \ln\!\left(\frac{T_{\rm snow}}{T}\!\right)
  \nonumber \\
\!\!\!
      & \quad
  +  c_{pd} \: (q_l+q_{\rm rain}) \, \gamma \ln\!\left(H_l\right)
  +  c_{pd} \: (q_i+q_{\rm snow}) \, \gamma \ln\!\left(H_i\right)
   .
\label{def_ths_e} 
\end{align}

The relationship (\ref{eq_thetas}) for $\theta_s$ results from the definition (\ref{eq_def_s_thetas}) 
$s 
= 
s_{d0}(T_0,p_0) 
+
c_{pd} \: \ln\left({\theta_{s}}/{T_0}\right)
= 
s_{d0}(T_0,p_0) 
-
c_{pd} \: \ln\left(T_0\right)
+
c_{pd} \: \ln\left(\theta_{s}\right)
$, 
namely with the same first two terms 
$s_{d0}(T_0,p_0)$ 
and 
$- c_{pd} \: \ln\left({T_0}\right)$
as in (\ref{def_ths_e}),
and then via an identification of $\theta_{s}$ 
with all other terms of 
(\ref{def_ths_e}) expressed as 
$\:c_{pd} \, \ln[\,(...)^{(...)}\,]$
or
$\:c_{pd} \, \ln[\,\exp(...)\,]$.
\vspace*{-2mm}

 \begin{center} \rule[0mm]{7.cm}{0.1mm} \end{center}
\vspace*{-2mm}

The absolute (third-law) standard entropies 
$s_{d0}(T_0,p_0)$, $s_{v0}(T_0, p_0)$, $s_{l0}(T_0)$ and $s_{i0}(T_0)$
appearing in (\ref{defAPPsdr})-(\ref{defAPPsir})
are computed at the standard values $T_0=273.15$~K and $p_0=1000$~hPa. 

The same standard values considered in \citet{Marquet_QJ_2011} were first published in atmospheric science by \citet[``HH87'', p.2891, however without references]{Hauf_Hoeller_1987}, who explained that:
``{\it the values of the zero-entropies have to be determined experimentally or by quantum statistical considerations,\/}''
leading to
\vspace{-2mm}
\begin{align}
& 
\;\;\;\;\;\;
\mbox{HH87}
\;\;\;\;\;\;
\mbox{M83}
\;\;\;\;\;\;\;
\mbox{L00}
\;\;\;\;\;\;\:
\mbox{MS22}
\;\;\;\;\;\;\;\:
\mbox{M22}
\nonumber \\
s_{d0}(T_0,p_0) & \: \approx \; \; 6775 \;
 \;\;\; \;
 [ \: -- \: ] 
 \;\;\; \;
 [ \: 6776 \: ] 
 \;\: 
 [ \: 6776.2  \: ] 
 \;\; 
 [ \: 6776.3 \: ] 
 \;\: \mbox{J/K/kg}
\label{eq_entropies_HH87_d} \: 
\\
s_{v0}(T_0,p_0) & \: \approx \; 10320 
 \;\;\; \;
 [ \: -- \: ] 
 \;\;\; \;\:
 [ \: -- \: ] 
 \; 
 [ \: 10319.7 \: ] 
 \; 
 [ \: 10318.1 \: ] 
 \; \mbox{J/K/kg}
\label{eq_entropies_HH87_v} \: 
\\
s_{l0}(T_0) & \: \approx \; \; 3517  
 \;\;\;
 [ \: 3516.2 \: ] 
 \;\;\:
 [ \: -- \: ] 
 \;\;\: 
 [ \: 3516.7 \: ] 
 \;\; 
 [ \: 3514.7 \: ] 
 \;\, \mbox{J/K/kg}
\label{eq_entropies_HH87_l} \: 
\\
s_{i0}(T_0) & \: \approx \; \; 2296  \;
 \;\;\; \;
 [ \: -- \: ] 
 \;\;\; \; \:
 [ \: -- \: ] 
 \;\;\; \;\:
 [ \: -- \: ] 
 \;\;\;\; 
 [ \: 2291.9  \: ] 
 \;\: \mbox{J/K/kg}
\label{eq_entropies_HH87_i} \: 
\end{align}
The other statistical-physics (standard and absolute) values indicated into brackets in (\ref{eq_entropies_HH87_d})-(\ref{eq_entropies_HH87_i}) are from 
\vspace{-4mm}
\begin{itemize}
\item 
\citet[][``M83'']{Millero_1983} for liquid-water only ($s_{l0}$) and from the value $70.00$~J/K/mol at $298.15$~K of \citet{Lewis_Randall_1961};
\vspace{-3mm}
\item 
\citet[][``L00'']{Lemmon_al_2000} for dry air only ($s_{d0}$) and for dry-air composed of N${}_2$+O${}_2$+Ar (without CO${}_2$);
\vspace{-3mm}
\item 
\citet[][``MS22'']{Marquet_Stevens_JAS_2022} without the ice value ($s_{i0}$);
\vspace{-3mm}
\item 
from recent full computations made in 
Marquet (2022, ``M22'', to appear on arXiv) 
based on the NIST-JANAF4 equations of \citet[][]{Chase_1998} 
for H${}_2$O (vapour+liquid+ice) and for dry-air composed of 
N${}_2$ $\;+\;$ O${}_2$ + Ar + $400$~ppm of CO${}_2$.
\end{itemize}
\vspace{-4mm}
The numerical values listed in (\ref{eq_entropies_HH87_d})-(\ref{eq_entropies_HH87_i}) show that the uncertainties on dry-air and water-vapour standard entropies $s_{d0}$ and $s_{v0}$ are small: about $1$ and $2$~J/K/kg, respectively.

It is worthwhile to remember that the standard entropies
computed at $298.15$~K and $1000$~hPa (or $1013.25$~hPa) for 
N${}_2$, O${}_2$, Ar, CO${}_2$ and H${}_2$O
were already fully available since 
\citet{Kelley_1932},
and then in the Thermodynamic Tables of
\citet{Rossini_al_1952}, 
\citet{Lewis_Randall_1961},
\citet{Wagman_al_1965} and
\citet{Robie_al_1978},
thus long before the computations made in atmospheric science by
\citet{Betts_QJ_1973}, 
\citet{IribarneGodson1973,IribarneGodson1981},
\citet{Emanuel_1994}, 
\citet{Pauluis_Czaja_Korty_2010}
and others,
with (very) little changes in comparison to  
the more modern computations available in
the Thermodynamic Tables and book of
\citet{Gokcen_Reddy_1996},
\citet{Chase_1998} and
\citet{Atkins_Paula_2014}.

\begin{figure*}[hbt]
\centering
\noindent
\includegraphics[width=0.75\linewidth]{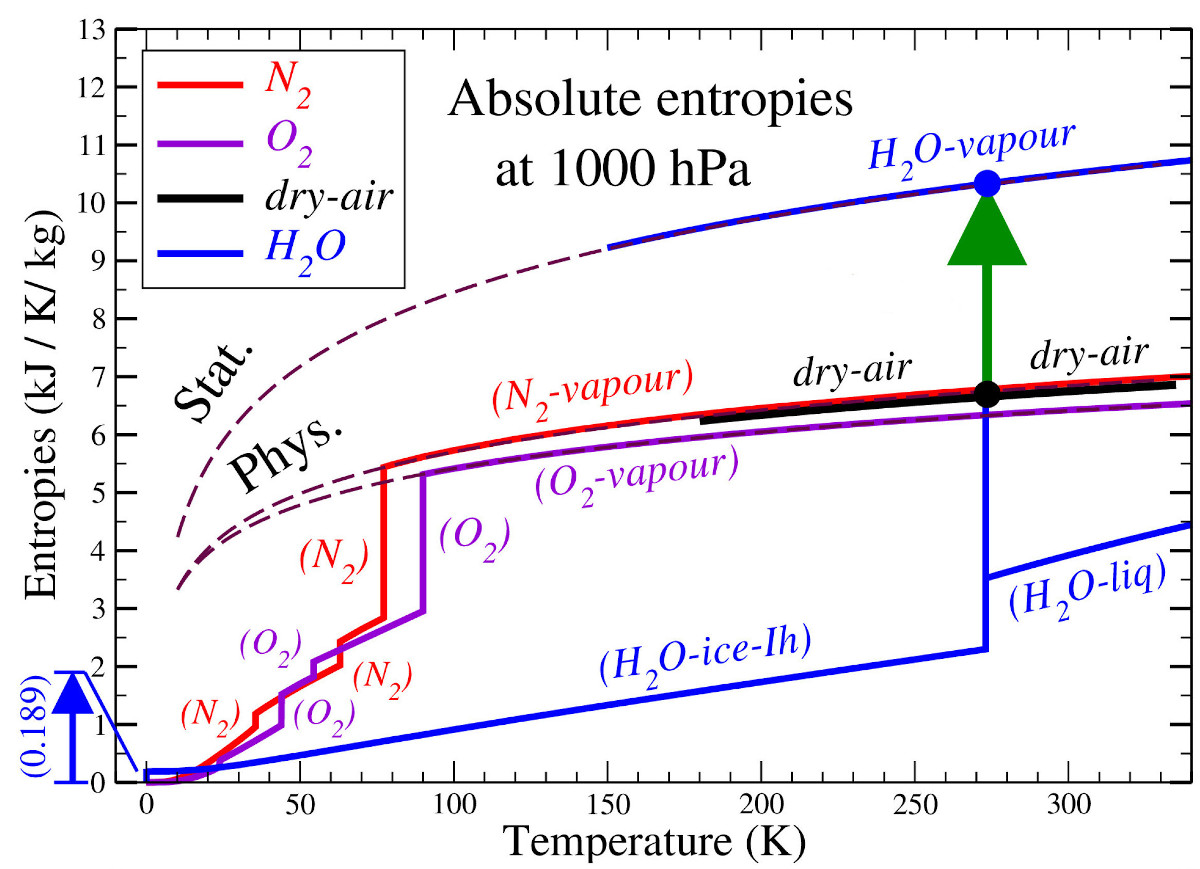} 
\\
\noindent
\includegraphics[width=0.75\linewidth]{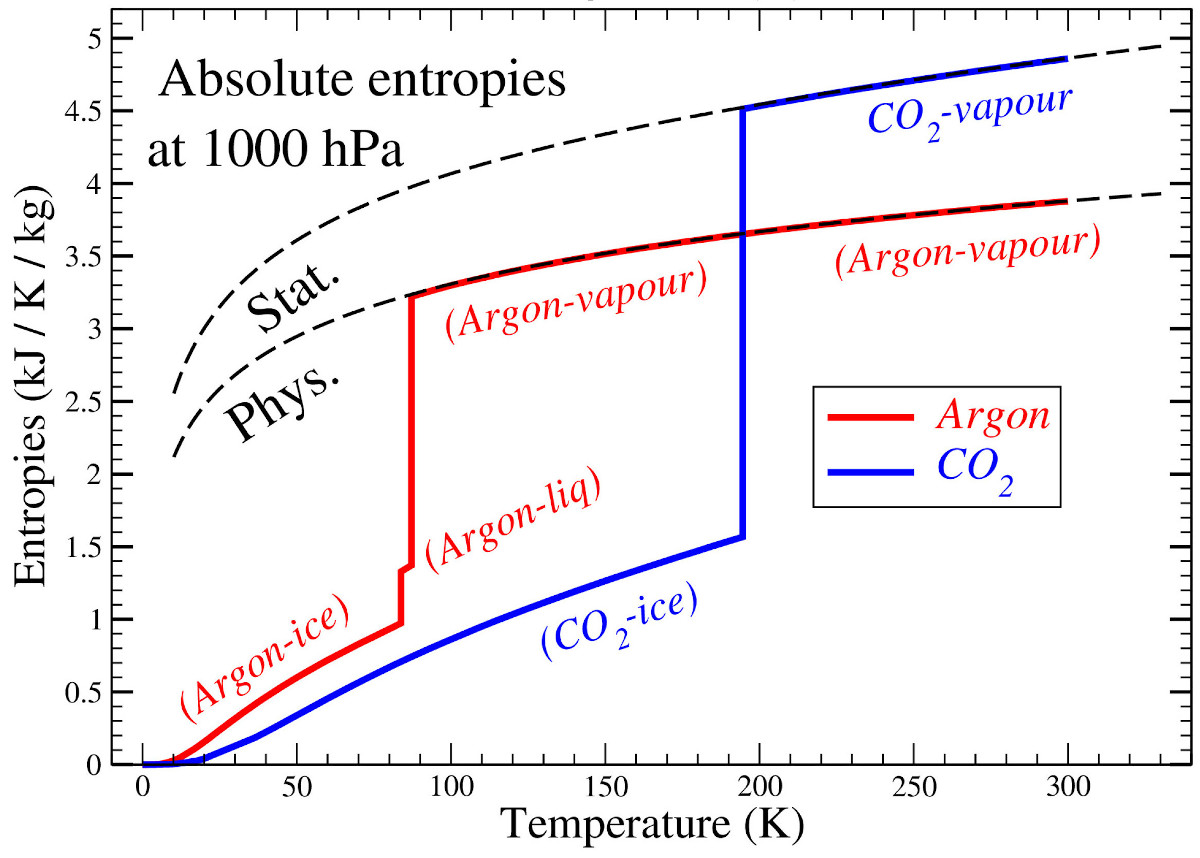}
\vspace*{-3mm}
\caption{\it \small Entropies for dry-air 
(N${}_2$, O${}_2$, Ar, CO${}_2$ and water H${}_2$O)
species plotted against the absolute temperature and computed at $1000$~hPa. 
The calorimetric method 
$\left(\int_0^T c_p(T') \, d\ln(T') + \sum_j L(T_j)/T_j \right)$ 
corresponds to the coloured solid lines. 
The third-law hypothesis is applied at $0$~K with zero entropies for all the solid phases, but with the residual entropy of 
$189$~J/K/kg for ice-Ih. 
The vertical jumps correspond to phase changes at $T_j$ with the phase-change enthalpies $L(T_j)$ between solids phases 
(for N${}_2$ and O${}_2$), 
then from solid to liquid phases, then from liquid to vapour phases. 
The statistical-physics values (black dashed lines) are computed from $S = k\: \ln(W)$ and $F = - \: k \: T \: \ln(Z)$ for the vapour phases according to the method described in \cite{Chase_1998} for translational, rotational, vibrational and electronic partition functions ($Z$).}
\label{figs-entropies}
\end{figure*}

Figure~\ref{figs-entropies} summarizes the way all absolute entropies can be computed with either the calorimetric or statistical-physics methods, leading to the same values at $273.15$~K.
The third law must be applied to the ``more stable crystalline states'' at $0$~K 
\citep[and not to the gases as wrongly stated in the Appendix of ][where criticisms of the third law of thermodynamics were unfounded]{Pauluis_Czaja_Korty_2010}. 
It is also needed to take into account the special residual entropy of $189$~J/K/kg for ice-Ih, according to \cite{Pauling1935} and \cite{Nagle1966}.

Similarly, IAPWS and TEOS10 computations of the entropies of ocean and moist-air
\citep{Wagner_Pruss_1995,Feistel_al_2008,IAPWS10,Feistel_2010,
Feistel_2012,Feistel_2019,Feistel_Hellmuth_2020}
may have use the values of \citet{Millero_1983} for sea water and \citet{Lemmon_al_2000} for dry air, which are moreover available (but commented and unused) in the TEOS10 FORTRAN software.
Differently, IAPWS and TEOS10 computations are made by canceling some standard entropies of dry air and water species at $273.15$~K, in that being in contradiction with the third law of thermodynamics and with statistical physics principles, like in \cite{IribarneGodson1973,IribarneGodson1981}, \cite{Emanuel_1994}, \cite{Pauluis_Czaja_Korty_2010} and others.

It was similarly assumed by \cite{Emanuel_1994}, \cite{Pauluis_Czaja_Korty_2010} and many others that $s_d(T_0,p_0)$ and either $s_v(T_0,p_0)$ for $\theta_l$ or $s_l(T_0)$ for $\theta_e$ can be freely set to zero, in clear disagreement with the third law of thermodynamics and differently from  \cite{Hauf_Hoeller_1987}, \cite{Millero_1983}, \cite{Lemmon_al_2000} and \cite{Marquet_QJ_2011,Marquet_JAS_2017}, 
where the third law is fully taken into account for the solid states of all dry-air and water species (N${}_2$, O${}_2$, Ar, CO${}_2$, H${}_2$O).

Moreover, in fact, there is no need to set to zero these standard values $s_d(T_0,p_0)$, $s_v(T_0,p_0)$ or $s_l(T_0)$ to defined $\theta_e$ and $\theta_l$ from the formulation $\theta_{s/\rm{HH87}}$ of \cite{Hauf_Hoeller_1987}, and the same should have been for \cite{Emanuel_1994}, \cite{Pauluis_Czaja_Korty_2010} and others.

The only (however big) price to pay for leaving terms depending on the variable water content $q_t$ outside the logarithm in (\ref{eq_HH87}) and (\ref{eq_M93}), including the terms depending on the non-zero entropies $s_d(T_0,p_0)$, $s_v(T_0,p_0)$ and $s_l(T_0)$, is that none of $\theta_e$, $\theta_l$, $\theta_{s/\rm{HH87}}$ and $\theta_{s/\rm{M93}}$ can represent the moist-air entropy in the real world, where $q_t$ is neither a constant due to evaporation, precipitations, entrainment, detrainment and turbulent-mixing processes.

Anyhow, since the entropy is a state function, there is only one definition of it, and only $\theta_s$ can represent the moist-air entropy in all conditions.
In order to understand more precisely this property, let us imagine two parcels of the moist-air fluid labeled by $A$ and $B$, with the thermodynamic conditions 
($T_{\small A}$, $p_{\small A}$, 
$q_{v{\small A}}$, 
$q_{l{\small A}}$, $q_{i{\small A}}$, 
...) and 
($T_{\small B}$, $p_{\small B}$, 
$q_{v{\small B}}$, 
$q_{l{\small B}}$, $q_{i{\small B}}$, 
...), respectively.  
The total water contents 
$q_{t{\small A}} = q_{v{\small A}} + q_{i{\small A}} + ...$ 
and
$q_{t{\small B}} = q_{v{\small B}} + q_{i{\small B}} + ...$ 
are thus different.
In that case, according to (\ref{eq_HH87}), (\ref{eq_M93})  or (\ref{eq_HH87_theta_l}), the difference in specific entropy 
$\Delta s = s_{\small B} - s_{\small A}$
between the two points $A$ and $B$ depends on differences like  
$\left[ \: s_l(T_0) - s_d(T_0,p_0) \: \right] \Delta q_t$ for (\ref{eq_HH87}),
where $\Delta q_t = q_{t{\small B}} - q_{t{\small A}}$.

\clearpage

Therefore, since the specific entropy ``$s$'' is a state function,  then the difference between two points $\Delta s = s_{\small B} - s_{\small A}$ must have a unique (non-arbitrary) value independent on any thermodynamic path leading from $A$ to $B$,
and then it is not possible to modify at will and arbitrarily 
the difference 
$\left[ \: s_l(T_0) - s_d(T_0,p_0) \: \right]$,
because it is in factor of the variable term $\Delta q_t$ and any arbitrarily change in $s_l(T_0)$ or $s_d(T_0,p_0)$ would generate arbitrary changes in $\Delta s$ associated with $\Delta q_t$.
This would be in contradiction with the second law of thermodynamics and the need to know precisely if the (moist-air) entropy decreases, increases or remain constant.
It is therefore needed to rely on the third law of thermodynamics (applied at $0$~K for the more stable crystalline states) and the absolute definitions of the entropies for dry-air and water species (N${}_2$, O${}_2$, Ar, CO${}_2$, H${}_2$O).
\vspace*{-2mm}

 \begin{center} \rule[0mm]{7.cm}{0.1mm} \end{center}
\vspace*{-2mm}

Another way to avoid arbitrarily canceling the reference entropies is to keep the terms depending on $q_t$ and the reference entropies in (\ref{eq_HH87}), (\ref{eq_M93}) or (\ref{eq_HH87_theta_l}) as they are, in order to define $\theta_l$ and $\theta_e$ without trying to make them synonymous with the specific entropy $s$.

This approach was chosen in \cite{Marquet_Stevens_JAS_2022}, where $\theta_l$, $\theta_s$ and $\theta_e$ are defined (for ice-free conditions $q_i=q_{snow} = 0$) in a similar way without canceling any of the reference partial entropies.
The new method is the one previously described in \citet[][from a course with the same name given at MPI]{stevenssiebesma2020} for deriving both $\theta_l$ and $\theta_e$, but generalizes in \cite{Marquet_Stevens_JAS_2022} to the definition of $\theta_s$.

\citet[][after Eq.~2.30]{stevenssiebesma2020} reinterpret entropy equations like (\ref{eq_HH87}) in the following way:
``{\it Given a completely specified reference state it provides an expression for the entropy. 
This was the sense in which it was derived. 
Alternatively, one can use this equation to ask what would the
reference state temperature need to be, for the system in
the reference state configuration (as specified through the
pressure, amount and distribution of water mass) to have
the same entropy as in the given state.\/}''
 
Accordingly, the three equations
\vspace*{-1mm}
\begin{align}
s(T, p, q_v, q_l) & = \, 
   c_{pe} \, \ln\left(\frac{\theta_e}{T_r}\right)
   \: + \: 
  \left[ \: (1-q_t) \: s_d(T_r,p_0) 
             +  q_t \: s_l(T_r) \: \right]
\label{eq_MS22_def_thetae} \: ,
\\
s(T, p, q_v, q_l) & = \, 
   c_{p\ell} \, \ln\left(\frac{\theta_\ell}{T_r}\right)
   \: + \: 
  \left[ \: (1-q_t) \: s_d(T_r,p_0-e_0) 
             + q_t  \: s_v(T_r,e_0) \: \right]
\label{eq_MS22_def_thetal} \: ,
\\
\mbox{and} \;\;\;
s(T, p, q_v, q_l) & = \, 
   c_{pd} \, \ln\left(\frac{\theta_s}{T_r}\right)
   \: + \: 
  \left[ \: s_d(T_r,p_0) \: \right]
  \: = \, 
   c_{pd} \, \ln\left(\frac{\theta_s}{T_0}\right)
   \: + \: 
  \left[ \: s_d(T_0,p_0) \: \right]
\label{eq_MS22_def_thetas} \: 
\end{align}
provide the new definitions:
\vspace*{-3mm}
\begin{quote}
{\it 
$\theta_e$, $\theta_\ell$ and $\theta_s$ are
the reference temperature $T_r$ for which the actual entropy $s(T, p, q_v, q_l)$ on the left-hand-sides of (\ref{eq_MS22_def_thetae})-(\ref{eq_MS22_def_thetas}) is equal to the reference entropies into brackets on the right-hand-sides of (\ref{eq_MS22_def_thetae})-(\ref{eq_MS22_def_thetas}).
}
\end{quote}
\vspace*{-3mm}
These definitions avoid canceling any of the reference partial entropies $s_d(T_r,p_0)$, $s_d(T_r,p_0-e_0)$, $s_v(T_r,e_0)$ or $s_l(T_r)$ for arbitrary values of $T_r$, 
as done in
\cite{IribarneGodson1973,IribarneGodson1981},
\cite{Emanuel_1994}, \cite{Pauluis_Czaja_Korty_2010} and others.

The peculiar feature valid for $\theta_s$ only is that the second part of (\ref{eq_MS22_def_thetas}) is also valid for $T_r$ replaced by the standard value $T_0=273.15$~K. 
This is simply due to the dry-air equation (\ref{defAPPsdr}) and the corresponding property
$s_d(T_0,p_0) - c_{pd} \, \ln\left(T_0\right)
= s_d(T_r,p_0) - c_{pd} \, \ln\left(T_r\right)$,
whatever $T_r$ may be.
This is precisely why $\theta_s$ given by (\ref{eq_thetas}) is independent on any value of $T_r$.

The method of \citet{stevenssiebesma2020} corresponding to (\ref{eq_MS22_def_thetae})-(\ref{eq_MS22_def_thetas}) is used in 
\cite{Marquet_Stevens_JAS_2022} to define and compute the three potential temperatures, leading to
\vspace*{-1mm}
\begin{align}
\theta_e & \, = \, 
T 
\: 
\left( \frac{p_0}{p}\right)^{\kappa_e}
\:
\exp \! \left( \frac{q_v \: L_v}{c_{pe} \: T}\right)
\;\: \Omega_e \: ,
\;\mbox{where}\;
\Omega_e = \left(\frac{R}{R_e}\right)^{\kappa_e} 
\left(H_l\right)^{-q_v R_v/c_{pe}} \: ,
\label{eq_MS22_thetae_Omegae} 
\\
\theta_\ell & \, = \, 
T \: 
\left(\frac{p_0}{p} \right)^{\kappa_\ell} 
\exp \!\left(-\frac{q_l \: L_v}{c_{p\ell} \: T}\right) 
\; \: \Omega_\ell
\: ,
\;\mbox{where}\;
\Omega_\ell 
\: = \: 
\left(\frac{R}{R_\ell} \right)^{\kappa_\ell} 
\left( \frac{q_t}{q_v} \right)^{q_t R_v/c_{p\ell}}
\: ,
\label{eq_MS22_thetal_Omegal} 
\\
\theta_s & = \, 
 \theta \:
 \exp\!\left( -\frac{ q_l \: L_v}{c_{pd} \: T}
 + 
 \Lambda_r \: q_t
\right) 
\: \Omega_s \: , 
\;\mbox{where}\;
\Omega_s \, = \, 
\left(\frac{R}{R_e}\right)^{\kappa_d} \!
\left(\frac{p-e}{p_0-e_0}\right)^{q_t \, \kappa_d} \!
\left(\frac{e_0}{e}\right)^{q_t \,\gamma}
\!\left(\frac{T}{T_0}\right)^{q_t \, \lambda}
\:
\label{eq_MS22_thetas_Omegas}
\end{align}
and with $\theta = T \: (p_0/p)^{\kappa_d}$ the usual dry-air potential temperature.

The moist-air exponent terms appearing in  
(\ref{eq_MS22_thetae_Omegae})-(\ref{eq_MS22_thetas_Omegas})
are
$R = R_d \, (1-q_t) + R_v \, q_v$
(i.e; the moist-air gas constant),
$R_e = R_d \: (1-q_t)$,
$c_{pe} = c_{pd} \:(1-q_t) + c_l \:q_t$,
$\kappa_e = R_e / c_{pe}$,
$R_\ell = R_d \: (1-q_t) + R_v \: q_t$, 
$c_{p\ell} =  c_{pd} \: (1-q_t) + c_{pv} \: q_t$
and
$\kappa_\ell = R_\ell / c_{p\ell}$.
\vspace*{-2mm}

 \begin{center} \rule[0mm]{7.cm}{0.1mm} \end{center}
\vspace*{-2mm}

The hope is that the Omega terms ($\Omega_e$, $\Omega_\ell$, $\Omega_s$) in  
(\ref{eq_MS22_thetae_Omegae})-(\ref{eq_MS22_thetas_Omegas})
may remain close to unity in all atmospheric conditions, with therefore the ``first-order approximations'' of 
$\theta_e$, 
$\theta_\ell$ and 
$\theta_s$ 
given by the other terms 
depending on $T$, $(p_0/p)$ and the exponentials. 
In particular, we can recognize for $\theta_s$ in (\ref{eq_MS22_thetas_Omegas}) the first line of (\ref{eq_thetas}) with $q_i=q_{snow}=0$, this first line being interpreted as the ``first-order approximation of $\theta_s$'' hereafter.

Indeed, the vertical profiles of the FIRE-I (RF03B) observed radial flight studied in \cite{Marquet_QJ_2011} plotted between $0$ and $1.5$~km are shown in the Fig.~(\ref{fig_FIRE_MS22}) below, together with the vertical profiles plotted between $0$ and $4$~km 
in the Fig~1 of \cite{Marquet_Stevens_JAS_2022}, 
which are the average for $757$ drop-sounds observed during the EUREC${\:}^4$A-Circle.
It is possible to check with these vertical profiles that the dashed or dotted lines (first-order approximations) remain close to the solid lines (exact formulations), up to about $0.6$~K for $\theta_e$ and $\theta_s$ and better than $0.1$~K for $\theta_\ell$.
Moreover, these differences are by far smaller than the differences between $\theta_\ell$, $\theta_s$ and $\theta_e$, this comforting the status of ``first-order approximations''.
\vspace*{-2mm}

 \begin{center} \rule[0mm]{7.cm}{0.1mm} \end{center}
\vspace*{-2mm}

A rigorous method is described in
\cite{Marquet2015WGNEthetas2,Marquet_thetas2_2019}
to derive not only the first-order approximation $(\theta_s)_1$, but also a second-order approximation $(\theta_s)_2$

The relationship already suggested in \cite{Marquet_QJ_2011} is indeed a first-order approximation:
it corresponds to the first line of (\ref{eq_thetas})
\vspace{-1mm}
\begin{equation}
\theta_s  
\; \approx \; 
\boxed{ \;
(\theta_s)_1
\; = \; 
\theta \:
\exp\!\left[ \:
- \: \frac{
     L_v \: 
   \left( q_l + q_{\rm rain} \right)
   + L_s \: 
   \left( q_i + q_{\rm snow} \right)
   }{c_{pd} \: T} 
\: \right]
\exp\!\left( \Lambda_r \: q_t \right)
\; \approx \; 
\theta_{il} \:
\exp\!\left( \Lambda_r \: q_t \right)
\; }
\: ,
\label{eq_theta_s1}
\end{equation}
where 
\vspace{-1mm}
\begin{equation}
\theta_{il}
\; = \;
\theta \:
\exp\!\left[ \:
- \: \frac{
     L_v \: 
   \left( q_l + q_{\rm rain} \right)
   + L_s \: 
   \left( q_i + q_{\rm snow} \right)
   }{c_{pd} \: T} 
\: \right]
\:
\label{eq_theta_il}
\end{equation}
is close to the ice-liquid formulation of \cite{Tripoli_Cotton_1981}, except the constant values $T_0$, $L_v(T_0)$ and $L_v(T_0)$ are replaced by the  variable values $T$, $L_v(T)$ and $L_v(T)$ that depend on $T$ in the exponential of (\ref{eq_theta_il}), and with the rain and snow contents added from (\ref{eq_theta_il_a}).
The second-order relationship derived in 
\cite{Marquet2015WGNEthetas2,Marquet_thetas2_2019} 
corresponds to:
\vspace{-2mm}
\begin{equation}
\theta_s  
\; \approx \; 
\boxed{ \;
(\theta_s)_2
\; = \; 
(\theta_s)_1 \;
\exp\!\left[
 \: - \; \gamma \; q_t \: 
 \ln\!\left( \frac{r_v}{r_{\ast}} \right) 
 \: - \: \gamma \; 
 (q_l + q_{\rm rain} + q_i + q_{\rm snow})
\: \right]
\; }
\: ,
\label{eq_theta_s2}
\end{equation}
where $r_{\ast} \approx 12.4$~g/kg is a tuning parameter.

The reference entropies are computed with the standard temperature $T_r=T_0=273.15$~K and pressures $p_r=p_0=1000$~hPa and $e_r = e_{sw}(T_r)\approx 6.11$~hPa, with
Eq.~(\ref{defAPPsdr}) and (\ref{defAPPsvr})
leading to
$s_{dr}(T_r,p_r-e_r) \approx 6777$~J/K/kg,
$s_{vr}(T_r,e_r) \approx 12673$~J/K/kg,
$r_r = \varepsilon \: e_r \, / \,  (p_r - e_r) 
\approx 3.8214 $~g~kg${}^{-1}$ and 
$q_r=r_r/(1+r_r) \approx 3.8068$~g~kg${}^{-1}$

These values of the dry air and water vapor reference entropies lead to
\vspace{-1mm}
\begin{equation}
\boxed{\;
\Lambda_r \; = \; \frac{s_{vr}(T_r, e_r) - s_{dr}(T_r , p_r - e_r)}{c_{pd} }
\; \approx \; 5.869 \pm 0.003 \: .
\;}
\label{eq_lambda_r_value}
\end{equation}
The accuracy of $\Lambda_r$ can be computed from 
$\Delta s_{vr} \approx 2$~J/K/kg 
and
$\Delta s_{dr} \approx 1$~J/K/kg,
leading to
$\Delta \Lambda_r \approx 3/1000 = 0.003$.


Furthermore, for ice-free conditions ($q_i=0$), $\theta_{s1}$ can be expressed in terms of the \citet{Betts_QJ_1973} ``conservative variables'' ($q_t$, $\theta_l$ and $\theta_e$), according to
\vspace{-2mm}
\begin{align}
\theta_{s1}   
& \: \approx \; 
\theta_l \:
\exp\!\left( \Lambda \: q_t \right)
\; \approx \; 
\theta_e \:
\exp\!\left[ \:
  - \: 
  \left(
 \frac{L_v}{c_{pd} \: T} - \Lambda_r
 \right) \: q_t
\: \right]
\: ,
\\
\mbox{and thus} \;\;\;\;
\theta_{s1}  
& \: \approx \; 
\theta_l \:
\exp\!\left( 6 \: q_t \right)
\; \approx \; 
\theta_e \:
\exp\!\left( - \: 3 \: q_t \: \right)
\;\;\;\;
\mbox{}
\: 
\end{align}
because ${L_v}/(c_{pd} \: T) \approx 9$ and $\Lambda_r \approx 6$.

This result shows that $\theta_s$ is different from both $\theta_l$ and $\theta_e$ if $q_t \neq 0$, and is in a two-third position in between these two variables, a property indeed valid in Fig.~(\ref{fig_FIRE_MS22}):
\\ \hspace*{8mm}
    (i) for the FIRE-I (RF03B) observed radial flight 
    \citep{Marquet_QJ_2011};
\\ \hspace*{8mm}
    (ii) for drop-sounds observed during the 
    EUREC${\,}^4$A~-~Circle 
    \citep{Marquet_Stevens_JAS_2022}; 
\\ \hspace*{8mm}
    (iii) for ($2.5$~km) DYAMOND simulation outputs
    \citep{Marquet_Stevens_JAS_2022}; and
\\ \hspace*{8mm}
    (iv) for all studied undergone since 
    December 2008 
    (the date of the discovery of $\theta_s$).

\begin{figure*}[hbt]
\centering
\noindent
\includegraphics[width=0.85\linewidth]{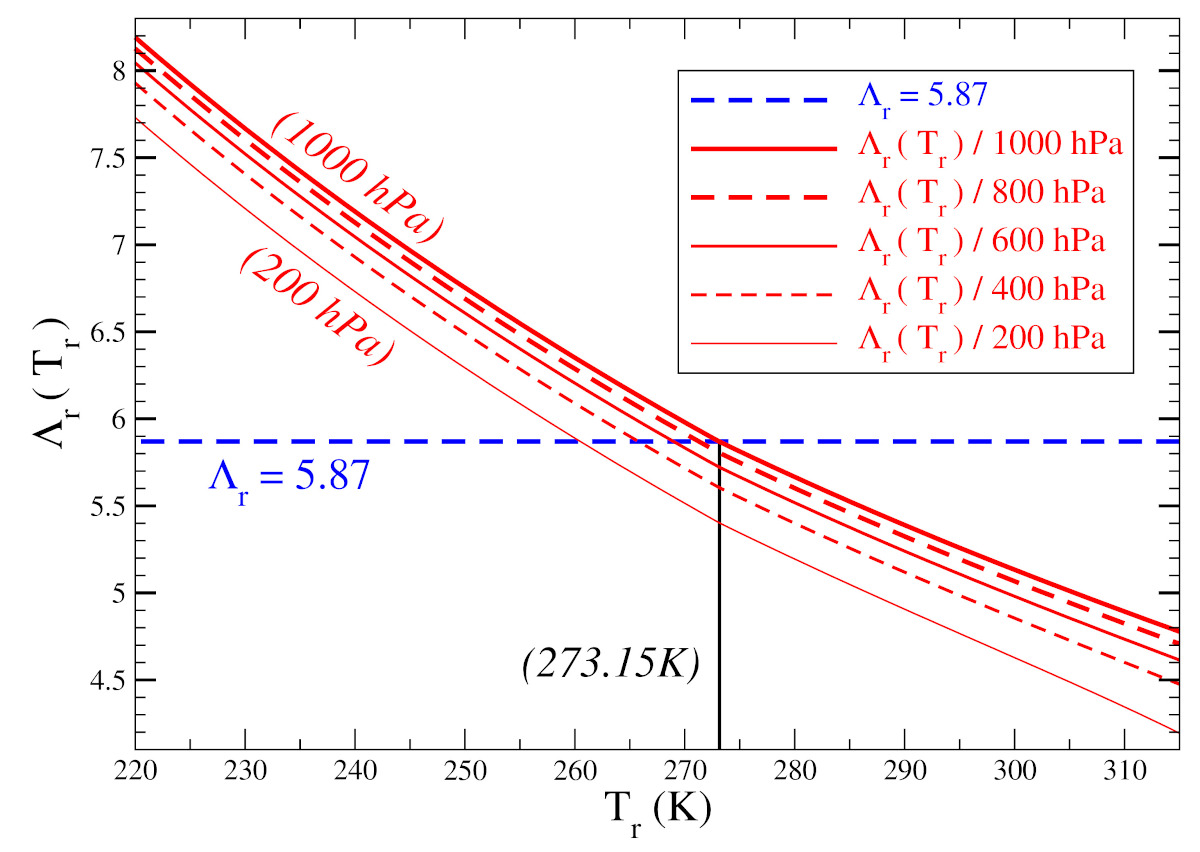} 
\vspace*{-4mm}
\caption{\it \small 
An unpublished plot of $\Lambda(T_r,p_r)$ for $T_r$ from $220$~K to $315$~K and $p_r$ from $1000$~hPa to $200$~hPa.
}
\label{fig_Lambdar}
\vspace{-5mm}
\end{figure*}

 \begin{center} \rule[0mm]{7.cm}{0.1mm} \end{center}
\vspace*{-2mm}

An important question is now to explain and justify the choices 
$T_r=T_0=273.15$~K and $p_r=p_0=1000$~hPa
for the reference values equal to the standard values for temperature and pressure.
Indeed, if $\theta_s$ given by (\ref{eq_thetas}) is independent on any values of $T_r$ and $p_r$, 
both $(\theta_s)_1$ and $(\theta_s)_2$ given by 
(\ref{eq_theta_s1}) and (\ref{eq_theta_s2}) do
depend on $\Lambda_r$ given by (\ref{eq_lambda_r_value}),
and thus on $T_r$ and $p_r$.

Fig.~\ref{fig_Lambdar} shows that $\Lambda(T_r,p_r)$ varies relatively little with the reference pressure $p_r$.
Also, why not choose the standard value $p_r=p_0=1000$~hPa?
And among all the possible values of $T_r$, the zero Celcius seems relevant since it represents a kind of average value according to the variations of $T$.
Also, why choose another value than $273.15$~K?

Moreover, it is the product of $\Lambda_r$ with $q_t$ that influences $\theta_s$, and therefore with a maximum impact in the lower layers and temperatures and pressures close to $p_0=1000$~hPa and $T_0=273.15$~K.

\begin{figure*}[hbt]
\centering
\noindent
\includegraphics[width=0.85\linewidth]{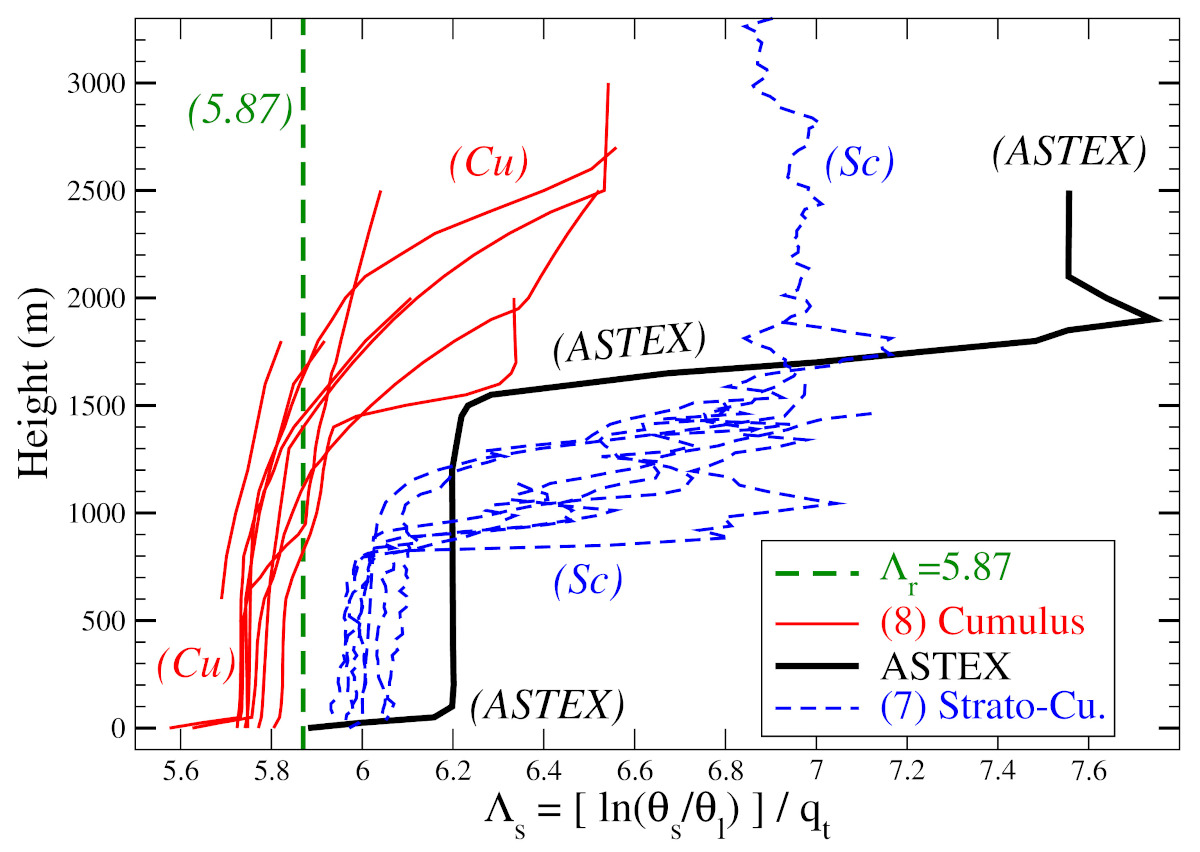} 
\\
\includegraphics[width=0.85\linewidth]{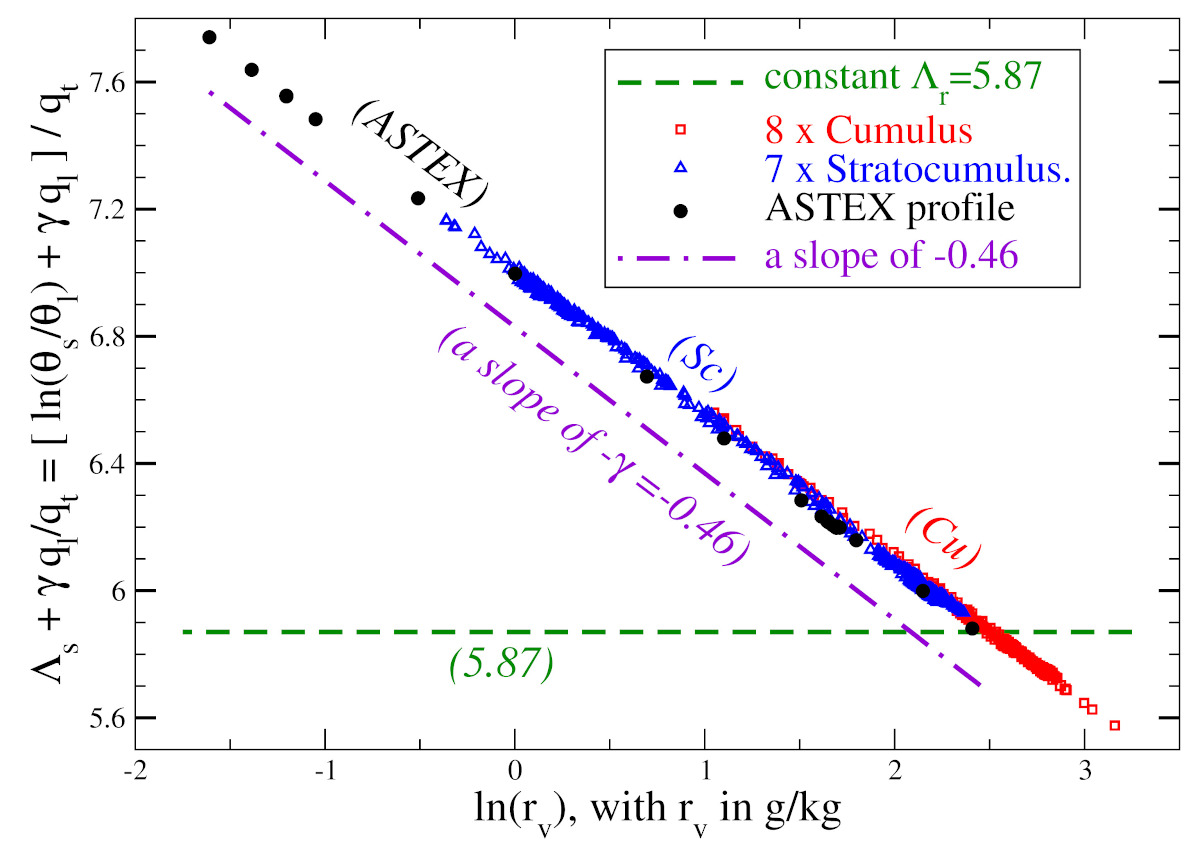} 
\vspace*{-3mm}
\caption{\it \small 
{\bf Top:}
an unpublished plot of $\Lambda(T_r,p_r)$ for $T_r$ from $220$~K to $315$~K and $p_r$ from $1000$~hPa to $200$~hPa.
{\bf Bottom:}
a plot from \cite{Marquet_thetas2_2019} showing $\Lambda_s$ a function of $\ln(r_v)$.
}
\label{fig_Lambda_r_s_a}
\end{figure*}

In any case, if the aim is to go beyond the first order to the second order, the impact of humidity must be taken into account.
Indeed, let us first define (without approximation) the more general variable $\Lambda_{s}$ by the formula ${\theta}_{s}=   \theta_{il} \;  \exp\! \left(  \Lambda_s \: q_t  \right)$, where  ${\theta}_{il}$, $q_t$ and ${\theta}_{s}$ given by (\ref{eq_thetas}) are three known quantities, and thus with $\Lambda_{s}$ given by the reciprocal relationship valid for the liquid case (with $q_i=0$ and ${\theta}_{il}={\theta}_l$):
\vspace*{-2mm} 
\begin{equation} 
\Lambda_s(T, p, q_v, q_l)  
\; = \; \frac{1}{q_t} \; \ln\!\left(  \frac{{\theta}_{s}}{{\theta}_{l}}   \right).
\label{def_lambda_s}
\end{equation}
Fig.~\ref{fig_Lambda_r_s_a}~(top) shows that, as a first guess, the term $\Lambda_s$ varies from about $5.7$ to $7.6$ for 16 vertical profiles of cumulus and stratocumulus (FIRE-I RF02/03/04/08/10, EPIC, DYCOMS-II, ARM-Cu, ASTEX, ATEX, GATE, BOMEX, SCMS-RF12).
A detailed study indicates that $\Lambda_s$ is:
\\ 
\hspace*{5mm} -
smaller for the moister (solid-red) cumulus 
($q_t$ and $q_v$ from $9$ to $23$~g/kg);
\\
\hspace*{5mm} -
larger for the moderate (dashed-blue) stratocumulus 
($q_t$ and $q_v$ from $1$ to $15$~g/kg); and
\\
\hspace*{5mm} -
even larger for the dryer (solid-black) ASTEX case 
($q_t$ and $q_v$ from $0.3$ to $11$~g/kg).

\clearpage

As a consequence, we are encouraged to seek for a value of $\Lambda_s$ which would decrease with $q_t$ or $q_v$.
To do so, according to the second-order formulations derived in  
\cite{Marquet2015WGNEthetas2} and \cite{Marquet_thetas2_2019} and leading to (\ref{eq_theta_s2}),
the dependence of $\Lambda_s$ with moisture variables could be of the form
\vspace*{-2mm} 
\begin{align} 
\Lambda_s 
& 
\: = \;
\frac{1}{q_t} \; 
\ln\!\left(  
  \frac{{\theta}_{s}}{{\theta}_{l}}
\right)
\; \approx \;
\Lambda_r 
\: - \: \gamma \: \ln\,( r_v )
\: - \: \gamma \: \frac{q_l}{q_t}
\: + \: Cste 
\label{def_lambda_star1} \: ,
\end{align}
where $\Lambda_r=5.87$ and $\gamma \approx 0.46$.
Fig.~\ref{fig_Lambda_r_s_a}~(bottom) clearly shows that, for the same 16 vertical profiles of cumulus and stratocumulus as in the top figure,
the term $\Lambda_s$ indeed varies like 
$- \, \gamma \ln(r_v) = - \, 0.46 _: \ln(r_v)$.
This linear law appears to be valid for a large range of $r_v$ (from $0.2$ to $24$~g~kg${}^{-1}$).
This corresponds to a linear decrease of
\vspace*{-2mm} 
\begin{align} 
\Lambda_s \: + \: \gamma \: \frac{q_l}{q_t}
& 
\: = \;
\frac{1}{q_t} \; 
\left[ \: 
\ln\!\left(  
  \frac{{\theta}_{s}}{{\theta}_{l}}
\right)
\: + \: \gamma \: q_l
\: \right]
\; \approx \;
\Lambda_r 
\: - \: \gamma \: 
\ln\,\left( \frac{r_v}{r_{\star}} \right)
\label{def_lambda_star2} \:
\end{align}
with $\ln(r_v)$, where the constant in (\ref{def_lambda_star1}) is written without loss of generality as $- \: \gamma \: \ln(r_{\star})$ in (\ref{def_lambda_star2}) in terms of a tuning parameter $r_{\star}$ to be determined.

\begin{figure*}[hbt]
\centering
\noindent
\includegraphics[width=0.85\linewidth]{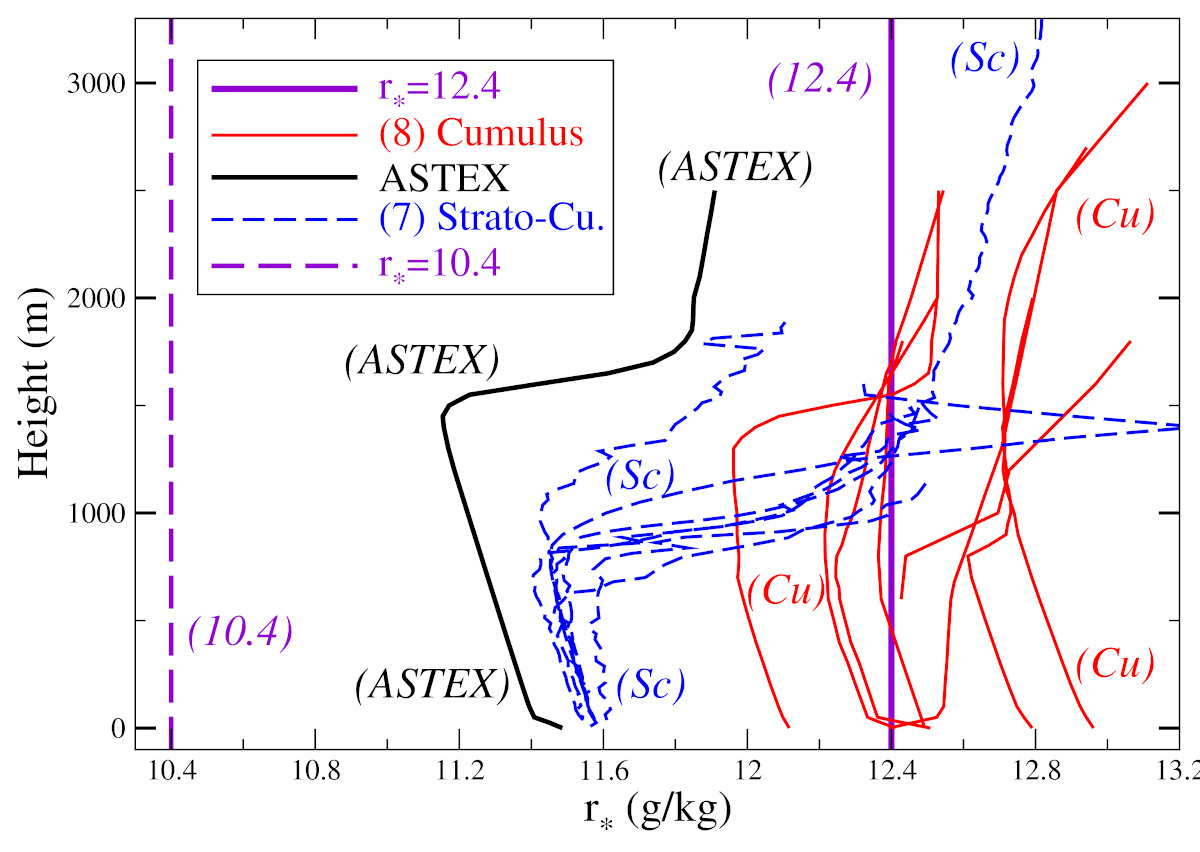}
\\
\includegraphics[width=0.85\linewidth]{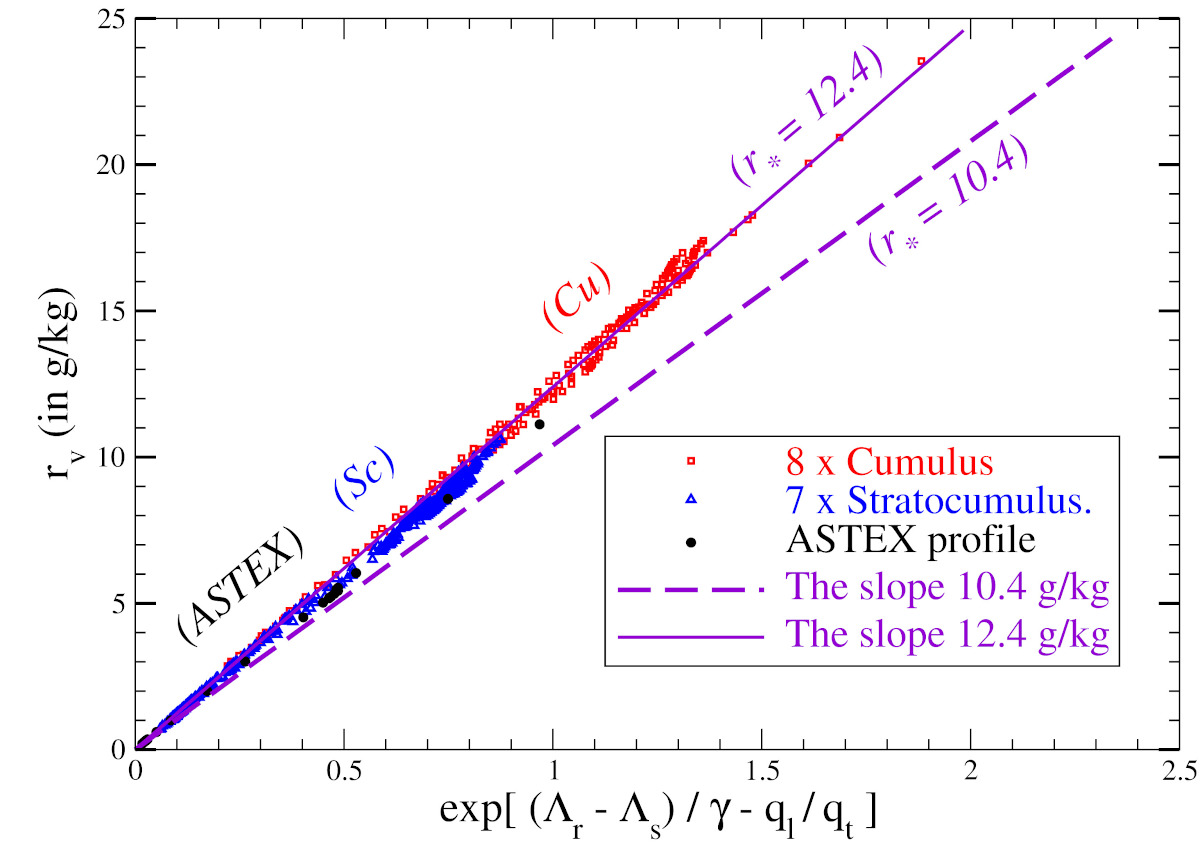}
\vspace*{-3mm}
\caption{\it \small 
{\bf Top:}
{\bf Bottom:}
}
\label{fig_Lambda_r_s_b}
\end{figure*}

It is then possible to plot in Fig.~\ref{fig_Lambda_r_s_b}~(top) the vertical profiles of this tuning parameter $r_{\star}$ for the same 16 vertical profiles of cumulus and stratocumulus as in the Figs.~\ref{fig_Lambda_r_s_a}, with $r_{\star}$ obtained from (\ref{def_lambda_star2}) and written as
\vspace*{-2mm} 
\begin{align} 
r_{\star}
& 
\: \approx \;
r_v \;
\exp\left\{\: 
\frac{1}{\gamma \: q_t} \; 
\left[ \: 
\ln\!\left(  
  \frac{{\theta}_{s}}{{\theta}_{l}}
\right)
\: + \: \gamma \: q_l
\: \right]
\: - \: \frac{\Lambda_r}{\gamma}
\: \right\}
\label{def_lambda_star3} \: .
\end{align}

A first-guess value for $r_{\star}$ derived in \cite{Marquet2015WGNEthetas2} and \cite{Marquet_thetas2_2019}
is 
$r_r \times \exp(1) \approx 3.82 \times 2.718 \approx 10.4$~g/kg.

-------------------------

According to the second-order formulation derived in \cite{Marquet_thetas2_2019}, it is then possible to find the tuning mixing ratio $r_{\star}$ for which
hold true, where $r_{\star}$ will play the role of positioning the dashed-dotted thick black line of slope $- \,\gamma \approx - 0.46$ in between the cumulus and stratocumulus profiles.
This corresponds to a linear fitting of  $r_v$ against 
$\exp[\: (\Lambda_r - \Lambda_s)/\gamma 
+ \gamma \: q_l \:]$, 
$r_{\star}$ being the average slope of the scattered data points.
It is shown in 
Fig.~\ref{fig_Lambda_r_s_b}~(top) 
that the value $r_{\star} \approx 12.4$~g~kg${}^{-1}$ corresponds to a relevant fitting of all cumulus and stratocumulus vertical profiles for a range of $r_v$ from $0.2$ up to $24$~g~kg${}^{-1}$.

It is finally shown in
Fig.~\ref{fig_Lambda_r_s_b}~(bottom)
that  $\Lambda_s$ can indeed be approximated by $\Lambda_{\star}(r_v, r_{\ast})$ given by (\ref{def_lambda_star1}), with a clear improved accuracy in comparison with the constant value $\Lambda_r \approx 5.87$ for a range of $r_v$ between $0.2$ and $24$~g~kg${}^{-1}$.
Curves of $\Lambda_{\star}(r_v, r_{\ast})$ with $r_{\ast} = 10.4$ and $12.4$~g~kg${}^{-1}$ (solid black lines) both simulate with a good accuracy the non-linear variation of $\Lambda_s$ with $r_v$, and both simulate the rapid increase of $\Lambda_s$ for $r_v < 2$~g~kg${}^{-1}$.

\clearpage

\begin{figure*}[hbt]
\centering
\noindent
\includegraphics[width=0.98\linewidth]{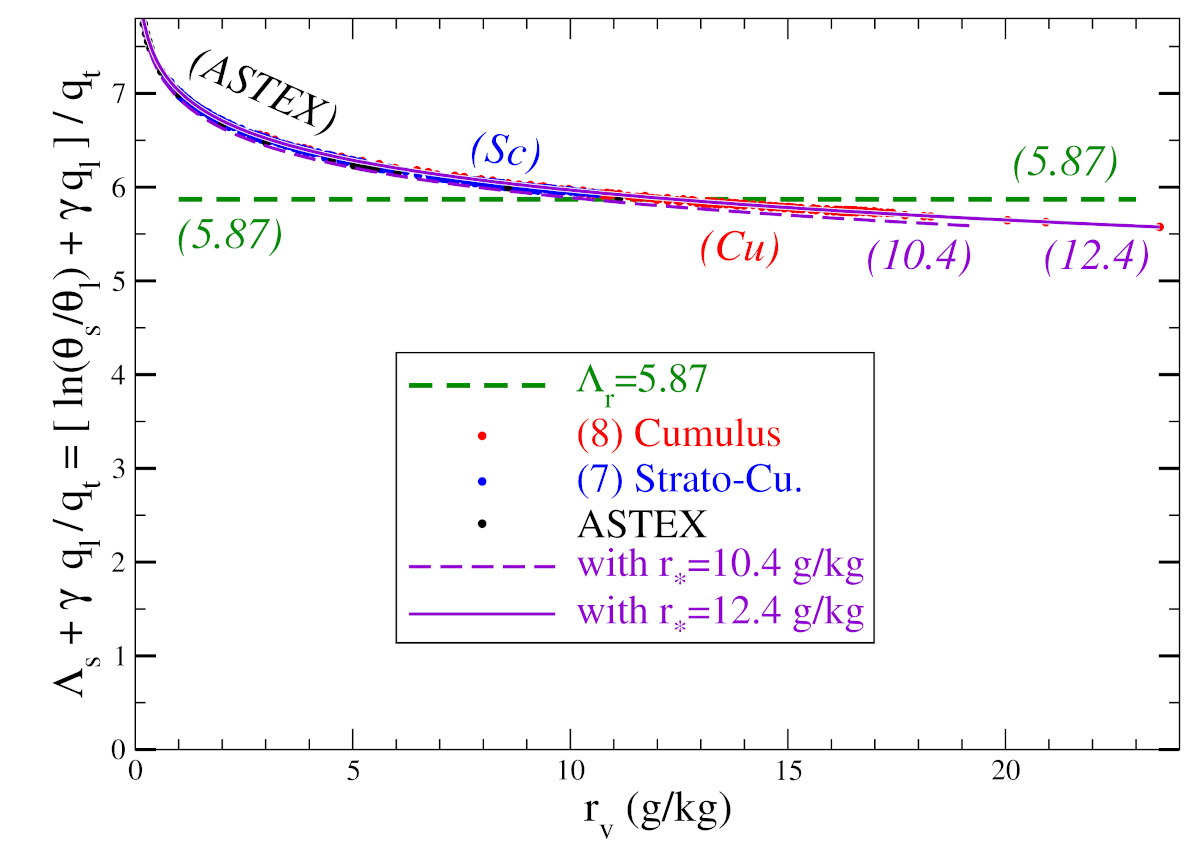}
\vspace*{-3mm}
\caption{\it \small 
{\bf Top:}
{\bf Bottom:}
}
\label{fig_Lambda_r_s_c}
\end{figure*}
\clearpage

\vspace{-2mm}

 \begin{center} \rule[0mm]{7.cm}{0.1mm} \end{center}
\vspace*{-2mm}

All definitions of 
$\theta_{s/\rm{HH87}}$ or $\theta_l$  by \cite{Hauf_Hoeller_1987},
$\theta_e$ by \cite{Betts_QJ_1973}
recalled in (\ref{eq_HH87}),
(\ref{eq_theta_e})

Previously, the relationship (\ref{eq_HH87}) for $\theta_{s/\rm{HH87}}$ is usually approximated by the first-order approximation $\theta_e$ of \cite{Betts_QJ_1973} given by (\ref{eq_theta_e}).
Similarly, $\theta_l$ defined by (\ref{eq_thetal_HH87}) is usually approximated by the first-order approximation of \cite{Betts_QJ_1973} given by (\ref{eq_thetal_B73}).

By doing this, the first-order approximation neglects the impact of the ratio $e/e_0$ for $\theta_l$ defined by (\ref{eq_thetal_HH87}), and also the impact of the choice of $T_0$ and $e_0$ in the additional terms left outside the logarithms in (\ref{eq_HH87}) and (\ref{eq_thetal_HH87}), which depends on the reference values
$s_d(T_0,p_0)$, $s_v(T_0,e_0)$ and $s_l(T_0)$.

\vspace{-2mm}

 \begin{center} \rule[0mm]{7.cm}{0.1mm} \end{center}
\vspace*{-2mm}

The moist-air entropy variables $\theta_s$, $\theta_{s1}$ and $\theta_{s2}$ have been studied in several papers since 2011:
\vspace*{1mm}
\\ $\bullet$ 
for computing the moist-air Brunt-V\"ais\"al\"a frequency
\citep{Marquet_Geleyn_QJ_2013};
\vspace*{2mm}
\\ $\bullet$ 
for computing and studying the associated Potential Vorticity
$PV(\theta_s)$ \citep{Marquet_QJ_2014};
\vspace*{2mm}
\\ $\bullet$ 
in a synthetic chapter about moist-air thermodynamics
\citep{Marquet_Geleyn_2015};
\vspace*{2mm}
\\ $\bullet$ 
for computing improved adjustment of marine bulk formulas
\citep{Marquet_Belamari_WGNE_Lewis_2017};
\vspace*{2mm}
\\ $\bullet$ 
for computing the work- and stream-functions for the hurricane Dumil\'e
\citep{Marquet_JAS_2017};
\vspace*{2mm}
\\ $\bullet$ 
to answers O. Pauluis's criticisms with Hector-the-Convector
\citep{Marquet_Thibaut_JAS_2018};
\vspace*{2mm}
\\ $\bullet$ 
to explain in French and English the third-law of thermodynamics
\citep{Marquet_LM_2019a,Marquet_LM_2019b};
\vspace*{2mm}
\\ $\bullet$ 
to improve a $\theta_s$-based E.I.S. in the IFS-ECMWF model
\citep{Marquet_Bechtold_WGNE_2020};
\vspace*{2mm}
\\ $\bullet$ 
to compare the H2O pathways with lines of constant $\theta_e$ or $\theta_s$
\citep{Marquet_Bailey_WGNE_2021};
\vspace*{2mm}
\\ $\bullet$ 
to provide a common method to derive $\theta_l$, $\theta_e$ and $\theta_s$
\citep{Marquet_Stevens_JAS_2022};
\vspace*{2mm}
\\ $\bullet$ 
to use $\theta_s$ and $q_t$ in a 1D-Var assimilation scheme
\citep{Marquet_al_AMT_2022}.
\vspace{2mm}

\newpage
 \section{\underline{\large Comparison of dry and moist-air potential temperature}}
 \label{==Compare_ThetaX==}
\vspace{-2mm}

\begin{figure}[htb]
\centering
\includegraphics[width=0.419\linewidth]{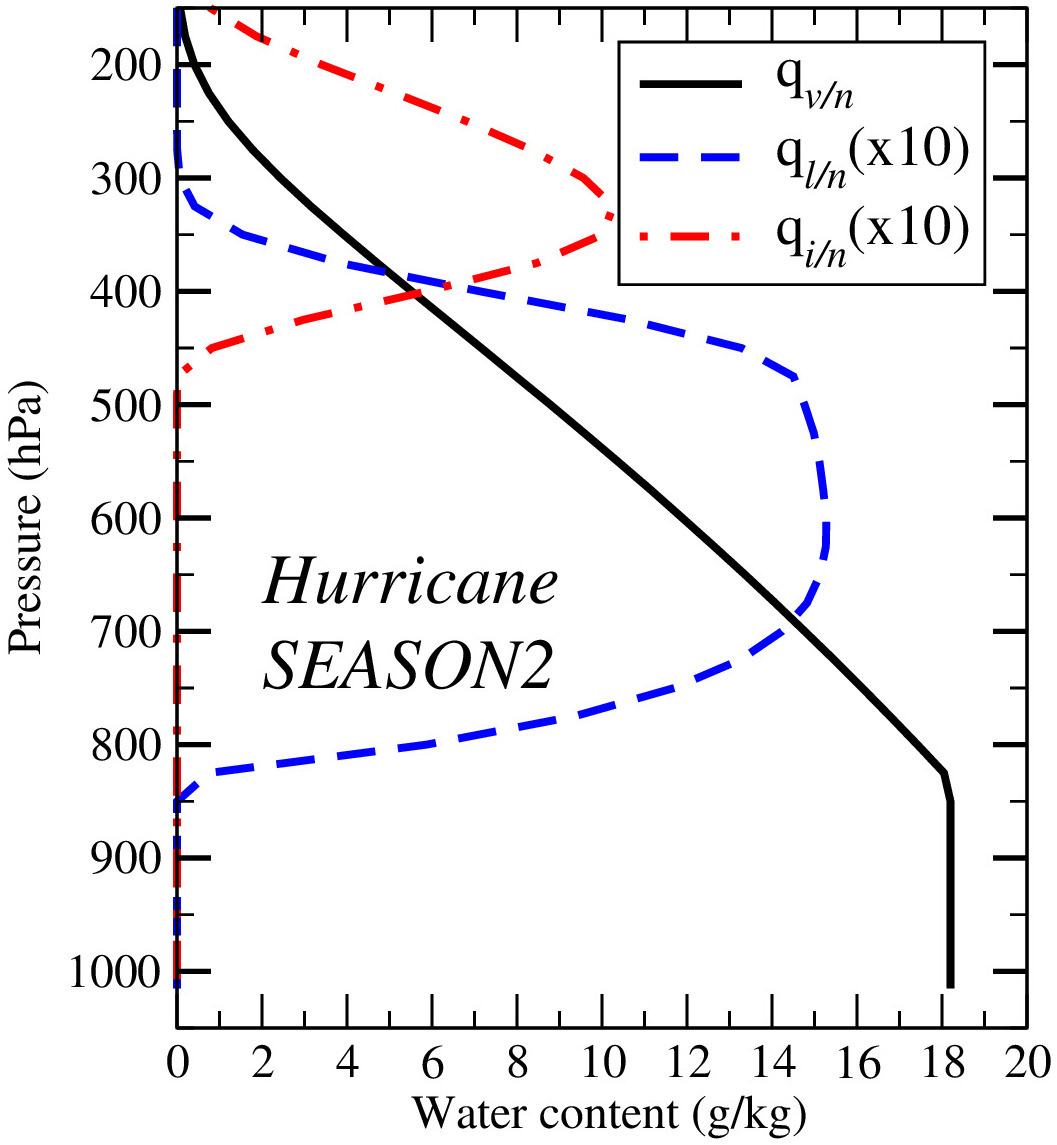}
\!\!
\includegraphics[width=0.57\linewidth]{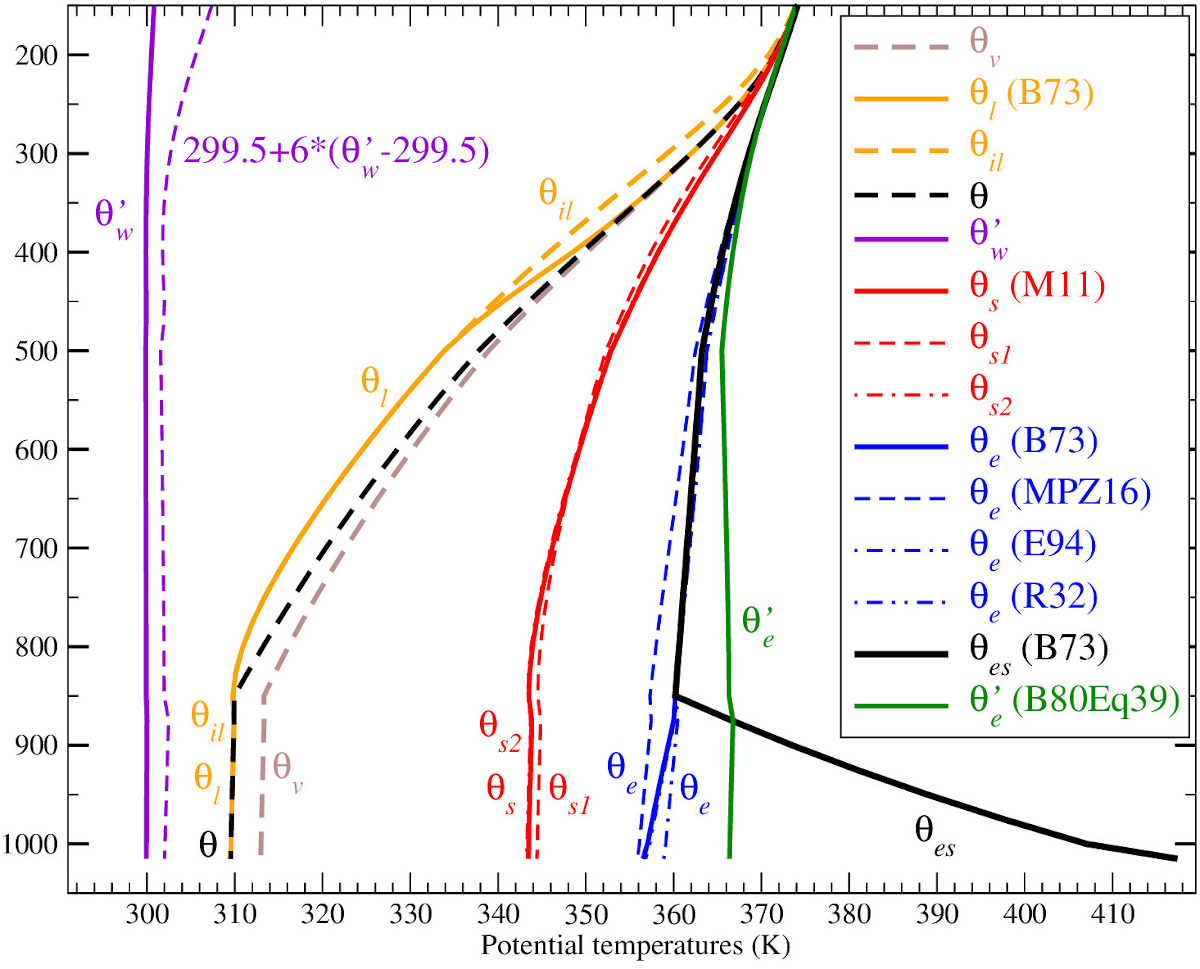}
\vspace{-8mm}
\caption{\small \it
Vertical profiles for an updraft based on the Hurricane ``Season'' \citep{Jordan_1958}. 
Left: water vapor ($q_v$),
liquid water ($\,10 \times q_l$) and 
ice ($\,10 \times q_i$) 
specific contents (in g/kg).
Right: potential temperatures (in K).
}
\label{fig_qvil_Theta}
\end{figure}

All previous dry- and moist-air potential temperatures 
($\theta'_w$, $\theta_v$, $\theta_l$, $\theta_{il}$, 
$\theta$,  $\theta_s$, $\theta_e$, $\theta_{es}$ and 
$\theta'_e$)
have been computed for a realistic updraft based on (a slightly modified version of) the vertical profile of the Hurricane ``Season'' described in Table~5 of \cite{Jordan_1958}, with precipitating and entrainment processes and with liquid water, ice and mixed phases.
Figs~\ref{fig_qvil_Theta} shows that: 
\vspace*{1mm} \\ $\bullet$
a clear-air region extends from the surface up to about $820$~hPa;
\vspace*{1mm} \\ $\bullet$
a mixed-phase exists between about $500$~hPa and $300$~hPa; 
\vspace*{1mm} \\ $\bullet$
$\theta_l$ and $\theta_{il}$ (in orange) remain close to $\theta$ (dashed black), with differences (in the cloud region between $850$ and $200$~hPa) of the same order of magnitude than the difference between $\theta$ and $\theta_v$ (dashed grey);
\vspace*{1mm} \\ $\bullet$
the four versions of $\theta_e$ (in blue) are not so different from each others (they all increases with height with discrepancies of less than $1$ to $2$~K);
\vspace*{1mm} \\ $\bullet$
the vertical changes in $\theta'_w$ (in violet) and $\theta_e$ (in blue) are very different ($\theta'_w$ is almost conservative whereas the four $\theta_e$ increase with height);
\vspace*{1mm} \\ $\bullet$
the first- and second-order approximations $\theta_{s1}$ and $\theta_{s2}$ (dashed and dotted-dashed red) remain very close to the exact version $\theta_s$ (solid red), up to less than $0.6$~K for $\theta_{s1}$ and $0.05$~K for $\theta_{s2}$  ;
\vspace*{1mm} \\ $\bullet$
the entropy potential temperature $\theta_s$ (in red) is indeed in a two-third position in between $\theta$ or $\theta_l$ (dashed black or orange) and $\theta_e$ (in blue), and is significantly different from all the other potential temperatures;
\vspace*{1mm} \\ $\bullet$
the saturated-equivalent temperature $\theta_{es}$ (in black) is very different from the four $\theta_e$ (in blue) in the unsaturated low-levels region below the cloud ($\theta_{es}$ is much larger by more than $50$~K);
\vspace*{1mm} \\ $\bullet$
the pseudo-adiabatic equivalent potential temperature $\theta'_e$ (in green) computed from 
Eqs.~(\ref{eq_theta_pe1})-(\ref{eq_theta_pe3}), 
which looks like $\theta'_w$ but with enhanced variations close to those of the (about $6$ times) enhanced (dashed violet) vertical profile of 
$299.5+6  \times (\theta'_w - 299.5)$.
\vspace*{-0mm}

\begin{figure}[htb]
\centering
\includegraphics[width=0.94\linewidth]{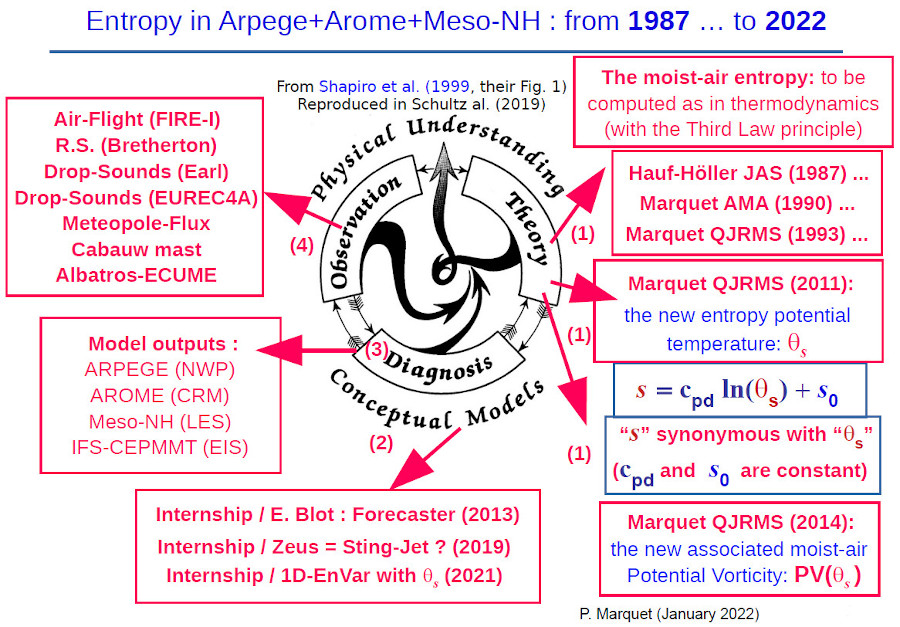}
\\
\vspace{-2mm}
---------------------------------------------------------------------------------------------------------
\\
\includegraphics[width=0.94\linewidth]{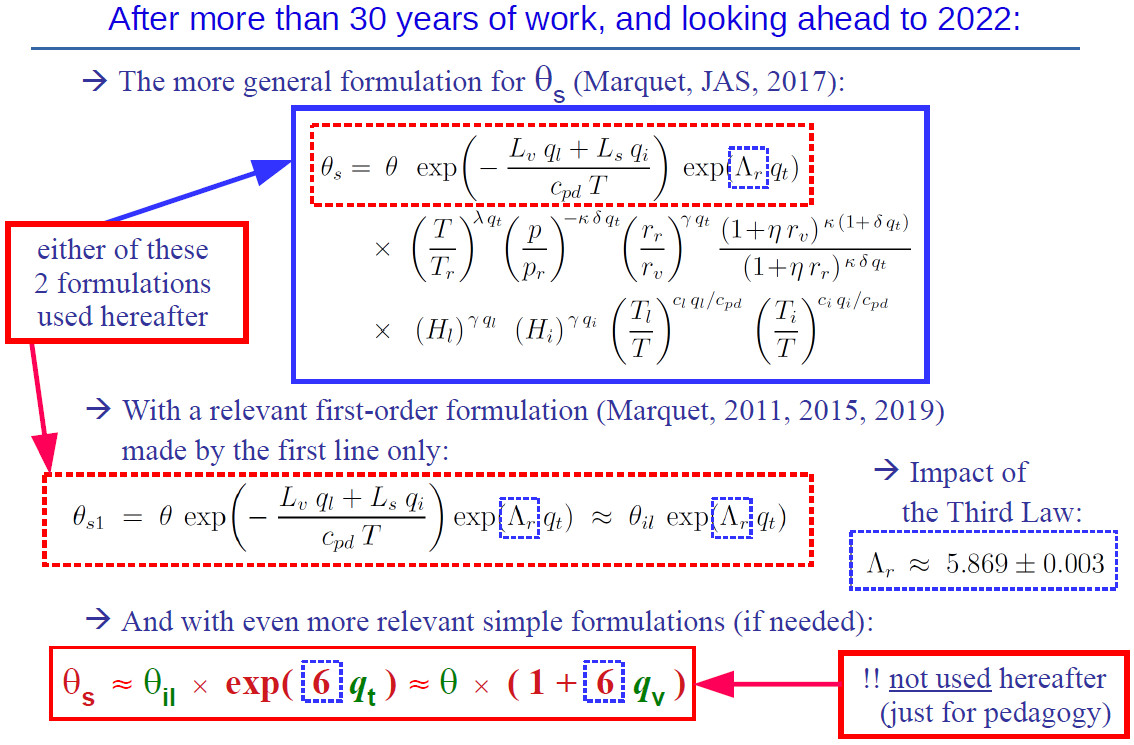}
\vspace{-4mm}
\caption{\small \it
Two slides from 
\cite{Marquet_2022},
where the main properties of the moist-air specific entropy 
$s=c_{pd} \: \ln(\theta_s) + s_0$, 
the associated potential temperature $\theta_s$ and the potential vorticity $PV(\theta_s)$ are summarized.
}
\label{fig_}
\end{figure}

\begin{figure}[htb]
\centering
\vspace{-10mm}
\includegraphics[width=0.7\linewidth]{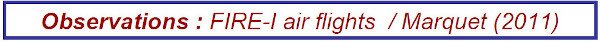}
\\
\includegraphics[width=0.47\linewidth]{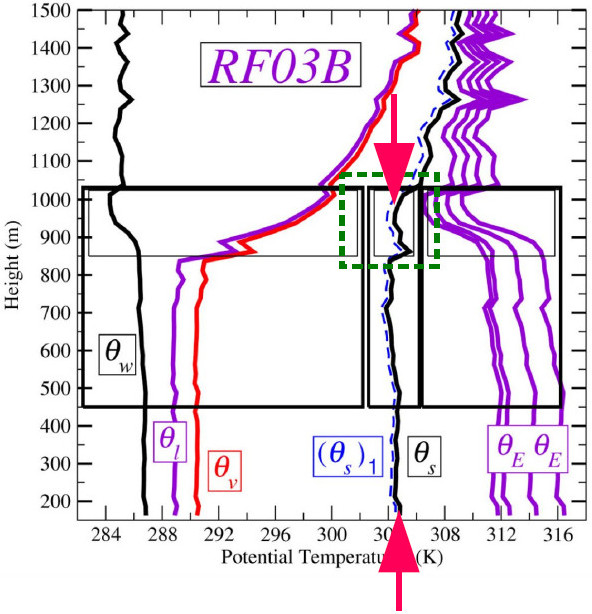}
\includegraphics[width=0.45\linewidth]{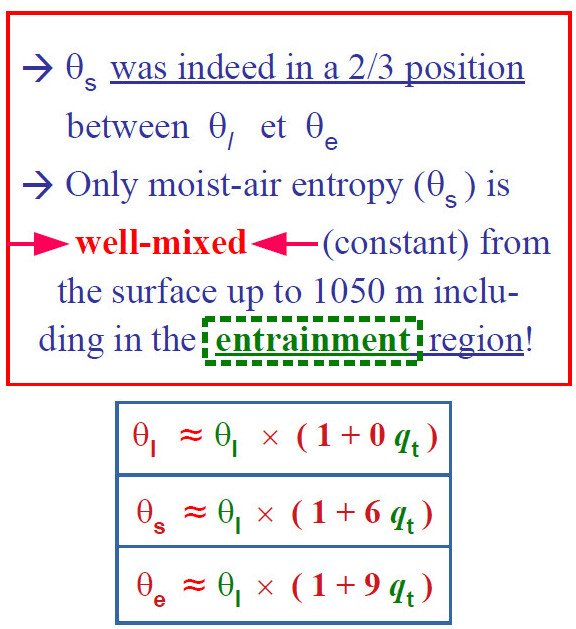}
\\
\vspace{-2mm}
---------------------------------------------------------------------------------------------------------
\\
\vspace*{2mm}
\includegraphics[width=0.95\linewidth]{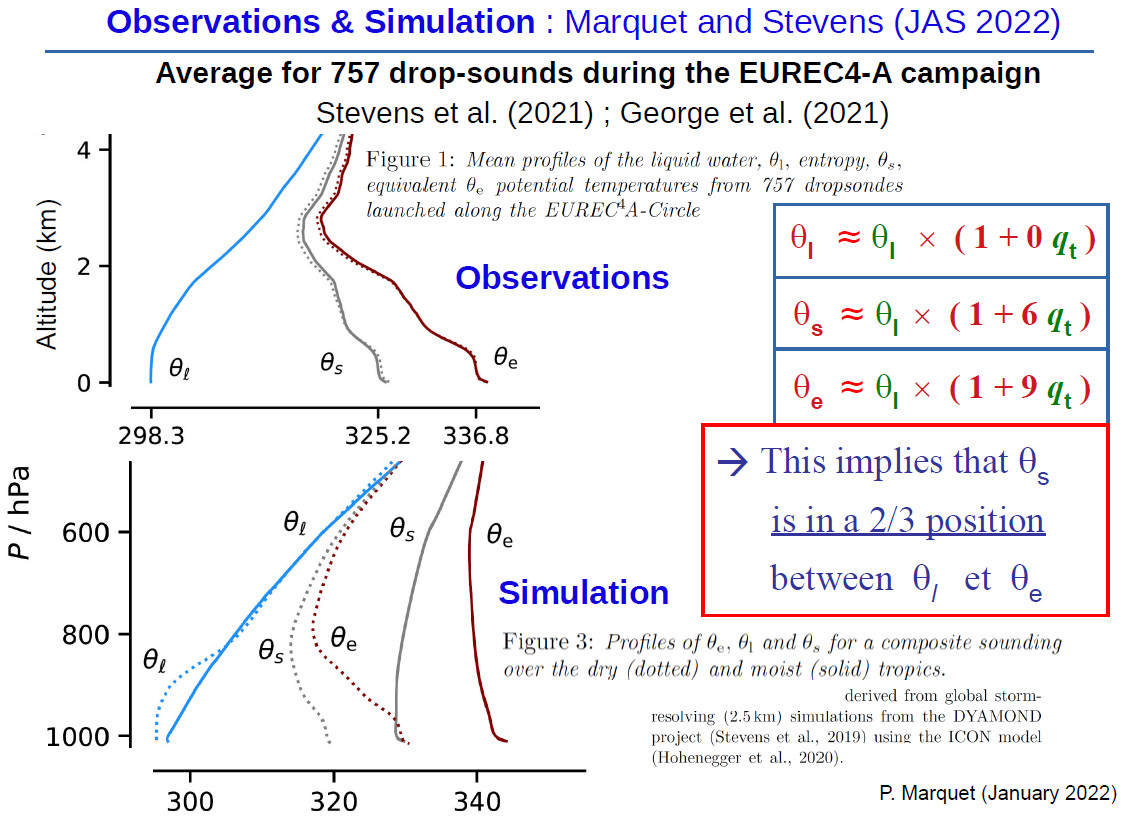}
\vspace{-3mm}
\caption{\small \it
From \cite{Marquet_2022}.
{\bf Top:} 
vertical profiles of potential temperatures from the Radial-Flight 03B of FIRE-I \citep{Marquet_QJ_2011}.
{\bf Bottom:} 
vertical profiles of potential temperatures from dropsondes along the EUREC${\:}^4$A-Circle and from tropical simulated  composite soundings \citep{Marquet_Stevens_JAS_2022}
}
\label{fig_FIRE_MS22}
\end{figure}

\begin{figure}[htb]
\centering
\vspace{-9mm}
\includegraphics[width=0.83\linewidth]{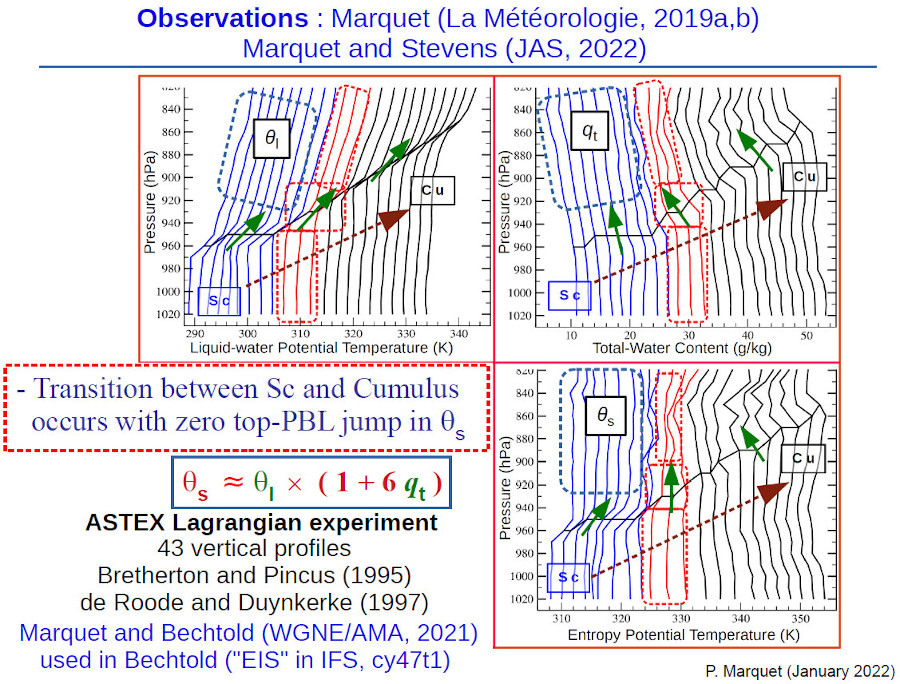}
\\
\vspace{-2mm}
---------------------------------------------------------------------------------------------------------
\\
\includegraphics[width=0.83\linewidth]{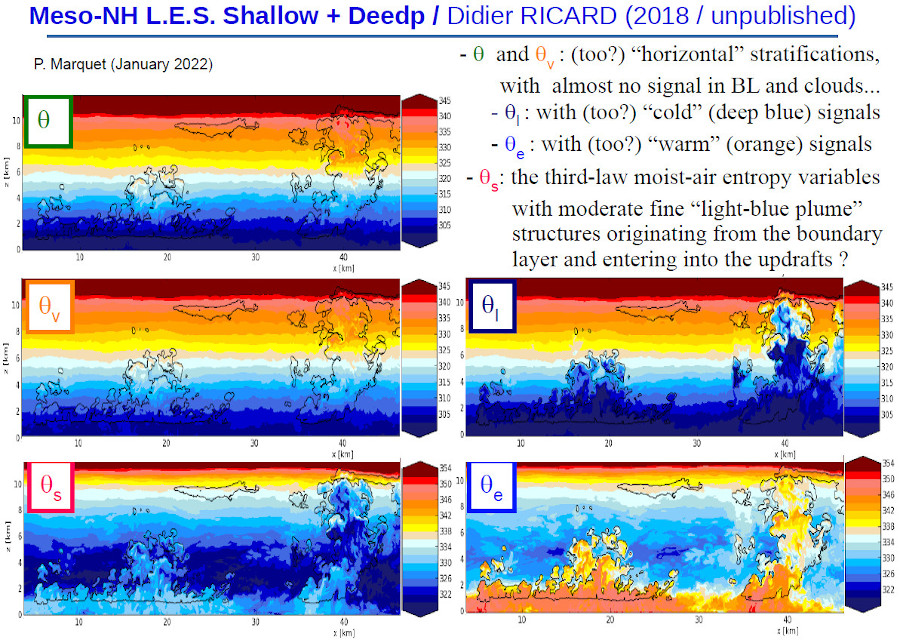}
\vspace{-4mm}
\caption{\small \it
From \cite{Marquet_2022}.
{\bf Top:}
half of the ASTEX Lagrangien vertical profiles for $\theta_l$, $q_t$ and $\theta_s$ computed from the radio-sounding dataset for $T$, $q_v$, $q_l$
\citep{Marquet_LM_2019a,Marquet_LM_2019b,
Marquet_Stevens_JAS_2022}.
The special property $\partial \theta_s / \partial p \approx 0$ at the transition between Stratocumulus and Cumulus regimes has been used to built a new Estimated-Inersion-Strength in the IFS model at ECWF \citep{Marquet_Bechtold_WGNE_2020}.
{\bf Bottom:}
diagnostic outputs from the Meso-NH models (Ricard, 2018, unpublished), where the moist-air entropy (and $\theta_s$) entering the plumes of convective cells are more similar to mid-tropospheric values at $6$ to $8$~km (differently from colder PBL values for $\theta_l$ and warmer PBL values for $\theta_e$). 
}
\label{fig_LM19_MS22}
\end{figure}

\begin{figure}[htb]
\centering
\vspace{-9mm}
\includegraphics[width=0.98\linewidth]{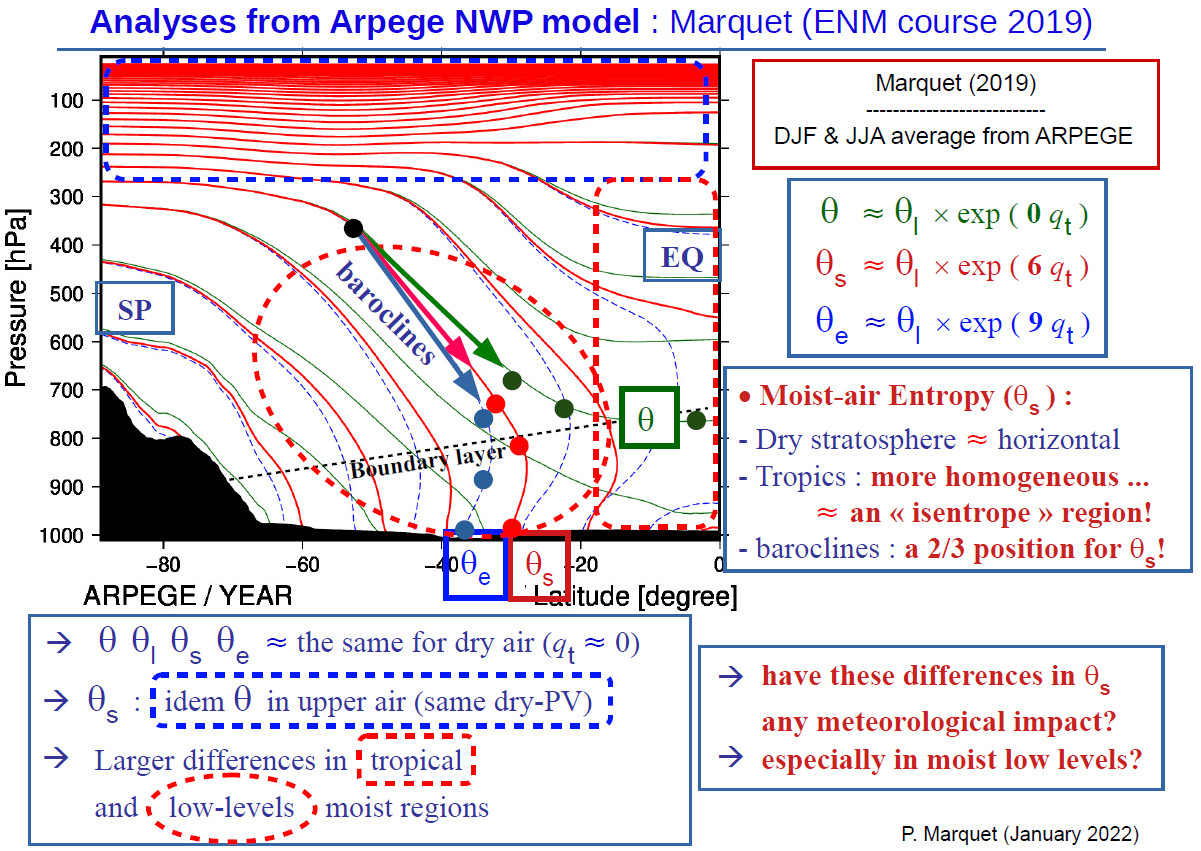}
\\
\vspace{-2mm}
---------------------------------------------------------------------------------------------------------
\\
\includegraphics[width=0.98\linewidth]{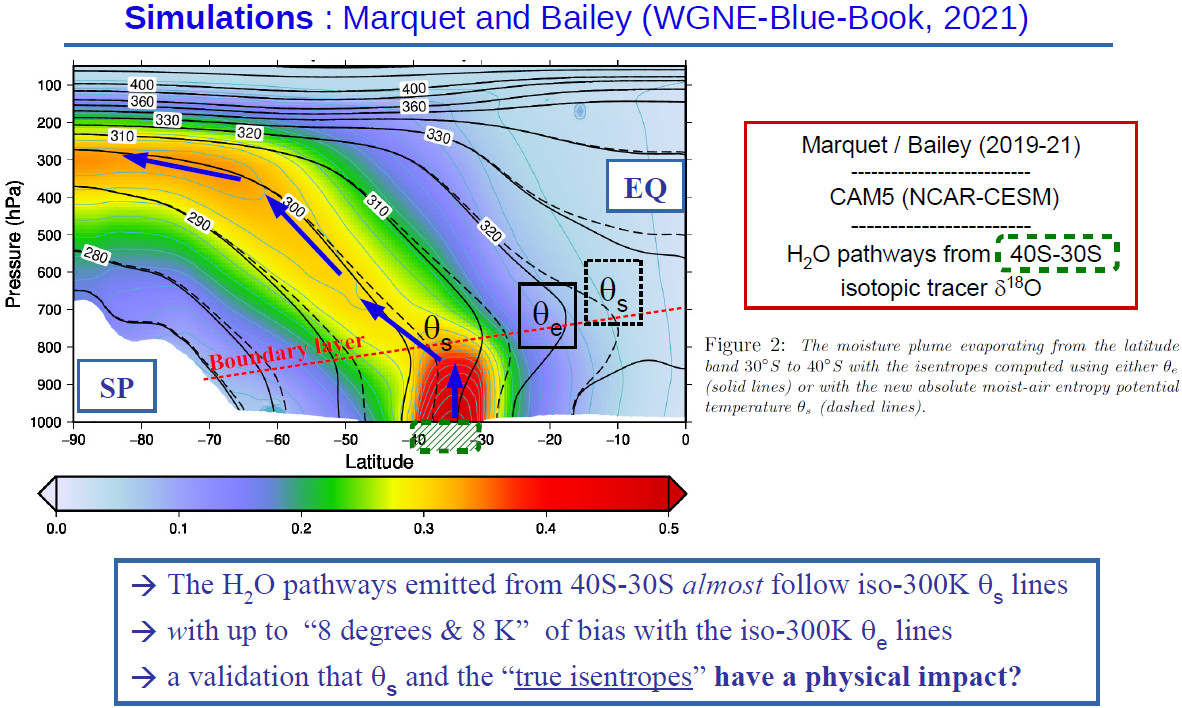}
\vspace{-3mm}
\caption{\small \it
From \cite{Marquet_2022}.
{\bf Top:}
the surface of equal values of $\theta$, $\theta_s$ and $\theta_e$ do not coincide in annual and zonal average, with a $2/3$ position of $\theta_s$ in between $\theta$ and $\theta_e$ (unpublished result; from a 2019 course at the ENM / French School of Meteorology).
Only $\theta_s$ represents the moist-air entropy and generates the true moist-air isentropes.
One of the consequence is an impact on the definition of moist-air baroclinic waves in mid-latitudes (different slopes of ``isentropic'' surfaces).
{\bf Bottom:}
the H${}_2$O pathways originating from a certain band of latitude seems to preferentially follow the moist-air isentropes labeled with $\theta_s$
\citep{Marquet_Bailey_WGNE_2021}.
}
\label{fig_Marquet_Bailey_WGNE_2021}
\end{figure}
\clearpage

\begin{figure}[htb]
\centering
\vspace{-10mm}
\includegraphics[width=0.98\linewidth]{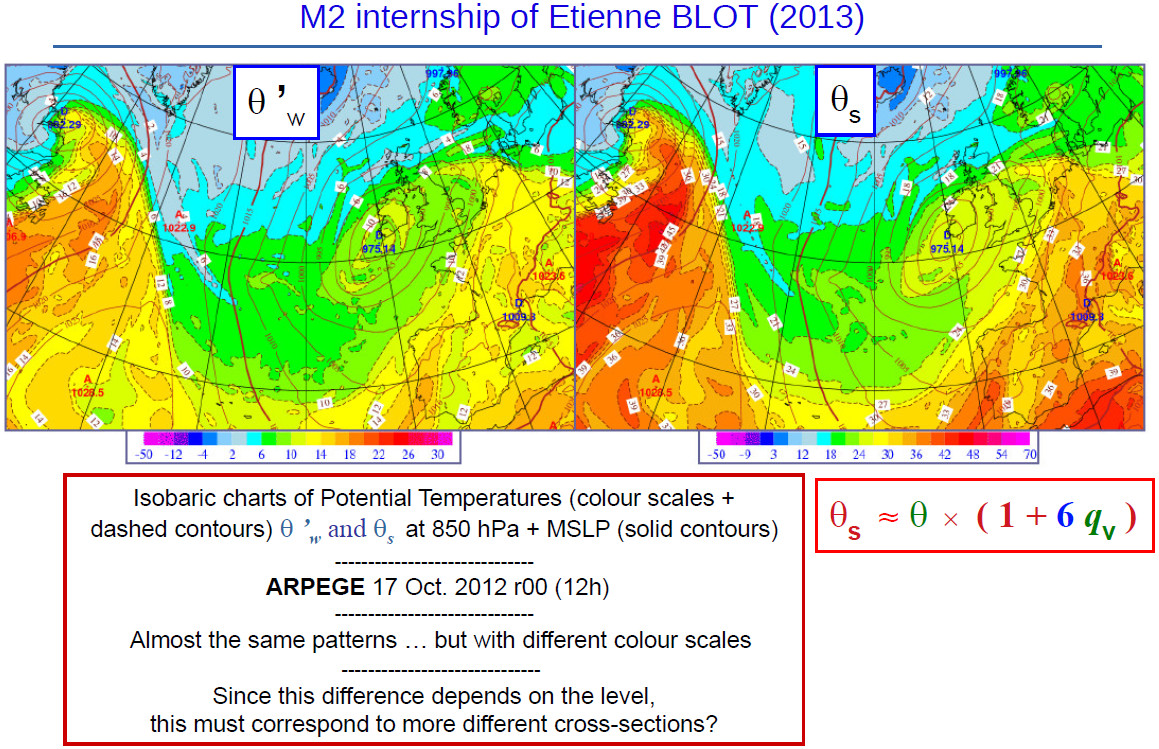}
\\
\vspace{-2mm}
---------------------------------------------------------------------------------------------------------
\\
\includegraphics[width=0.98\linewidth]{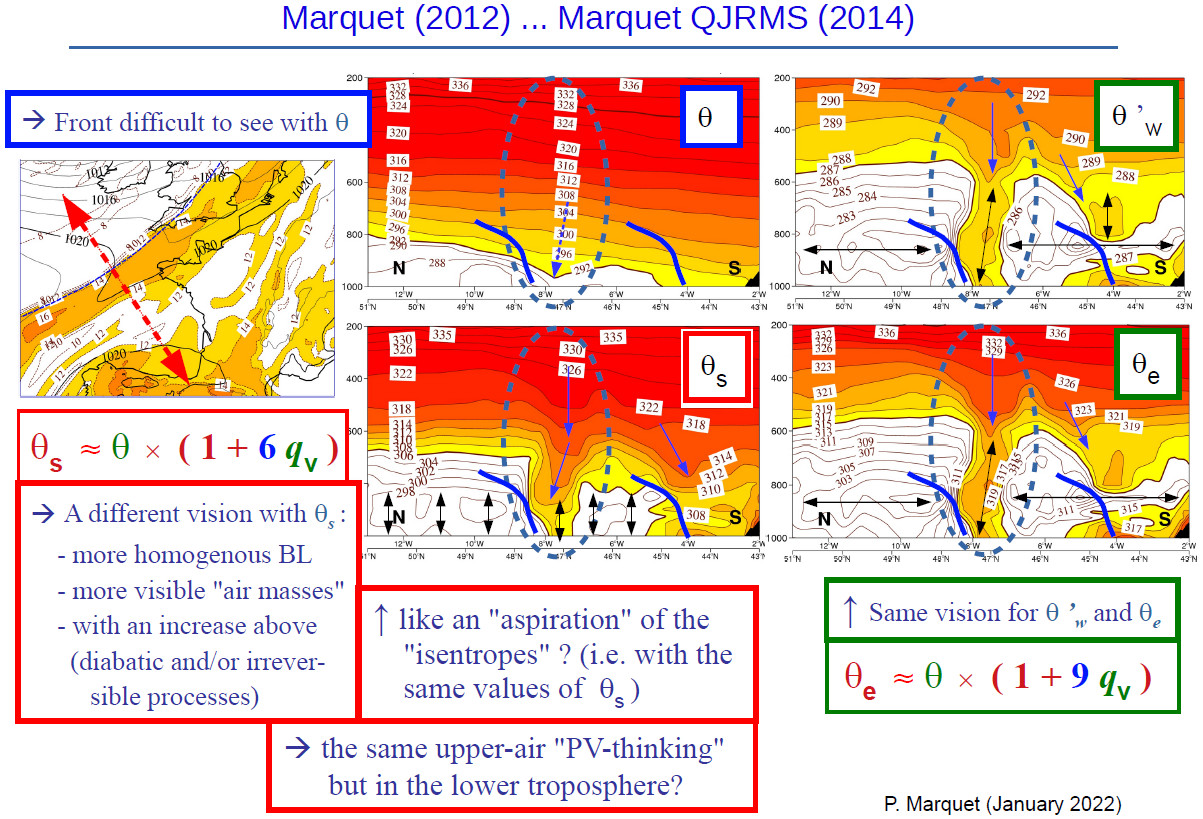}
\vspace{-4mm}
\caption{\small \it
From \cite{Marquet_2022}.
{\bf Top:}
It is possible to plot air-mass charts (here at $850$~hPa) by replacing the pseudo-adiabatic variable $\theta'_w$ (or $\theta_e$) by the moist-air entropy variable $\theta_s$ \citep{Blot_2013}. 
It is just necessary to change the color scale.
{\bf Bottom:}
Vertical cross-sections published in \cite{Marquet_QJ_2014} showing how a front can be described with the variables $\theta$, $\theta'_w$, $\theta_s$ and $\theta_e$.
The moist-air entropy variable $\theta_s$ allows a better view of the air masses in the lower layers in relation to the frontal surfaces, with a boundary layer that is better mixed (homogeneous) in $\theta_s$, and with the convective regions associated with some sort  of ``aspiration of moist-air isentropes''.
}
\label{fig_Blot13_Marquet14}
\end{figure}
\clearpage

\begin{figure}[htb]
\centering
\vspace{-9mm}
\includegraphics[width=0.98\linewidth]{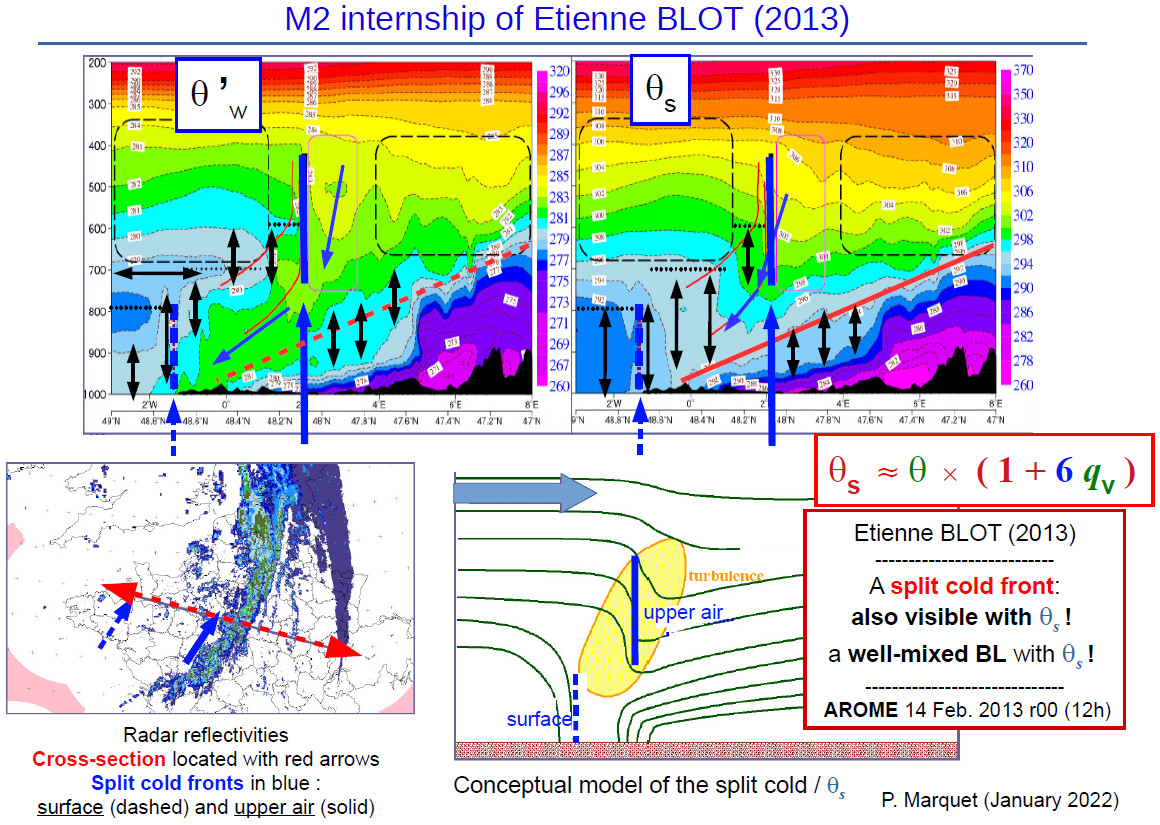}
\\
\vspace{-2mm}
---------------------------------------------------------------------------------------------------------
\\
\includegraphics[width=0.98\linewidth]{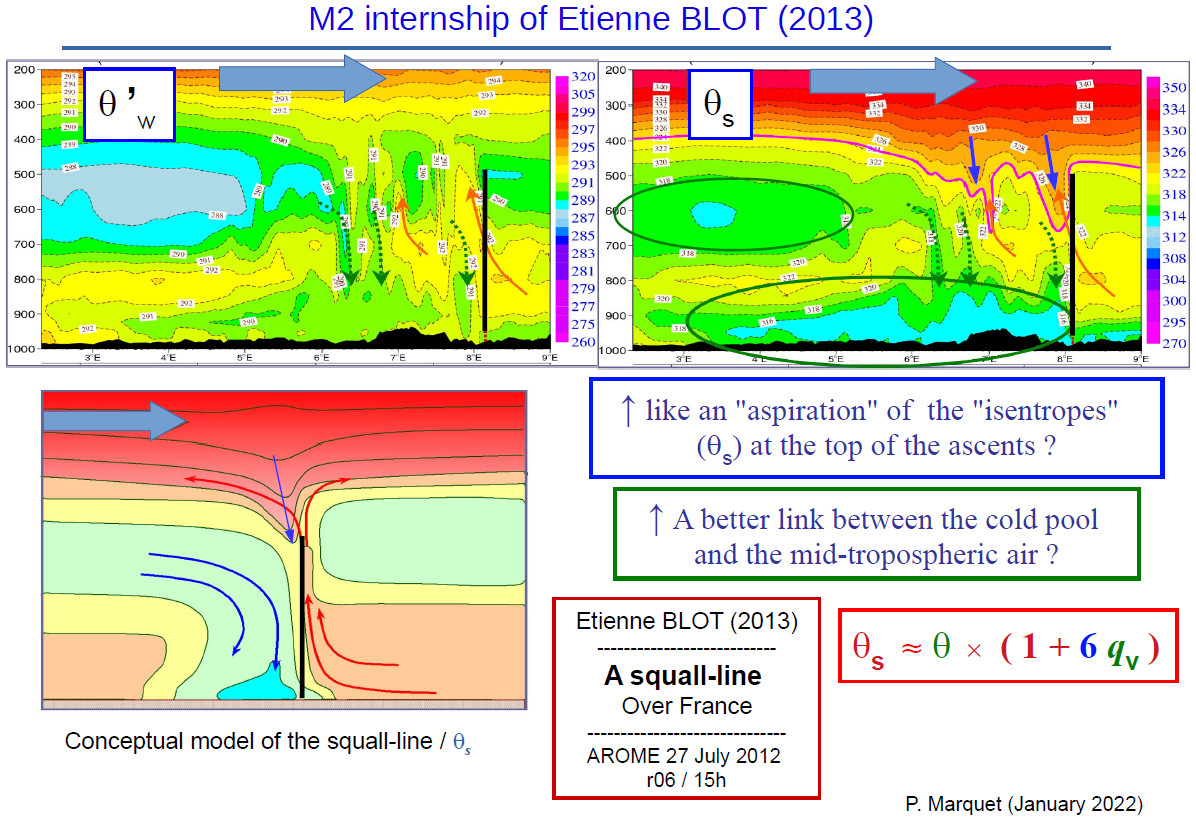}
\vspace{-4mm}
\caption{\small \it
From \cite{Marquet_2022} and \cite{Blot_2013}.
{\bf Top:}
the structure of a split cold front can be represented by the moist-air entropy variable $\theta_s$, as relevant as with the $\theta'_w$ variable.
{\bf Bottom:}
the same is true for a squall-line line, which is clearly visible in terms of $\theta_s$, with the convective regions associated with some sort of ``aspiration of moist-air isentropes''.
}
\label{fig_Blot_2013_split_front_squall_line}
\end{figure}
\clearpage

\begin{figure}[htb]
\centering
\vspace{-10mm}
\includegraphics[width=0.98\linewidth]{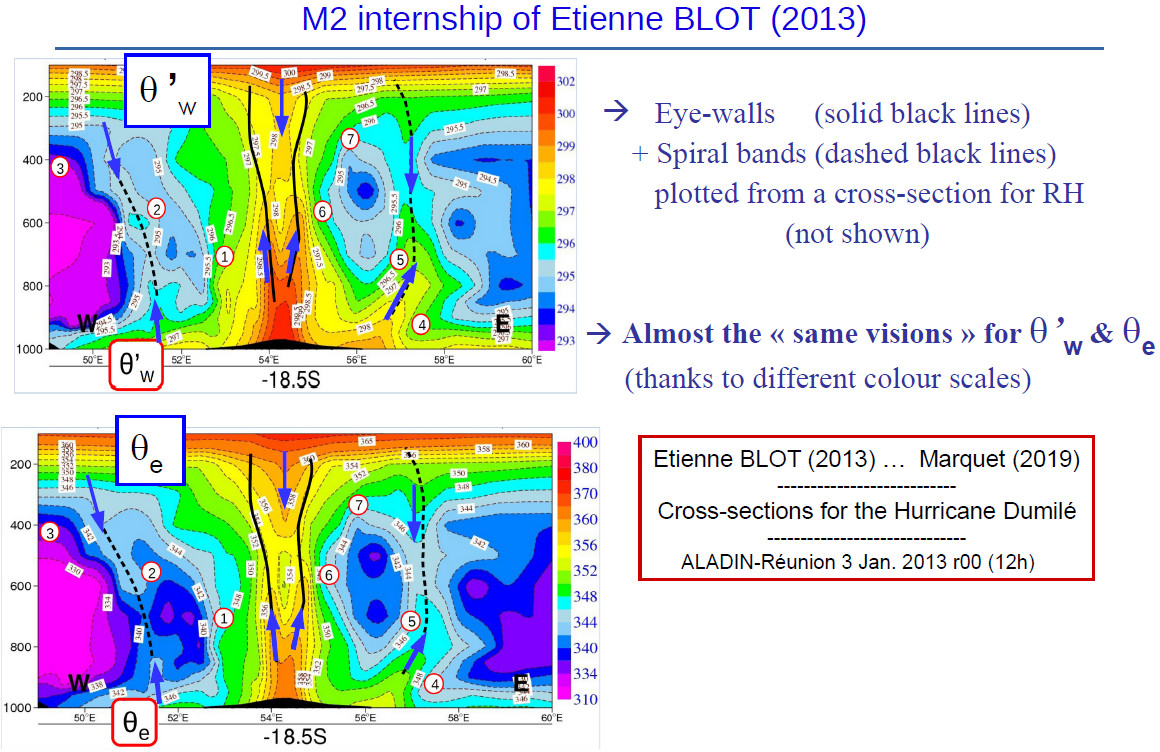}
\\
\vspace{-2mm}
---------------------------------------------------------------------------------------------------------
\\
\includegraphics[width=0.98\linewidth]{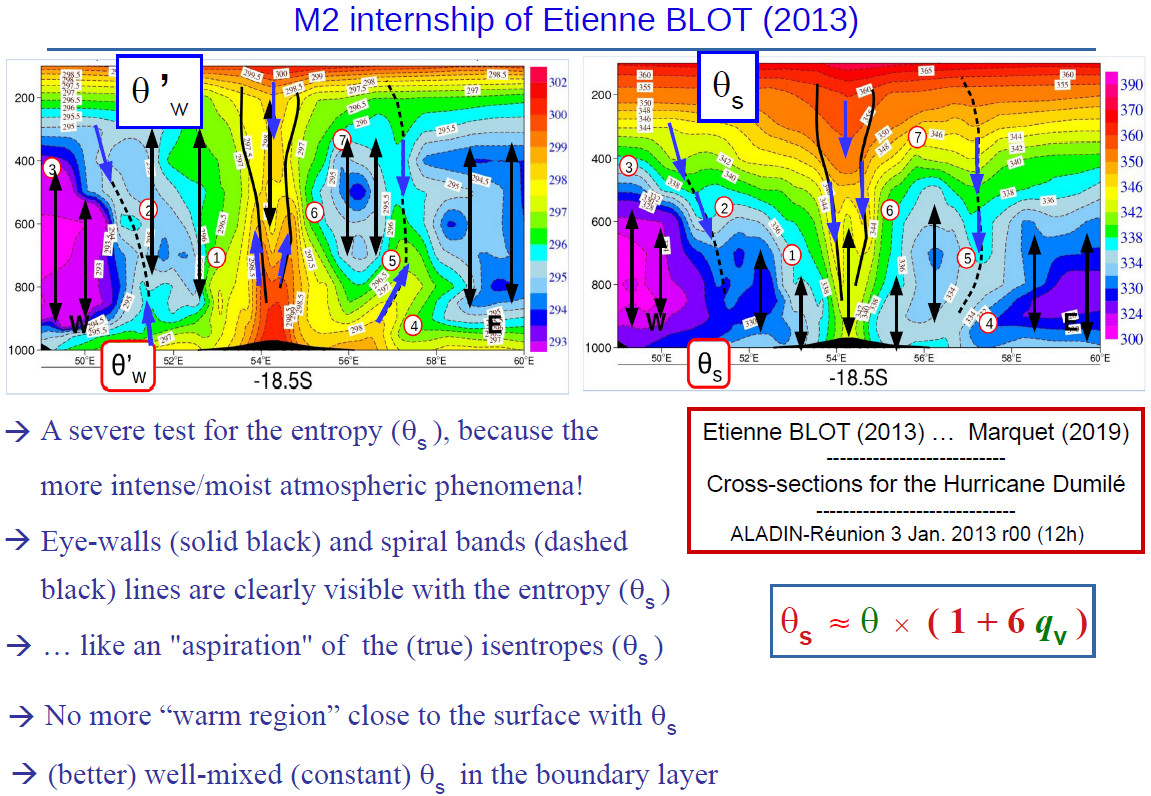}
\vspace{-2mm}
\caption{\small \it
From \cite{Marquet_2022} and \cite{Blot_2013}.
{\bf Top:}
the structure of a hurricane (here Dumil\'e, simulated by the Aladin $8$~km model) can be equally represented by the pseudo-adiabatic variables $\theta'_w$ and $\theta_e$ (thanks to different color scales).
{\bf Bottom:}
the moist-air entropy variable $\theta_s$ is as relevant as $\theta'_w$ and $\theta_s$ variable to see the core, the eye-walls and the spiral bands, with more homogeneous boundary layers and with the convective regions associated with some sort of ``aspiration of mid-troposphere moist-air isentropes'' labeled with $\theta_s$.
}
\label{fig_Blot_2013_Dumile}
\end{figure}
\clearpage

\begin{figure}[htb]
\centering
\vspace{-7mm}
\includegraphics[width=0.94\linewidth]{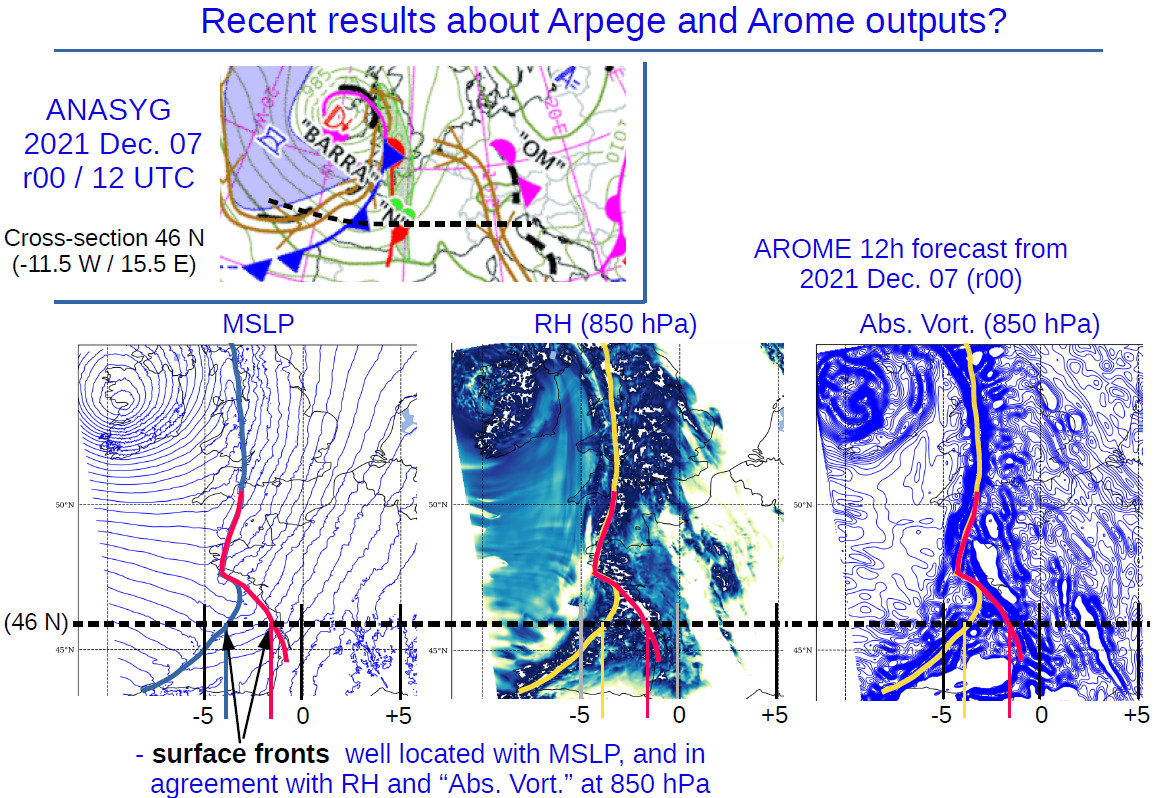}
\\
---------------------------------------------------------------------------------------------------------
\\
\includegraphics[width=0.94\linewidth]{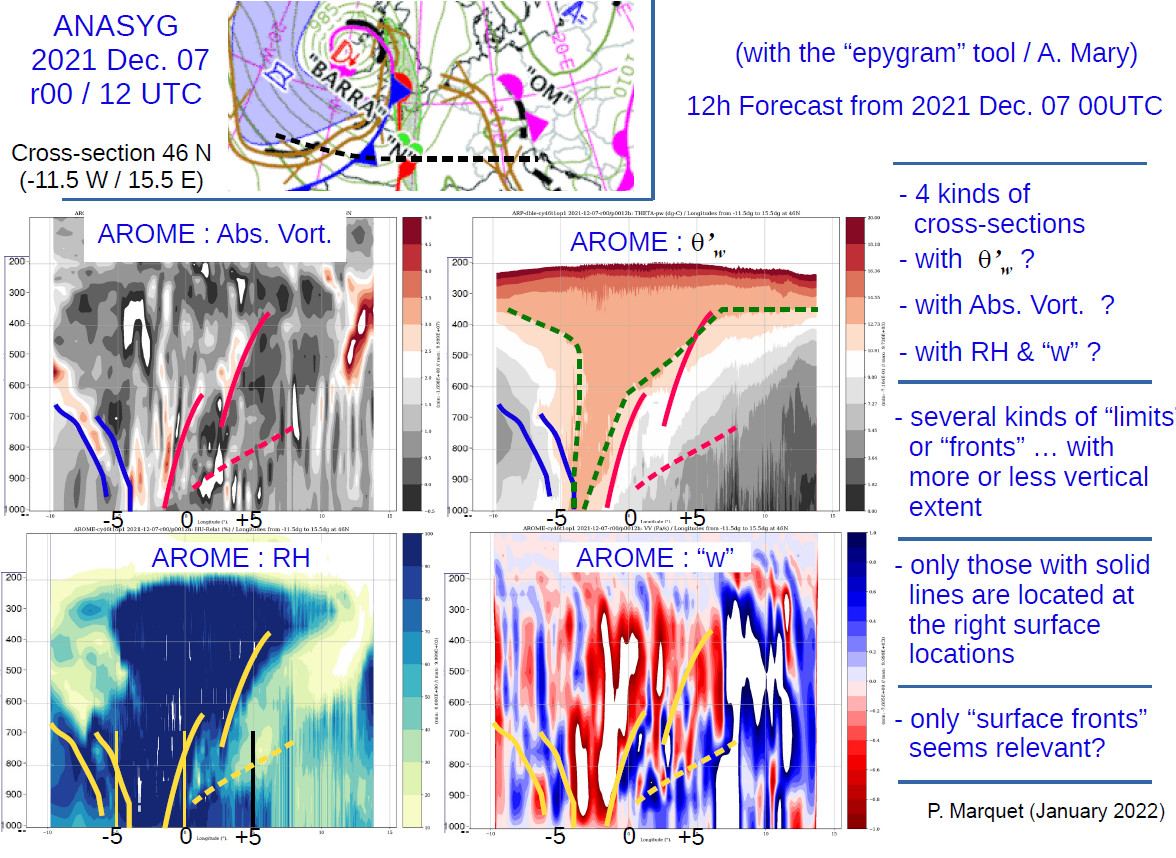}
\vspace{-2mm}
\caption{\small \it
From \cite{Marquet_2022}.
{\bf Top:}
The surface fronts for the storm ``Barra'' are clearly located with the MSLP simulated with the Arome $1.3$~km model.
{\bf Bottom:}
The cross-sections at $46$~N allow the plot of cold (solid blue) and warm (solid red) fronts up to $700$ to $400$~hPa.
}
\label{fig_Marquet_2022_MSLP_RH_TA_RH_W_Thetapw}
\end{figure}
\clearpage

\begin{figure}[htb]
\centering
\vspace{-8mm}
\includegraphics[width=0.92\linewidth]{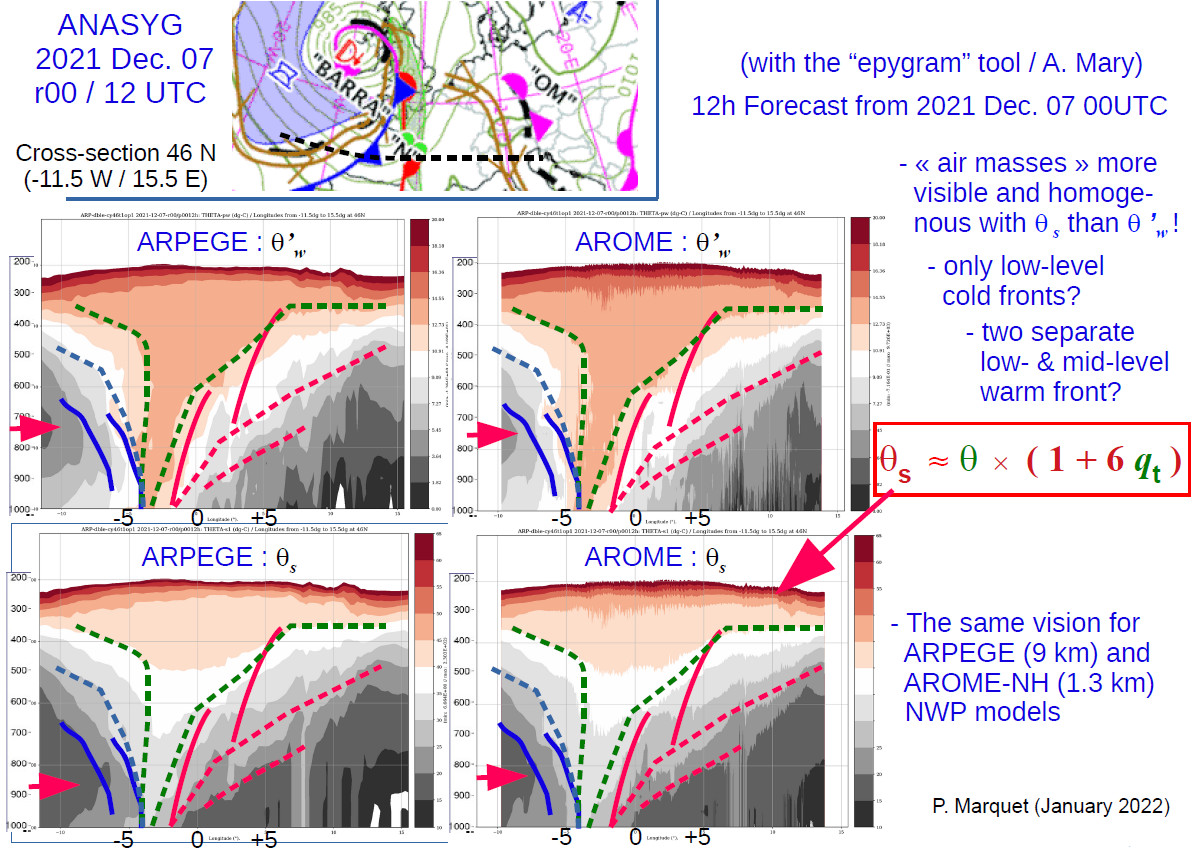}
\\
---------------------------------------------------------------------------------------------------------
\\
\includegraphics[width=0.92\linewidth]{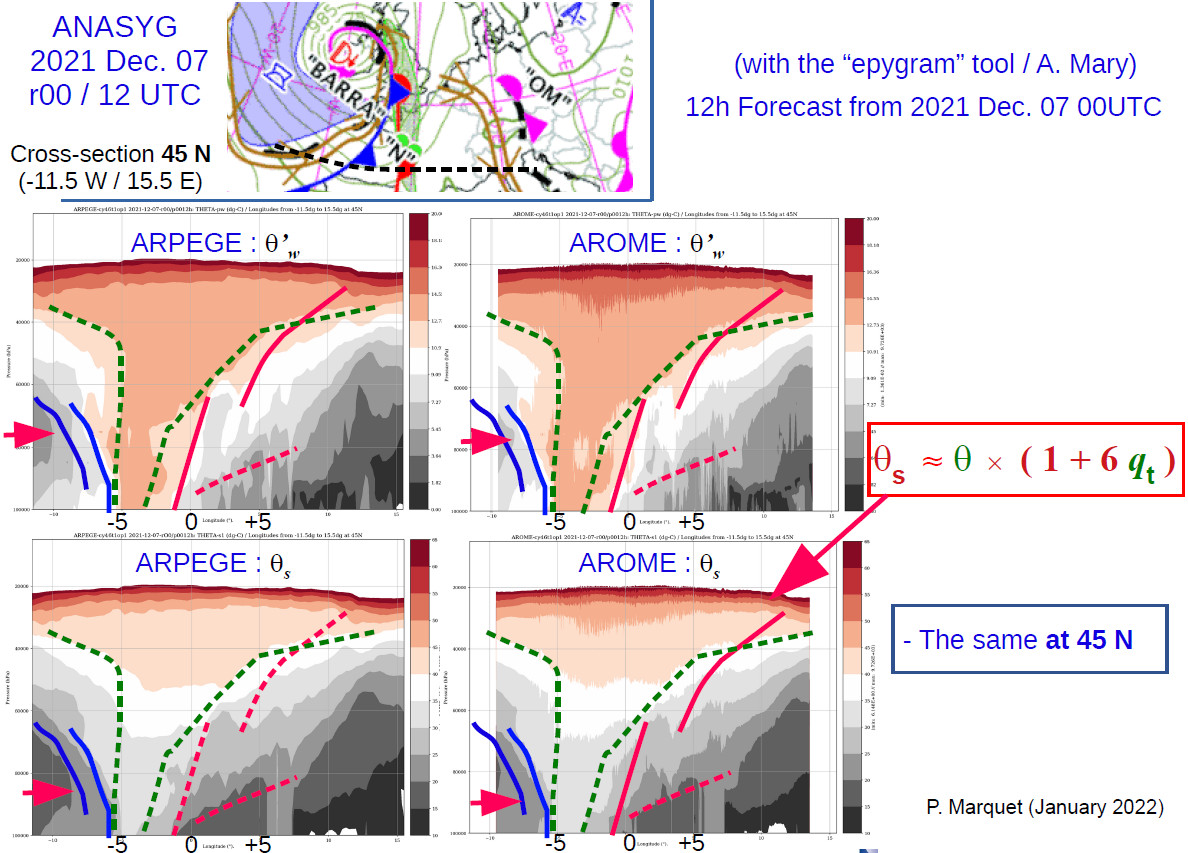}
\vspace{-2mm}
\caption{\small \it
From \cite{Marquet_2022}.
{\bf Top:}
The cold (solid blue) and warm (solid red) fronts seems to be better associated with ``air masses'' labeled with the moist-air entropy variable $\theta_s$ for the cross section at $46$~N (no more marked minimum at $850$ or $800$~hPa as for $\theta'_w$). 
This result is valid for the  Arome $1.3$~km model and the Arpege $8$~km models as well.
{\bf Bottom:}
The same is true for the cross section at $45$~N.
}
\label{fig_Marquet_2022_thetas}
\end{figure}
\clearpage

\begin{figure}[htb]
\centering
\vspace{-1mm}
\includegraphics[width=0.98\linewidth]{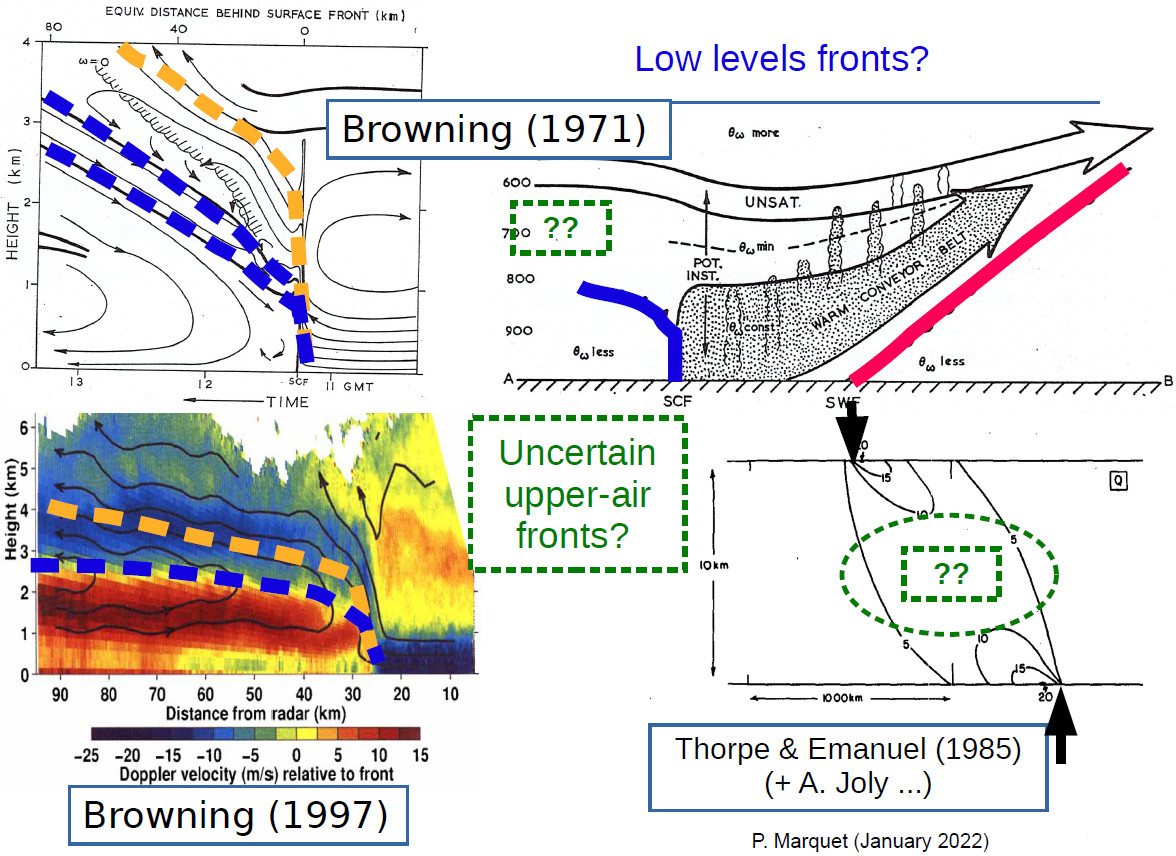}
\vspace{-2mm}
\caption{\small \it
From \cite{Marquet_2022}.
Historical examples justifying the plot of fronts with limited vertical extension on the previous cross-sections.
}
\label{fig_Marquet_2022_Browning}
\end{figure}
\clearpage

 \section{\underline{\large The dry- and moist-air potential vorticity formulations $PV(\theta)$}}
 \label{==Moist-PV==}
\vspace{-2mm}

The original formulation of ``potential vorticity'' by \cite{Ertel_1942a} 
\begin{align}
PV(\,\psi \,)   
 & = \, 
       \frac{1}{\rho}
       \: \; \boldsymbol{\zeta}_a\; . \: 
      \boldsymbol{\nabla}\!
      \left( \, \psi \, \right)
      \:
\end{align}
was suitable for any ``hydrodynamical invariant'' (i.e. ``conservative variable'') $\psi$,
with $\rho$ the density, $\boldsymbol{\zeta}_a$ the 3D absolute vorticity vector and $\boldsymbol{\nabla}$ the 3D gradient operator
\cite[see the English translations of all Ertel's papers in][]{Schubert_al_2004}.

Ertel suggested different kinds of such   ``conservative variable'' $\psi$:
\vspace*{-4mm}
\begin{itemize}
\item
the ``\underline{polytropic temperature} under polytropic conditions''
(for which the soleonidal term 
$N(p,\rho,\psi)$ cancels out), 
and thus the (dry-air) ``\underline{potential temperature}'' $\psi = \theta\,$, or the ``\underline{total water content}'' $\psi = q_t=q_v+q_l+q_i$ under conditions of no precipitation and without moisture input \citep{Ertel_1942a};
\vspace*{-3mm}
\item
the (dry-air) \underline{potential temperature} 
$\psi = \theta\,$, i.e. a special case of the 
\underline{polytropic temperature} 
$\psi = \Theta_x$ with
$\Theta_x = T \, (\rho_0/\rho)^{(1/x)}$
and with $x=c_v/(c_p-c_v)$
\citep{Ertel_1942b};
\vspace*{-3mm}
\item 
the (dry-air) \underline{polytropic temperature} or
the (dry-air) \underline{entropy} $\psi = s_d$
\citep{Ertel_1942c}, 
due to the \citet{Bauer_1908,Bauer_1910} relationship 
$s_d = c_{pd}\,\ln(\theta)+cste$, 
and thus with
$PV(s_d) =
PV[\,c_{pd}\,\ln(\theta)\,]
= (c_{pd}/\theta) \: PV(\theta)$
indeed
proportional to $PV(\theta)$.
\end{itemize}

After Ertel's papers, different dry- and moist-air ``(more or less) conservative'' quantities have been studied:
\vspace*{1mm}
\\
$\bullet$ $PV(\,\theta\,) $ in 
\cite{Hoskins_Bretherton_1972,
Hoskins_1974,
Hoskins_McIntyre_Robertson_1985},
\\ \hspace*{17mm}
\cite{Hoskins_Berrisford_1988,
Hoskins_1991,Hoskins_1997,Hoskins_2003},
\\ \hspace*{17mm}
 (...) \cite{Hoskins_2015};
\vspace*{1mm}
\\
$\bullet$ $PV(\,\theta'_w\,)$ in 
\cite{Bennetts_Hoskins_1979};
\vspace*{1mm}
\\
$\bullet$ $PV(\,\theta_{e}\,)$ in
\cite{Rotunno_Klemp_1985},
\cite{Cao_Cho_1995},
\cite{Persson_1995},
\\ \hspace*{19mm}
\cite{Gao_Lei_Zhou_2002},
\cite{Deng_Gao_2009};
\vspace*{1mm}
\\
$\bullet$ $PV(\,\theta_{es}\,)$ in 
\cite{Emanuel_al_1987};
\vspace*{1mm}
\\
$\bullet$ $PV(\,\theta_s\,) $ 
 in \cite{Soriano_Garcia_Diez_1997} /
for $\theta_s$ defined in \cite{Hauf_Hoeller_1987};
\vspace*{1mm}
\\
$\bullet$ $PV(\,\theta_{v}\,)$  in
\cite{Schubert_al_2001} 
and
\cite{Schubert_2004};
\vspace*{1mm}
\\
$\bullet$ $PV(q_v)$, $PV(\theta)$ and $PV(\theta_e)$  in \cite{Gao_Zhou_2008};
\vspace*{1mm}
\\
$\bullet$ $PV(\,\theta_{s}\,)$,
$PV(\,\theta_{v}\,)$ and
$PV(\,q_t\,)$ in
\cite{Marquet_QJ_2014} /
for $\theta_s$ defined in \cite{Marquet_QJ_2011}.


\newpage
 \section{\underline{\large Comparison of dry and moist-air Potential Vorticity}}
 \label{==Compare_PV==}
\begin{figure}[htb]
\vspace{-9mm}
\centering
\vspace{0mm}
\includegraphics[width=0.6\linewidth]{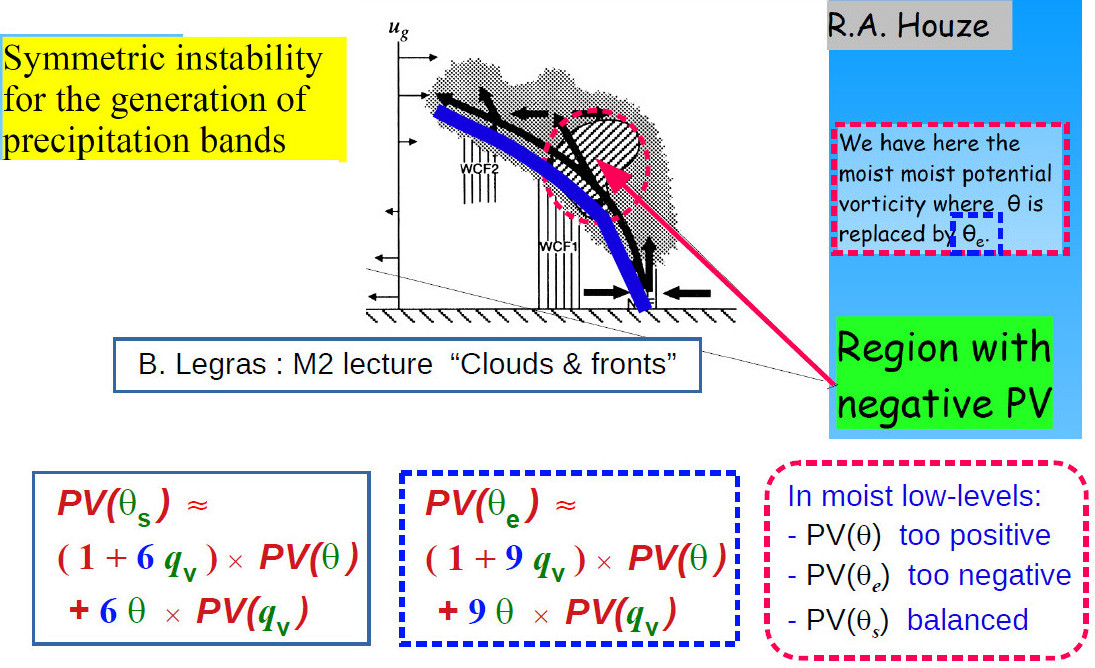}
\\
\includegraphics[width=0.92\linewidth]{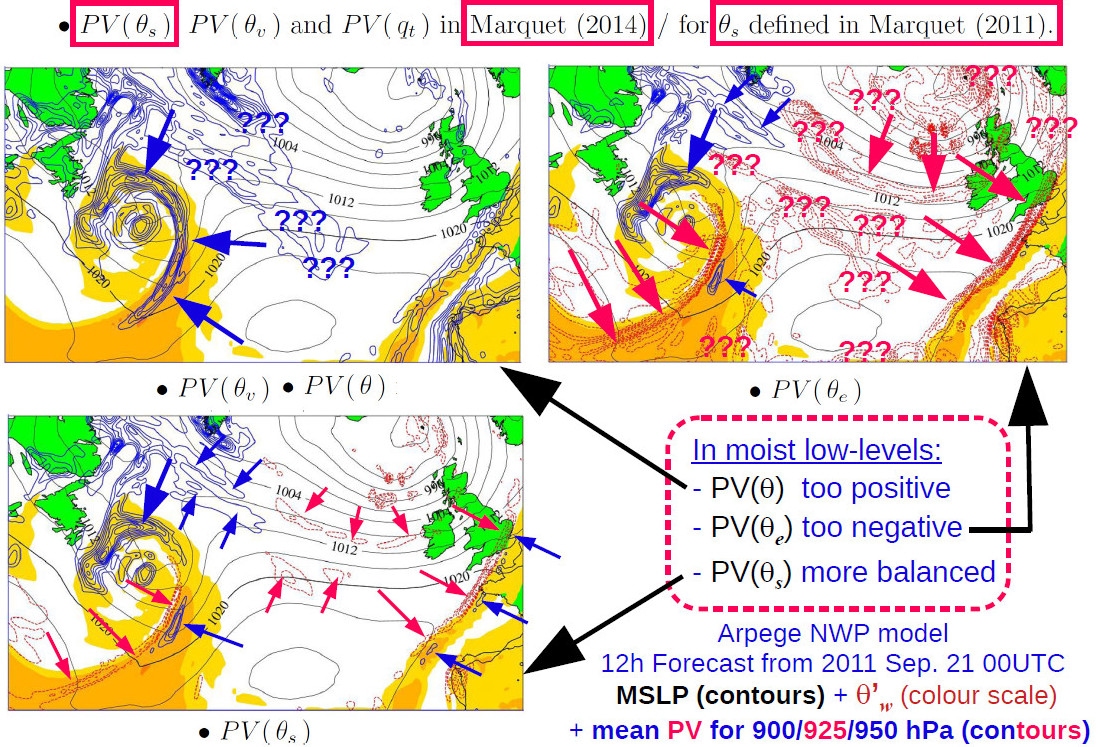}
\vspace{-4mm}
\caption{\small \it
From \cite{Marquet_2022}.
{\bf Top:}
It is recalled in this figure of R.A. Houze that regions with negative values of $PV(\theta_e)$ are associated with symmetric instability and slantwise convection.
The first-order formulations for $PV(\theta_s)$ and $PV(\theta_e)$ show that the impact of the moisture part $PV(q_v)$ in $PV(\theta_s)$ is about $2/3$ of the moisture part in $PV(\theta_e)$.
{\bf Bottom} \citep[from][]{Marquet_QJ_2014}:
Average low-level values seems 
too positive for $PV(\theta)\approx PV(\theta_v)$, 
too negative for $PV(\theta_e)$ and
well-balanced for $PV(\theta_s)$.
}
\label{fig_Sym_Instab_Marquet_2014_PV_PVs}
\end{figure}
\clearpage

\begin{figure}[htb]
\centering
\vspace{-9mm}
\includegraphics[width=0.9\linewidth]{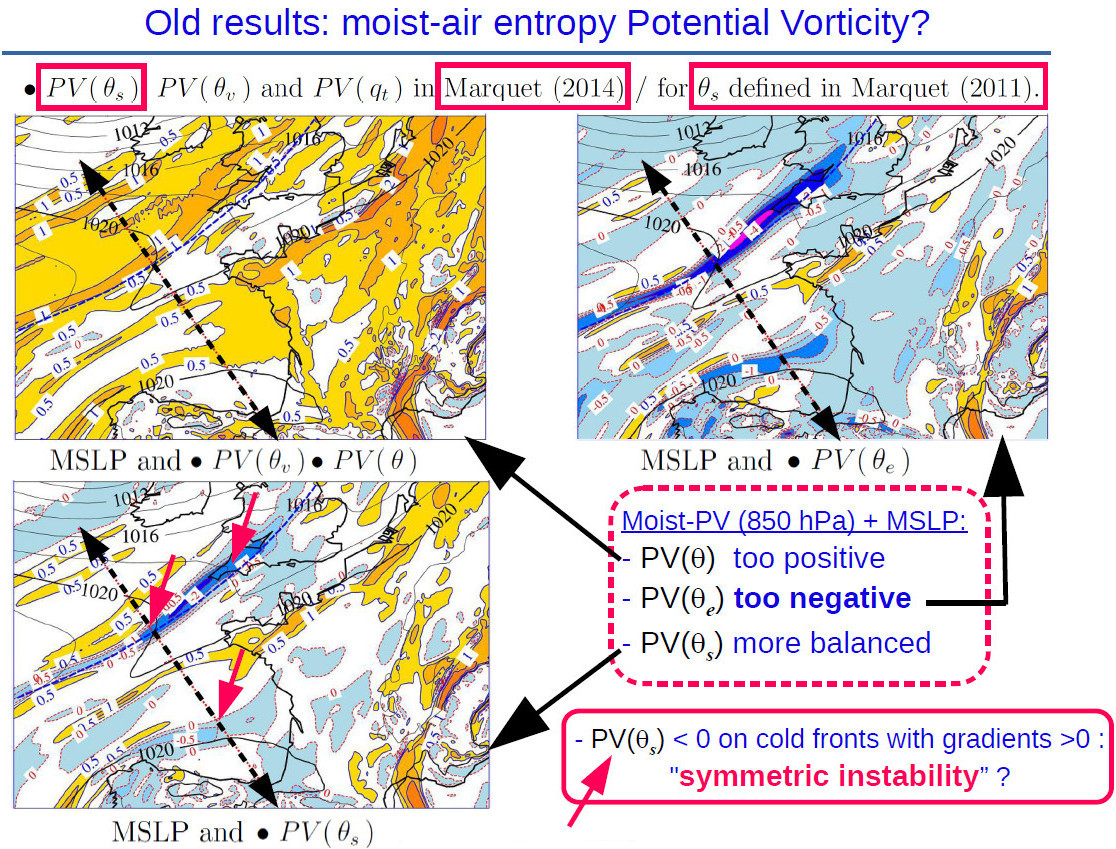}
\\
\vspace{-2mm}
-----------------------------------------------------------------------------------------------------------------
\\
\includegraphics[width=0.9\linewidth]{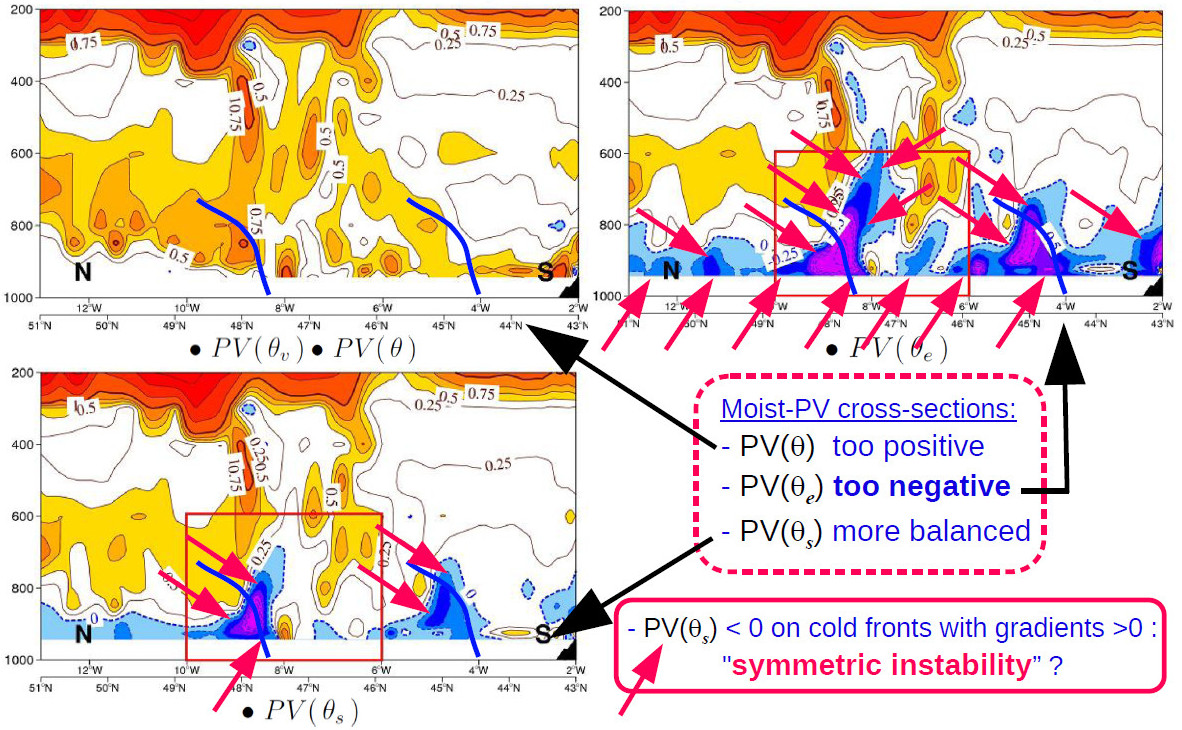}
\vspace{-3mm}
\caption{\small \it
From \cite{Marquet_2022} and \cite{Marquet_QJ_2014}.
{\bf Top:}
the low-level ($\,850$~hPa) cold fronts seems to be associated with the negative values of $PV(\theta_s)$, whereas $PV(\theta)$ is positive everywhere and $PV(\theta_e)$ negative almost everywhere.
{\bf Bottom:}
the negative values of $PV(\theta)$ seems clearly correspond to (are located with) the cold fronts in these cross-sections (same fronts as in Fig.~\ref{fig_Blot13_Marquet14}), and possibly to symmetric instability and slantwise convection.
On the one hand there is no negative values for $PV(\theta)$, whereas on the other hand there are negative values of $PV(\theta_e)$ almost everywhere in the low-level boundary  layer.
}
\label{fig_Marquet_2014_MSLP_PV850_Cross}
\end{figure}
\clearpage

\begin{figure}[htb]
\centering
\vspace{-15mm}
\includegraphics[width=0.98\linewidth]{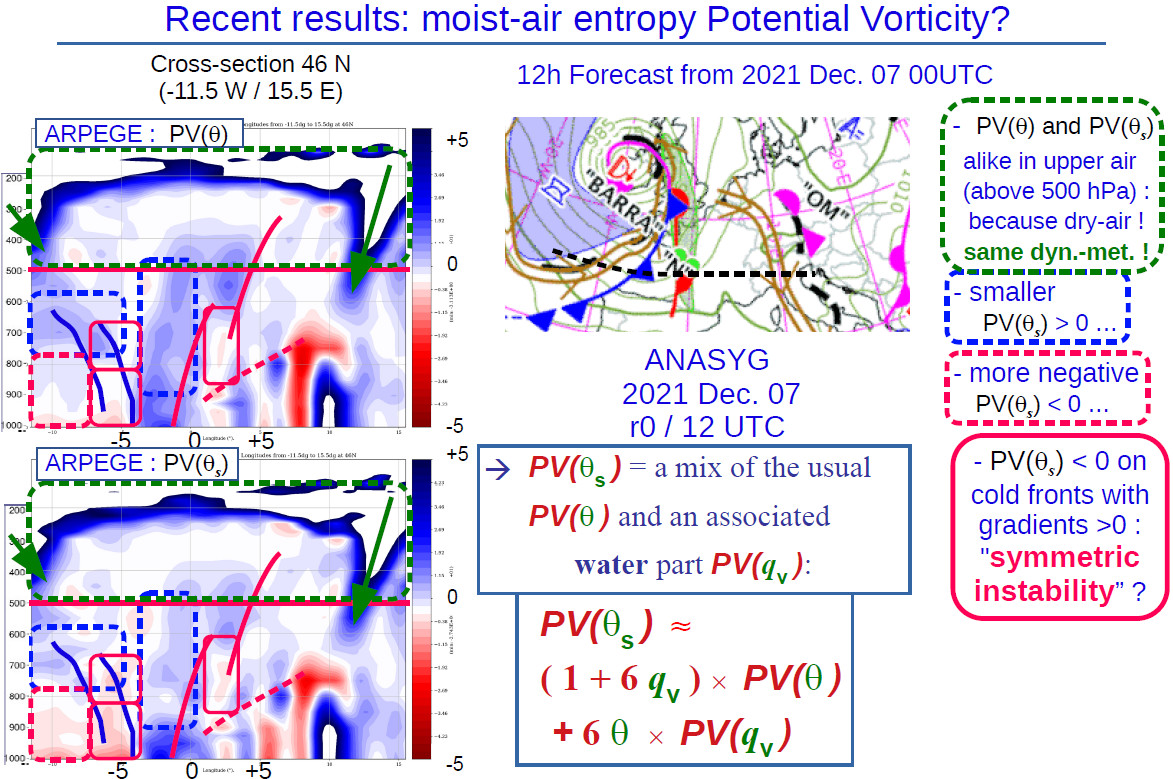}
\\
--------------------------------------------------------------------------------------------------------------------
\\
\includegraphics[width=0.98\linewidth]{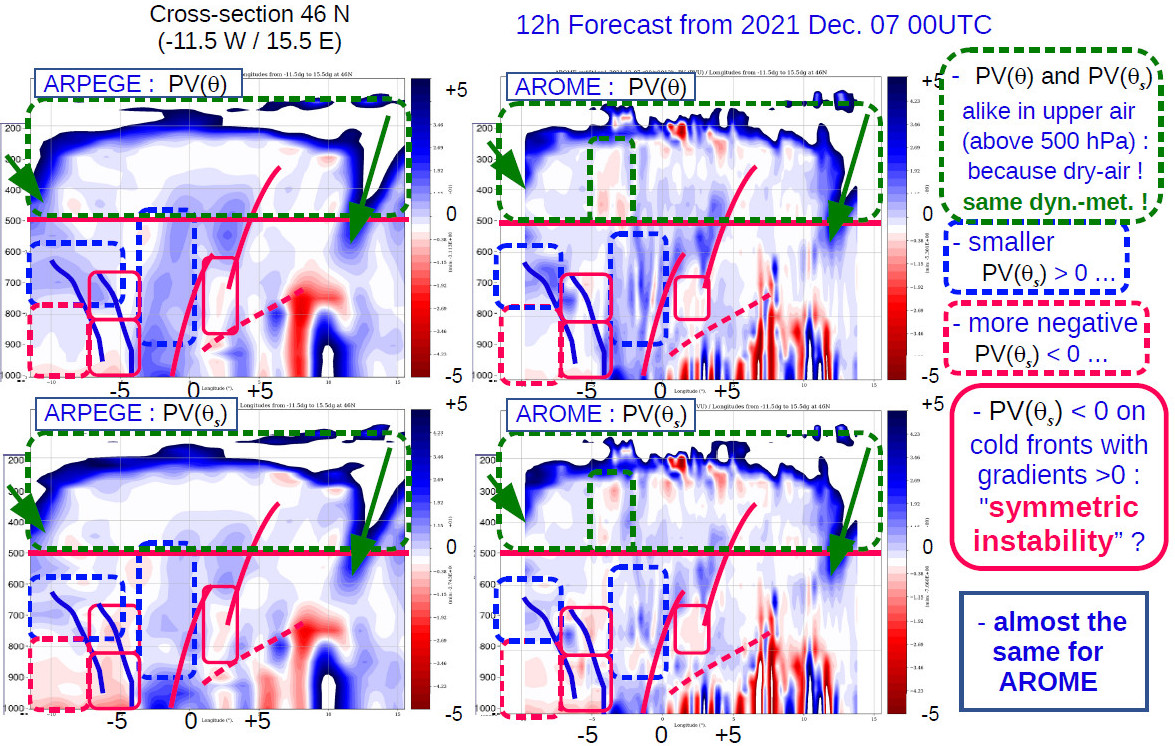}
\vspace{-2mm}
\caption{\small \it
From \cite{Marquet_2022}.
{\bf Top:}
The potential vorticity $PV(\theta)$ and $PV(\theta_s)$ for the storm ``Barra'' simulated with the global Arpege NWP ($\,8$~km) model (cross-sections at $46$~N).
Some negative values of the moist-air entropy variable $PV(\theta_s)$ appears close to the cold front (they already exist and are the same between the warm fronts), with almost unchanged upper-levels dynamic-meteorology signals above $500$~hPa in the dry regions (tropospheric intrusion of high stratospheric values).
{\bf Bottom:}
The same is true for the Arome $1.3$~km (NH) model.
}
\label{fig_Marquet_2022_PV_PVs}
\end{figure}
\clearpage

\begin{figure}[htb]
\centering
\vspace{-5mm}
\includegraphics[width=0.98\linewidth]{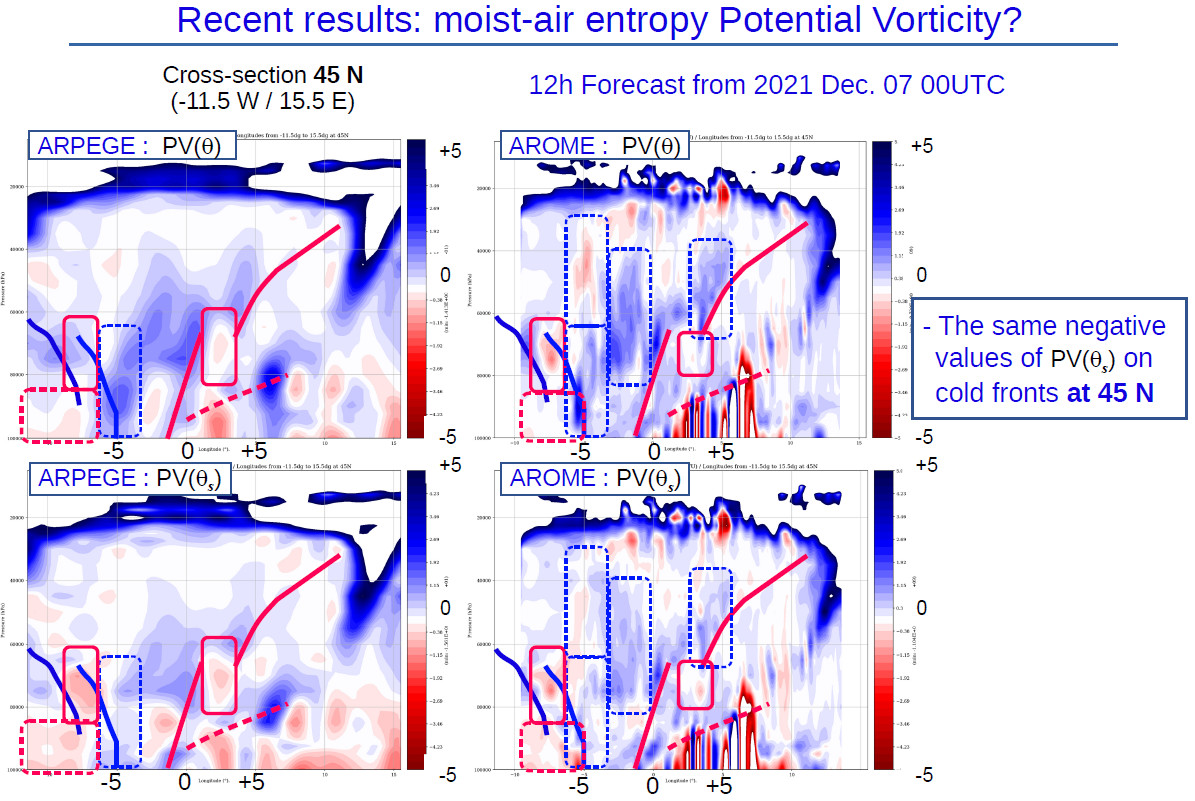}
\vspace{-2mm}
\caption{\small \it
From \cite{Marquet_2022}.
Same results as in Fig~\ref{fig_Marquet_2022_PV_PVs} but at $45$~N.
}
\label{fig_Marquet_2022_PV_PVs_45N}
\end{figure}

 \section{\underline{\large Conclusions}}
 \label{==Conclusions==}
\vspace{-2mm}

This note is a mere synthesis of the most important results I have discovered since 2009, with the aim to mainly show Figures instead of formulas (which can be found in my papers published since 2011 and available on arXiv). 

However, I am now I am in retirement (since May, 2022) from the CNRM and M\'et\'eo-France, and I am preparing a (huge) book showing in detail why it is needed, and how it is possible to defined and compute the third-law (absolute) definitions of not only the moist-air entropy for the atmosphere, but also the sea-salt entropy for the ocean.

In this book I will recall (and show) almost all the bibliography I have studied since 1988 about the atmosphere and ocean thermodynamics (even though many of them, but not all of them, have been recalled in my papers published between 2011  and 2022).

\clearpage


\newpage
\bibliographystyle{ametsoc2014}
\bibliography{Marquet_Thetas_PV_Thetas_v1}

 \end{document}